\documentclass[twocolumn,aps,superscriptaddress,showpacs,floatfix,prc]{revtex4-1}
\usepackage{mathrsfs}
\usepackage{amssymb}
\usepackage{amsmath}
\usepackage{graphicx}
\usepackage[normalem]{ulem}
\usepackage[dvips]{color}
\usepackage{bm}
\usepackage{longtable}\usepackage{slashed}

\renewcommand\theequation{\arabic{section}.\arabic{equation}}

\usepackage{times}
\usepackage{fouriernc}

\renewcommand\sout{\bgroup \color{red} \ULdepth=-.5ex \ULset}

\renewcommand{\v}[1]{\textbf{#1}}
\renewcommand{\rm}[1]{\textrm{#1}}
\renewcommand{\d}{\mathrm{d}}

\begin{document}

\title{Nuclear Equation of State and Single-nucleon Potential from Gogny-like Energy Density Functionals Encapsulating Effects of Nucleon-nucleon Short-range Correlations}

\author{Bao-Jun Cai\footnote{bjcai87@gmail.com}}
\affiliation{Quantum Machine Learning Laboratory, Shadow Creator Inc., Shanghai 201208, China} 
\author{Bao-An Li\footnote{Bao-An.Li$@$tamuc.edu}}
\affiliation{Department of Physics and Astronomy, Texas A$\&$M
University-Commerce, Commerce, TX 75429-3011, USA}
\date{\today}

\begin{abstract}
It is well known that nucleon-nucleon short-range correlations (SRCs) induce a high momentum tail (HMT) in the single-nucleon momentum distribution function $n_{\v{k}}^J(\rho,\delta)$ in cold neutron-rich matter of density $\rho$ and isospin asymmetry $\delta$. While there are clear experimental evidences that the SRC/HMT effects are different for neutrons and protons and their strengths depend strongly on the isospin asymmetry of finite nuclei mostly based on electron-nucleus scattering experiments, much less is known experimentally about the SRC/HMT effects in dense neutron-rich matter. Predictions based on microscopic nuclear many-body theories are still rather model dependent as the spin-isospin dependence of tensor forces and properties of the repulsive core thought to be responsible for the SRC/HMT are still poorly known. To facilitate further explorations of SRC/HMT effects in dense neutron-rich matter especially with heavy-ion reactions involving high-energy radioactive beams as well as multimessenger observations of neutron stars and their mergers, by incorporating the SRC-induced HMT in $n_{\v{k}}^J(\rho,\delta)$ into a Gogny-like energy density functional we study SRC/HMT effects on the equation of state (EOS) especially its symmetry energy term and single-nucleon potential in dense asymmetric nucleonic matter (ANM). Using a parametrization as a surrogate for the momentum-dependent kernel in the Gogny-like energy density functional (EDF) we derive analytical expressions for all components of the ANM EOS and their characteristics (e.g., magnitude, slope and curvature as well as nucleon effective mass) at saturation density $\rho_0$ as well as the momentum-dependent single-nucleon optical potential in neutron-rich matter using parameters characterizing nuclear interactions as well as the size, shape and isospin dependence of the HMT at $\rho_0$. Available constraints on the EOS of symmetric nuclear matter (SNM), symmetry energy, nucleon optical potential and SRC/HMT all around $\rho_0$ are used to fix or limit the parameters involved. We found that the SRC/HMT enhances the kinetic EOS but reduces the single-nucleon potential in SNM. It also reduces the kinetic symmetry energy but enhances the momentum-dependent part of symmetry/isovector potential while the total symmetry energy at supra-saturation densities is strongly softened. We also examined consequences of these effects on the density profile of proton fraction, crust-core transition density and pressure as well as the mass-radius correlation of neutron stars while their effects on heavy-ion reactions induced by high-energy radioactive beams are left for future studies.
\end{abstract}

\maketitle


\setcounter{equation}{0}
\section{Introduction}

The momentum-dependence of single-nucleon mean-field potentials is one of the fundamental features of finite-range nucleon-nucleon interactions in nuclear matter.
A moving nucleon experiences different potentials from its surroundings compared to static ones, either under the mean fields or those considering correlations.
Momentum-dependent nuclear potentials in fact play very important roles in heavy-ion collisions\,\cite{Gale1987,Bertsch1988,Pra88,Welke,Gale90,Aichelin1990,Pan,Zhang94,Hart,Pawel,Das2003,LiBA04}.
The mean-field potential $U_J$ (with the subscript $J$ denoting protons (p) or neutrons (n)) of a nucleon with momentum $\v{k}$ at zero temperature is generally a function of neutron density $\rho_{\rm{n}}$, the proton density $\rho_{\rm{p}}$, i.e., $U_J=U_J(\rho_{\rm{n}},\rho_{\rm{p}},|\v{k}|)$.
It can be written equivalently in the form of $U_J(\rho,\delta,|\v{k}|)$ with $\rho=\rho_{\rm{n}}+\rho_{\rm{p}}$ the total density and $\delta=(\rho_{\rm{n}}-\rho_{\rm{p}})/\rho$ the isospin asymmetry of the nucleon system under consideration.
The equation of state (EOS) of cold asymmetric nucleonic matter (ANM) can be conveniently denoted by the nucleon specific energy $E(\rho,\delta)$. The latter could be obtained by certain integration of $U_J(\rho_{\rm{n}},\rho_{\rm{p}},|\v{k}|)$ over $\v{k}$, i.e., $\sim\int^{\v{k}}U_J(\rho,\delta,|\v{k}|)\d\v{k}+\rm{``kinetic\;term''}\to E(\rho,\delta)$.
Effects of the momentum $\v{k}$ dependence of the single-nucleon potential on the EOS are implicit as the momentum is integrated out.
Connecting the momentum-dependence and the correlations/fluctuations of the interactions/potentials with the EOS as well as understanding how they influence each other are fundamentally important and interesting for resolving many related issues in both nuclear physics and astrophysics.

The EOS $E(\rho,\delta)$ of ANM can conventionally be expanded around the symmetric nuclear matter (SNM) with $\delta=0$, leading to the approximation,
\begin{equation}
E(\rho,\delta)\approx E_0(\rho)+E_{\rm{sym}}(\rho)\delta^2+E_{\rm{sym},4}(\rho)\delta^4+\mathcal{O}(\delta^6),
\end{equation}
here $E_{\rm{sym}}(\rho)=2^{-1}\partial^2E(\rho,\delta)/\partial\delta^2|_{\delta=0}$ is the conventional nuclear symmetry energy besides the EOS $E_0(\rho)\equiv E(\rho,0)$ of SNM,  $E_{\rm{sym},4}(\rho)\equiv 24^{-1}\partial^4E(\rho,\delta)/\partial\delta^4|_{\delta=0}$ is the fourth-order symmetry energy, etc. Owing to its fundamental importance in nuclear
physics\,\cite{LiBA98,Dan02,Bar05,Ste05,Che07a,LCK08,Tsa12,Che14}
and astrophysics\,\cite{Gle00,Lat04,Lat12,Lat14,Oze16,Oer17}, 
understanding nuclear symmetry energy $E_{\rm{sym}}(\rho)$,
especially its density dependence (both at sub- and supra-saturation densities), is currently one of the most
important goals in contemporary nuclear astrophysics\,\cite{Hor14,EPJA}. 

Thanks to the significant efforts made by many people in the nuclear astrophysics community during
the last two decades, much knowledge on the symmetry energy around the saturation density
$\rho_0\approx0.16\,\rm{fm}^{-3}$ has been obtained. However, the density dependence of
$E_{\rm{sym}}(\rho)$ especially at supra-saturation densities is still
poorly known. In particular, several nuclear structure observables are known as sensitive probes of symmetry energy at sub-saturation
densities, e.g., the correlation between the slope parameter $L(\rho)$ of the symmetry energy and the
neutron skin thickness of heavy nuclei\,\cite{Cen09,Zha13} as well
as the correlation between $L(\rho)$ and the electric dipole $\alpha_{\rm{D}}$
polarizability in $^{208}\rm{Pb}$\,\cite{Maz13,Zha14,Zha15}; the
correlation between the isobaric analog states (IAS) and the
symmetry energy\,\cite{Dan14}, to name a few. While analyses of neutron star observables (mostly the radii and tidal polarizability) especially since GW170817 
have provided strong constraints on the symmetry energy around $1\rho_0\mbox{$\sim$}2\rho_0$\,\cite{Ste10,LiBA21,Huth22}, the situation at higher densities is unclear\,\cite{ZL21,Fr22} from studying the mass and radius data of the currently known most massive neutron star 
PSR J0740+6620 from the NICER Collaboration\,\cite{Miller21,Riley21}. On the other hand, intermediate-relativistic energy heavy-ion collisions (HICs) have been playing a significant role in determining the symmetry energy from sub-saturation to supra-saturation densities, see, e.g., Refs.\,\cite{LCK08,Tsa12,Xia09}. Unfortunately, various analyses of heavy-ion reaction observables based on transport model simulations have not reached a firm conclusion regrading the high-density behavior of nuclear symmetry energy, 
for a recent review, see e.g., Ref.\,\cite{EPJA-review}. It is mostly because of the still existing relatively large model dependences in simulating heavy-ion collisions especially above the pion production threshold\,\cite{Xu16,Zhang18,Ono19,Colonna21,TMEP}.  The single-nucleon potential especially its momentum dependence in dense neutron-rich matter is among the most uncertain but very critical physics inputs of transport models. It is known to have significant impact on observables of heavy-ion reactions\,\cite{LiBA04,LiChen05,Cozma17,WangLi,Cozma21}. 

The symmetry energy can conventionally be decomposed into a kinetic and a potential part in the quasi-particle picture, i.e.,
$E_{\rm{sym}}(\rho)=E_{\rm{sym}}^{\rm{kin}}(\rho)+E_{\rm{sym}}^{\rm{pot}}(\rho)$.
Each individual term of this decomposition may have its own distinct and important physical ramifications on different problems. For instance, the critical
formation density of the charged $\Delta(1232)$ states in neutron
stars depends on $E_{\rm{sym}}^{\rm{kin}}(\rho)$ and
$E_{\rm{sym}}^{\rm{pot}}(\rho)$ separately\,\cite{Cai15a}; while on
the other hand, the potential part $E_{\rm{sym}}^{\rm{pot}}(\rho)$
is extremely relevant to certain dynamical processes and observables in
HICs\,\cite{Li15,Li15a,Hen15b,Yon17}. Thus it is not only important
to explore the density dependence of the total symmetry energy, but also its
individual terms and their physical origins. Since the total symmetry energy at the saturation density $\rho_0$, as mentioned above, is relatively well constrained\,\cite{LiBA13}, the decomposition of symmetry energy has to satisfy an asymptotic sum rule at $\rho_0$. In the traditional free Fermi gas (FFG) model of nuclear matter,
where the single-nucleon momentum distribution function
$n_{\v{k}}^J$ takes the form of a step function, the kinetic
symmetry energy is simply
$E_{\rm{sym}}^{\rm{kin}}(\rho)=k_{\rm{F}}^2/6M$ with $k_{\rm{F}}$ being
the nucleon Fermi momentum in SNM, and $M$ is the average static mass of nucleons. It will be used as a reference of kinetic symmetry energy in our following discussions.
In this work, we use the abbreviation ``FFG'' (``HMT'') to denote the calculation/model in which the momentum distribution $n_{\v{k}}^J$ is a step function (the function containing the HMT), even in cases where a nuclear potential is included.

In most calculations of cold nuclear matter EOSs especially within phenomenological models and/or input EOSs as well as initializations of transport model simulations of nuclear reactions, often a step function is used for the $n_{\v{k}}^J$ as if the system is a FFG at zero temperature. However, this assumption has to be modified if one considers the HMT induced by the SRC in nuclear matter. Moreover, this modification is known to have significant ramifications on the EOS especially the kinetic symmetry  energy\,\cite{CXu10,Lee11,Lee14,Wang12}. 
In a more realistic picture of nuclear matter, the form of the
$n_{\v{k}}^J$ should be different from the step function, owing to, e.g., the (poorly known) spin-isospin dependence of the three-nucleon forces and/or the isospin dependence of SRC induced by the tensor forces and/or the hard repulsive core. 
Indeed, It has been known for a long time that the SRC leads to a high (low)
momentum tail (depletion) in the single-nucleon momentum
distribution above (below) the nucleon Fermi surface, in cold
nucleonic matter\,\cite{Bethe,Ant88,Arr12,Cio15}. Significant
efforts have been made in the past
years\,\cite{Wei15,Wei15a,Cruz2018,Weiss2019-1,Weiss2019-2,Weiss2021,Hen14,Hen15,Col15,Egi06,kk1,kk2,kk3,kk4,kk5,
Hen2017RMP,Duer2018,Schmookler2019,Schmidt2020,Li22} to constrain the
(isospin-dependent) parameters characterizing the SRC-modified
$n_{\v{k}}^J$ in neutron-rich matter using both
experimental data and microscopic model calculations. For example,
it has been found via analyzing electron-nucleus scattering data
that the percentage of nucleons in the HMT above the Fermi surface is as high as about 28\% in SNM but
decreases gradually to about only 1\% in a pure neutron matter
(PNM)\,\cite{Hen14,Hen15}, indicating a totally different momentum
behavior of the $n_{\v{k}}^J$ due to the strong isospin dependence of SRC. 

Despite of the long history and impressive progresses especially in recent years in studying various features and impacts of SRC/HMT, many interesting issues regarding the size, shape, density and isospin dependence of SRC/HMT remain to be clearly resolved. 
For example, although various many-body theories
have consistently predicted effects of the SRC on the HMT
qualitatively consistent with the experimental findings, the
predicted size of the HMT still depends on the model and interaction
used. For instance, the self-consistent Green's function (SCGF)
theory adopting the AV18 interaction predicts a 11\%$\sim$13\% HMT for
SNM at saturation density\,\cite{Rio09,Rio14}. Moreover, the latest
Bruckner-Hartree-Fock (BHF) calculations predict a HMT ranging from
about 10\% via the next-next-next-leading order 450 (N$^3$LO450)
to over 20\% using several microscopic interactions\,\cite{ZHLi},
the latest VMC calculations for $^{12}$C gives a 21\%
HMT\,\cite{VMC,Wir16}, etc. While based on the observation that the SRC
between a neutron-proton pair is about 18$\sim$20 times that of two
protons, the HMT in PNM was estimated to be about
1\%$\sim$2\%\,\cite{Hen15b}. Furthermore, some recent calculations
indicate a significantly higher HMT in PNM, e.g., the SCGF
predicted a 4\%$\sim$5\% HMT in PNM\,\cite{Rio09,Rio14}. 

The uncertain properties of SRC/HMT in cold dense neutron-rich matter may bring large uncertainties to both the kinetic and potential parts of the symmetry energy as well as the single-nucleon potentials. 
For example, because of the momentum-squared weighting in
calculating the average nucleon kinetic energy, the strong isospin
dependence of the HMT makes the kinetic symmetry energy dramatically
different from the FFG model
prediction\,\cite{Cai16c,Hen15b,CXu11,CXu13,Vid11,Lov11,Car12,Rio14,Car14,Cai15}.
Specifically, the kinetic symmetry energy is significantly reduced even
to negative values in some model studies. Roughly speaking, nuclear symmetry
energy is the energy difference between PNM and SNM within the parabolic
approximation of the total ANM EOS. It is also known that the SRC is dominated by the isosinglet neutron-proton tensor interaction. It thus increases significantly the average kinetic energy per nucleon
in SNM but has little effect on that in PNM, leading to a reduction
of the kinetic symmetry energy. Consequently, the potential part of the $E_{\rm{sym}}(\rho)$, i.e., $E_{\rm{sym}}^{\rm{pot}}(\rho)$, should be modified correspondingly to satisfy the sum rule of symmetry energy at $\rho_0$. Moreover, the 
SRC/HMT may affect directly the potential symmetry energy if nuclear interactions are momentum dependent. Of course, it is also expected to affect the momentum dependence of both isoscalar and isovector parts of the single-nucleon potential and the associated effective masses of nucleons in dense neutron-rich matter.

Interestingly, SRC/HMT effects on the ANM EOS and some properties of neutron stars have been studied recently within relativistic mean field (RMF) models by several groups\,\cite{Cai16b,Lou22a,Lu2022,Bur22,Hong22,Lou22,Lu21,Souza20}. 
Most of these studies were carried out using nonlinear RMF models where the interaction is momentum independent. Given the results of these studies and those mentioned earlier, there still exist many interesting/important questions to be  investigated/answered to deepen our understanding of the SRC physics and its ramifications, e.g.,
\begin{enumerate}
\item[(a)]How does the SRC/HMT affect the single-nucleon potential in neutron-rich matter, especially the momentum dependence of its isoscalar and isovector parts, in comparisons with existing experimental observations of nucleon optical potentials and the associated nucleon effective masses at $\rho_0$? Answers to this question are potentially important for modeling HICs involving high-energy rare isotope beams.
\item[(b)]How does the SRC/HMT affect the ANM EOS (including both the SNM EOS and the symmetry energy) qualitatively and quantitatively, if a non-relativistic potential/interaction is adopted? This question is relevant since many such kind of interactions and energy density functionals are widely used successfully in studying nuclear structures, calculating the EOS for astrophysical purposes and/or simulating heavy-ion reactions.
\item[(c)] Due to the SRC/HMT induced sizable enhancement of SNM pressure and the reduction of total symmetry energy in nonlinear RMF models, the maximum mass of cold neutron stars was found to increase, see, e.g., Ref.\,\cite{Cai16b}. One may then wonder whether this conclusion would still hold or not if a non-relativistic and/or momentum-dependent potential is used.
\end{enumerate}

In a Gogny-like energy density functional (EDF) that has been widely used in simulating heavy-ion reactions, see, e.g., Refs.\,\cite{Gale1987,Bertsch1988,Pawel,Das2003,LCK08,LiBA04,LiChen05,Cozma17,Cozma21}, the momentum-dependent part of the single-nucleon potential has the form of 
$\int\d\v{k}'f_{J'}(\v{r},\v{k}')\Omega(\v{k},\v{k}')$ while the corresponding potential energy in the ANM EOS has the form of
$\int\d\v{k}\d\v{k}'f_J(\v{r},\v{k})f_{J'}(\v{r},\v{k}')\Omega(\v{k},\v{k}')$ where the kernel $\Omega(\v{k},\v{k}')=[1+(\v{k}-\v{k}')^2/\Lambda^2)]^{-1}$ is from the Yukawa-type finite-range two-nucleon interaction, $\Lambda$ is the momentum-scale parameter while $f_J(\v{r},\v{k})$ and $f_{J'}(\v{r},\v{k}')$ are the momentum distribution functions of the two interacting nucleons $J$ and $J'$. Analytical integrations of the two integrals involved after incorporating the HMT in the $f_J(\v{r},\v{k})$ and $f_{J'}(\v{r},\v{k}')$ have been found formidable\,\cite{CLC18}, while they are necessary for fixing the model parameters using empirical properties of nuclear matter and nucleon optical potentials at $\rho_0$.

In this work, using a parametrized $\Omega(\v{k},\v{k}')$ as the surrogate of its original form we derive analytical expressions for all relevant terms and their characteristic parameters in the ANM EOS and single-nucleon potential. The surrogate approach adopted here not only grasps all major physical features of the momentum-dependence of the original Gogny-like potential but also enables analytical expressions for all the relevant physical quantities. There are different physical origins of the HMT in various many-body systems under different conditions\,\cite{Coleman15}. We notice that while during heavy-ion collisions, a HMT is automatically generated in the $f_J(\v{r},\v{k})$ and $f_{J'}(\v{r},\v{k}')$ due to finite temperatures. Nevertheless, most of the parameters of the input single-nucleon potentials in transport models have to be determined by the EOS properties of nuclear matter at zero temperature. Thus, the SRC/HMT in cold nuclear matter is important for determining the EOS of hot matter formed during heavy-ion reactions. 
While the EOS of hot dense neutron-rich matter is a necessary input for simulating neutron star mergers and understanding post-merger spectrum of high-frequency gravitational waves.
Moreover, the HMT at zero temperature is also useful for investigating properties of cold neutron stars, e.g., the proton fraction and the core-crust transition density/pressure in neutron stars at $\beta$-equilibrium.

With the motivations concerning the questions outlined above, the main contributions/findings of this work can be summarized as follows:
\begin{enumerate}
\item[(a)]Qualitatively, the pressure $P_0$ of SNM is enhanced while the $E_{\rm{sym}}(\rho)$ is reduced (both at sub- and supra-saturation densities) considering the SRC-induced HMT, which are consistent with the predictions from the nonlinear RMF models\,\cite{Cai16b}.
Quantitatively, however, the reduction of $E_{\rm{sym}}(\rho)$ is much stronger than that from the RMF models, while the enhancement of $P_0$ is weaker compared with the one predicted from the RMF models.
\item[(b)]Due to the much stronger reduction of $E_{\rm{sym}}(\rho)$ as pointed out in (a), the maximum mass of neutron stars is found to decrease correspondingly, compared with calculations using the FFG momentum distribution.
In this sense, it is not quite clear whether the SRC-induced HMT tends to help support heavy neutron stars (around $2M_{\odot}$).
More studies/possible mechanisms for stiffening high-density EOS within the Gogny-like EDFs are still needed.
\item[(c)]The surrogate Gogny-like EDF is shown to produce reasonable (kinetic) energy-dependence of the single-nucleon potential compared with known nucleon optical potentials at $\rho_0$,
thus it improves another aspect of the ANM EOS with the HMT, compared with the nonlinear RMF models\,\cite{Cai16b}.
\item[(d)]The isovector (symmetry) potential $U_{\rm{sym}}(\rho,|\v{k}|)$ is largely enhanced as the $E_{\rm{sym}}^{\rm{pot}}(\rho)$ is enhanced (while the $E_{\rm{sym}}^{\rm{kin}}(\rho)$ is reduced).
This is a new feature originated from the SRC/HMT within the Gogny-like EDF compared to the nonlinear RMF models where there is no momentum-dependence. It provides a novel connection linking the SRC-induced HMT with the dynamics of HICs involving neutron-rich nuclei.
\item[(e)]Parametrized forms of the EOS $E_0(\rho)$ and single-nucleon potential $U_0(\rho,\v{k}$) in SNM as well as the symmetry energy $E_{\rm{sym}}(\rho)$ and isovector potential $U_{\rm{sym}}(\rho,|\v{k}|)$ are provided to facilitate their future applications in both nuclear physics and astrophysics.
\end{enumerate}

A brief report on some preliminary results of our study on some issues listed above can be found in a conference proceedings\,\cite{CLC18}. The rest of this paper is organized as follows: In section \ref{sec2}, the single-nucleon momentum distribution function $n_{\v{k}}^J$ as well as other relevant parameters in the calculations are given. The
momentum-dependent nucleon potential used in the present work will also be discussed in this section.
In section \ref{sec4}, several analytical expressions for the characteristics of the NM EOS are given. 
Section \ref{sec5} is devoted to numerical demonstrations of the formulas developed in section \ref{sec4}. Section \ref{sec6} gives the summary of the present work. The detailed
derivations of most formulas are given in the appendix.

\setcounter{equation}{0}
\section{Gogny-like energy density functional encapsulating SRC/HMT effects}\label{sec2}

\subsection{Characteristics of the ANM EOS}

We first briefly recall definitions of
several quantities characterizing the EOS of ANM.
Around the saturation density $\rho _{0}$, the $E_{0}(\rho )$ can be
expanded up to the third-order in density as,
\begin{equation}
E_{0}(\rho )\approx E_{0}(\rho _{0})+\frac{1}{2}K_0\chi
^{2}+\frac{1}{6}J_0\chi ^{3}, \label{DenExp0}
\end{equation}%
where $\chi =(\rho -\rho _{0})/3\rho _{0} $ is a dimensionless
variable characterizing the deviations of the density from $\rho _{0}$. The first term $E_{0}(\rho _{0})$ on
the right-hand-side of Eq.\,(\ref{DenExp0}) is the binding energy per
nucleon in SNM at $\rho _{0}$ and,
\begin{align}
K_{0} =&\left. 9\rho _{0}^{2}\frac{\text{d} ^{2}E_{0}(\rho
)}{\text{d} \rho ^{2}}\right\vert _{\rho =\rho
_{0}},~~J_0=\left.27\rho_0^3\frac{\d^3E_0(\rho)}{\d\rho^3}\right|_{\rho=\rho_0},
\label{K0}
\end{align}
are the incompressibility coefficient and the
skewness\,\cite{Ste10,Cai14,Sel14} of the SNM, respectively.

Similarly, one can expand the $E_{\mathrm{sym}}(\rho )$ around the
saturation density as
\begin{equation}
E_{\text{sym}}(\rho)\approx
E_{\text{sym}}(\rho_0)+L\chi
+\frac{1}{2}K_{\rm{sym}}\chi^2+\frac{1}{6}J_{\rm{sym}}\chi^3, \label{EsymLKr}
\end{equation}
with the slope parameter $L$ defined as,
\begin{align}
L\equiv&\left.3\rho_0\frac{\text{d}E_{\mathrm{sym}}(\rho)}{\text{d}\rho}\right|_{\rho
=\rho_{0} }.
\end{align}
The curvature and the skewness\,\cite{ZL18,ZL19,ZL21,ZL22} of the symmetry energy are defined similarly as 
\begin{align}
K_{\rm{sym}} =&\left.9\rho _{0}^{2}\frac{\text{d} ^{2}E_{\rm{sym}}(\rho
)}{\text{d} \rho ^{2}}\right|_{\rho =\rho
_{0}},\\
J_{\rm{sym}}=&\left.27\rho_0^3\frac{\d^3E_{\rm{sym}}(\rho)}{\d\rho^3}\right|_{\rho=\rho_0}.
\end{align}

\subsection{Single-nucleon Momentum Distribution}

We describe the SRC-modified single-nucleon momentum
distribution function encapsulating a HMT used in the present work.
More details can be found in Ref.\,\cite{Cai15}. The single-nucleon
momentum distribution function in cold ANM takes the following
form\,\cite{Cai15},
\begin{equation}\label{MDGen}
n^J_{\v{k}}(\rho,\delta)=\left\{\begin{array}{ll}
\Delta_J,~~&0<|\v{k}|<k_{\rm{F}}^J,\\
\displaystyle{C}_J\left({k_{\rm{F}}^{J}}/{|\v{k}|}\right)^4,~~&k_{\rm{F}}^J<|\v{k}|<\phi_Jk_{\rm{F}}^J.
\end{array}\right.
\end{equation}
where $k_{\rm{F}}^J=k_{\rm{F}}(1+\tau_3^J\delta)^{1/3}$ is the nucleon Fermi
momentum with $k_{\rm{F}}=(3\pi^2\rho/2)^{1/3}$ and
$\tau_3^{\rm{n}}=+1$, $\tau_3^{\rm{p}}=-1$.
The above form of nucleon momentum distribution function is
consistent with the well-known predictions of microscopic nuclear
many-body theories\,\cite{Bethe,Ant88,Arr12,Cio15} and the recent
experimental
findings\,\cite{Hen14,Hen15,Wei15,Wei15a,Col15}.
Very recently, an analysis of the energy spectra of protons emitted in reactions of 47\,MeV/u  (several light to medium sized) projectiles on Sn and Au targets provides new evidence for the $1/k^4$ shape of the HMTs in the intrinsic momenta spectra of the projectiles\,\cite{Hag21}. 

\begin{figure}[h!]
\centering
  \includegraphics[width=7.cm]{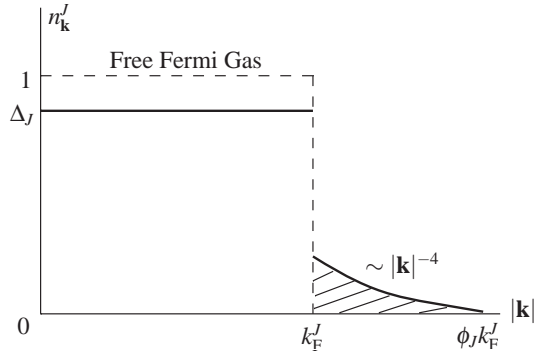}
  \caption{Sketch of the $n_{\v{k}}^J$ with a HMT. Taken from Refs.\,\cite{Cai15,Cai16b}.}
  \label{mom-dis}
\end{figure}

In (\ref{MDGen}), the $\Delta_J$ measures the depletion of the Fermi
sphere at zero momentum with respect to the FFG model. The sketch of
$n^J_{\v{k}}(\rho,\delta)$ is shown in FIG.\,\ref{mom-dis}. The
isospin structure of the parameters $C_J$ and $\phi_J$
is found to be $Y_J=Y_0(1+Y_1\tau_3^J\delta)$\,\cite{Cai15}. The
amplitude ${C}_J$ and high-momentum cutoff coefficient $\phi_J$
determine the fraction of nucleons in the HMT via
\begin{equation}\label{xPNM}
x_J^{\rm{HMT}}=3C_{{J}}\left(1-\phi_J^{-1}\right).
\end{equation}
The normalization condition between the density $\rho_J$
and the distribution $n_{\v{k}}^J$, i.e.,  $
[{2}/{(2\pi)^3}]\int_0^{\infty}n^J_{\v{k}}(\rho,\delta)\d\v{k}=\rho_J={(k_{\rm{F}}^{J})^3}/{3\pi^2}
$ requires that only two of the three parameters, i.e., ${C}_J$,
$\phi_J$ and $\Delta_J$, are independent. Here we choose the first
two as independent and determine the $\Delta_J$ by
\begin{equation}\label{DeltaJ}
\Delta_J=1-3{C}_J\left(1-\phi_J^{-1}\right)=1-x_J^{\rm{HMT}}.
\end{equation}

The ${C}/{|\v{k}|^4}$ shape of the HMT both for SNM and PNM is
strongly supported by recent studies both theoretically and
experimentally. Combining the results from analyzing cross sections
of $\rm{d}(\rm{e}, \rm{e}^{\prime}\rm{p})$ reactions\,\cite{Hen15}
and medium-energy photonuclear absorptions\,\cite{Wei15}, the $C_0$
was found to be $C_0\approx0.161\pm0.015$ in the HMT-exp
model\,\cite{Cai15}. With this $C_0$ and the value of
$x^{\rm{HMT}}_{\rm{SNM}}\approx28\%\pm4\%$\,\cite{Hen14,Hen15,Hen15b}
obtained from systematic analyses of inclusive (e,e$'$) reactions
and data from exclusive two-nucleon knockout reactions, the HMT
cutoff parameter in SNM is determined to be
$\phi_0=(1-x_{\rm{SNM}}^{\rm{HMT}}/3{C}_0)^{-1}\approx2.38\pm0.56$\,\cite{Cai15}.
Furthermore, the value of $C_{\rm{n}}^{\rm{PNM}}= C_0(1+C_1)$ was
extracted by applying the adiabatic sweep theorem\,\cite{Tan08} to
the EOS of PNM constrained by predictions of microscopic nuclear
many-body theories\,\cite{Sch05,Epe09a,Tew13,mm,Gez13,Gez10} and the
EOS of ultra-cold atoms under unitary
condition\,\cite{Tan08,Stew10,Kuh10}. And more specifically,
$C_{\rm{n}}^{\rm{PNM}}\approx0.12$ and $C_1\approx-0.25\pm0.07$ were
obtained\,\cite{Cai15}. Moreover, by inserting the values of
$x_{\rm{PNM}}^{\rm{HMT}}\approx1.5\%\pm0.5\%$\,\cite{Hen14,Hen15,Hen15b}
extracted in the same way as the $x_{\rm{SNM}}^{\rm{HMT}}$ and
$C_{\rm{n}}^{\rm{PNM}}$ into Eq.\,({\ref{xPNM}), the high momentum
cutoff parameter for PNM was determined to be
$\phi_{\rm{n}}^{\rm{PNM}}\equiv
\phi_0(1+\phi_1)=(1-x_{\rm{PNM}}^{\rm{HMT}}/3C_{\rm{n}}^{\rm{PNM}})^{-1}\approx1.04\pm0.02$\,\cite{Cai15},
thus $\phi_1\approx-0.56\pm0.10$\,\cite{Cai15} was obtained. 
See Ref.\,\cite{Cai15} for more details on these discussions.
On the other hand, if we take $x_{\rm{SNM}}^{\rm{HMT}}\approx12\%$ and
$x_{\rm{PNM}}^{\rm{HMT}}\approx4\%$ in the HMT-SCGF model, and using
a very similar scheme to determine the parameters in the HMT-exp
model, we obtain $\phi_0\approx1.49, \phi_1\approx-0.25$,
$C_0\approx0.121$ and $C_1\approx-0.01$\,\cite{Cai16c}.
In the following, the parameter sets adopting the experimental verification and the SCGF are denoted as HMT-exp and HMT-SCGF, respectively, and the corresponding calculations are made under the HMT-exp model or the HMT-SCGF model.

Besides the effects of the SRC-induced HMT on the kinetic symmetry
energy, several related interesting problems are explored recently,
including the enhancement of the isospin quartic term\,\cite{Cai15},
the predictions on the effective E-mass together with its isospin
splitting at the Fermi surface\,\cite{Cai16a,LiBA15,LiBA16}, the
correlation between the neutron skin thickness and the corresponding
kinetic energy density in the surface region of heavy
nuclei\,\cite{Cai16c}, and the effects on the neutron star
mass-radius relation\,\cite{Cai16b,Hen16}. 
Recently, the EOS of ANM considering the HMT in a general space of dimension $d$ is investigated\,\cite{CaiLi2022,CaiLi2022a}, see Ref.\,\cite{LiBA2018PPNP} for a recent review on more related issues\,\cite{Souza2020,YongGC2017PRC,YongGC2018PLB,
Wang2017PRC,Guo2021PRC,Yong2022,Yang2019,Bulgac2022,Bulgac2022a,Miller2019}.

\subsection{Surrogate Momentum-dependent Single-nucleon Potential in ANM}

The total EOS of ANM could be decomposed into its kinetic and potential parts as
\begin{equation}\label{Gen-EOS}
E(\rho,\delta)=E^{\rm{kin}}(\rho,\delta)+E^{\rm{pot}}(\rho,\delta)
\end{equation}
with the kinetic part given by
\begin{equation}\label{Ekin}
E^{\rm{kin}}(\rho,\delta)=\frac{1}{\rho}\frac{2}{(2\pi)^3}\sum_{J=\rm{n,p}}\int_0^{\infty}\frac{\v{k}^2}{2M}n_{\v{k}}^J(\rho,\delta)\d\v{k}
\end{equation}
where $n_{\v{k}}^J(\rho,\delta)$ is the single-nucleon momentum
distribution function, i.e., a step function in the FFG model or the one given in Eq.\,(\ref{MDGen}) in the HMT model. The potential part
in the present work adopts the following form\,\cite{CLC18},
\begin{align}\label{Epot}
E^{\rm{pot}}(\rho,\delta)=&\sum_{J,J'}\frac{C_{J,J'}}{\rho\rho_0}\int\d\v{k}\d\v{k}'f_J(\v{r},\v{k})f_{J'}(\v{r},\v{k}')\Omega(\v{k},\v{k}')\notag\\
&+\frac{A_\ell(\rho_{\rm{p}}^2+\rho_{\rm{n}}^2)}{2\rho\rho_0}
+\frac{A_{\rm{u}}\rho_{\rm{p}}\rho_{\rm{n}}}{\rho\rho_0}\notag\\
&+\frac{B}{\sigma+1}\left(\frac{\rho}{\rho_0}\right)^{\sigma}\left(1-x\delta^2\right),
\end{align}
here \begin{equation}
A_\ell=A_\ell^0+{2xB}/({1+\sigma}),~~A_{\rm{u}}=A_{\rm{u}}^0-{2xB}/({1+\sigma}).
\end{equation}
Besides the first line of (\ref{Epot}), the remaining terms are
totally the same as those in the original Momentum Dependent Interaction (MDI)\,\cite{Gale1987,Bertsch1988,Pawel,Das2003,LCK08,LiBA04,LiChen05,Cozma17,Cozma21,Che05,XuJ10,XuJ15}. 
Moreover, $A_\ell,
A_{\rm{u}},B,\sigma$ and $x$ are five phenomenological
parameters characterizing the density dependence of the nucleon
potential. In particular, if the replacements
$t_3=16B/(\sigma+1)\rho_0^{\sigma}$ and $x_3=(3x-1)/2$ are made,
then the parameters $B,\sigma$ and $x$ are reduced to the ones widely used to describe the
effective three-nucleon forces\,\cite{XuJ10}. Furthermore, the
parameter $x$ only affects the slope parameter of the symmetry energy. It does not affect the magnitude of the symmetry energy at $\rho_0$, providing a convenient way to investigate the density dependence of the symmetry energy\,\cite{Das2003,Che05}.
For the same type of nucleons, one has $C_{J,J}\equiv C_\ell$, while for the
unlike ones,  then $C_{J,\overline{J}}\equiv C_{\rm{u}}$,
here $\overline{\rm{n}}=\rm{p}$ and $\overline{\rm{p}}=\rm{n}$.

The single-nucleon potential corresponding to the EOS of (\ref{Gen-EOS}) is given by
\begin{align}\label{Gen-U}
U_J(\rho,\delta,|\v{k}|)=&\frac{A_\ell\rho_J}{\rho_0}+\frac{A_{\rm{u}}\rho_{\overline{J}}}{\rho_0}
+B\left(\frac{\rho}{\rho_0}\right)^{\sigma}\left(1-x\delta^2\right)\notag\\
&-4x\tau_3^J\frac{B}{\sigma+1}\frac{\rho^{\sigma-1}}{\rho_0^{\sigma}}\delta\rho_{\overline{J}}
\notag\\
&+\sum_{J'}\frac{2C_{J,J'}}{\rho_0}\int\d\v{k}'f_{J'}(\v{r},\v{k}')\Omega(\v{k},\v{k}').
\end{align}
 The momentum-dependence of the single-nucleon potential is controlled by
the last term, which is composed of two intrinsically different
terms, i.e., $f_J(\v{r},\v{k})$ and $\Omega(\v{k},\v{k}')$, both are
functions of momentum, leading to a non-trivial momentum behavior of
the nucleon potential $U_J(\rho,\delta,|\v{k}|)$ once the HMT in the $n_{\v{k}}^J$ is
considered.

The $f_J$ used in several integrations above is the phase space distribution function of nucleon $J$. It is related to the nucleon momentum distribution $n_{\v{k}}^J(\rho,\delta)$ by
\begin{equation}
f_J(\v{r},\v{k})=\frac{2}{h^3}n_{\v{k}}^J(\rho,\delta)=\frac{1}{4\pi^3}n_{\v{k}}^J(\rho,\delta),~~\hbar=1.
\end{equation}
For example, in the FFG model,
$n_{\v{k}}^J=\Theta(k_{\rm{F}}^J-|\v{k}|)$ with $\Theta$ the
standard step function, then
$f_J(\v{r},\v{k})=(1/4\pi^3)\Theta(k_{\rm{F}}^J-|\v{k}|)$. As mentioned briefly earlier, the $\Omega(\v{k},\v{k}')$ in (\ref{Epot}) is the function characterizing
the momentum dependence of the single-nucleon potential due to the finite-range of nuclear interactions.
For instance, for a Yukawa form of the two-nucleon interaction
$\exp[-\mu|\v{r}|]/|\v{r}|$ with $|\v{r}|=|\v{x}-\v{x}'|$
the distance between two nucleons,  one obtains $
\Omega(\v{k},\v{k}')=[1+(\v{k}-\v{k}')^2/\Lambda^2)]^{-1}$, which
is the familiar MDI interaction\,\cite{Gale1987,Bertsch1988,Das2003}. 
In applying the above formalism to transport model simulations of nuclear reactions, the $f_J (\v{r}, \v{k})$ and $n_{\v{k}}^J$ become time dependent and they are calculated self-consistently from solving
dynamically the coupled Boltzmann-Uehling-Uhlenbeck (BUU) transport equations for quasi-nucleons\,\cite{Bertsch1988,Aichelin1990,Sto86}. 
Similarly, in the original Gogny interaction\,\cite{GOG}, the finite-range two-nucleon interaction has the form of a Gaussian
regulator $\exp[-\mu^2|\v{r}|^2]$, and the corresponding $\Omega$ takes the form of $\exp[-\mu^2|\v{k}-\v{k}'|^2/4]$.

\begin{figure}[h!]
\centering
  \includegraphics[width=8.cm]{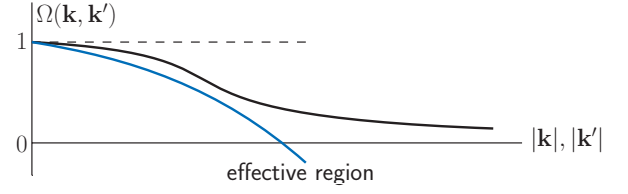}
  \caption{Sketch of the function $\Omega(\v{k},\v{k}')$, where the blue line is only effectively applicable.}
  \label{fig_Omega}
\end{figure}

It is useful to emphasize here the following basic properties of $\Omega(\v{k},\v{k}')$: 
\begin{enumerate}
\item[(a)] It is symmetric between the momentum $\v{k}$ and $\v{k}'$, i.e.,
$\Omega(\v{k},\v{k}')=\Omega(\v{k}',\v{k})$, reflecting the scalar
property of the nucleon potential. Thus $\Omega$ is a function of
$\v{k}+\v{k}'$ and $\v{k}\cdot\v{k}'$, and only the second term
couples the momenta of the two nucleons.
In general, one can show that it only depends on the magnitude of the difference $|\v{k}-\v{k}'|$.
Both the Yukawa potential and the Gogny force have such form.

\item[(b)] At zero momentum, $\Omega(\v{k},\v{k}')$ is a constant (which can
always be normalized to unity), indicating
$\Omega(\v{k},\v{k}')\approx1+\rm{``corrections''}$ when
$|\v{k}|\approx0,|\v{k}'|\approx0$. Furthermore, as the momentum
scale increases, the momentum-dependent interaction between nucleons
decreases, even to become free/negative, leading to that
$\Omega(|\v{k}|\to\infty,|\v{k}'|\to\infty)\to0$ or approaches a
small negative quantity $\mathcal{N}$. This property is relevant to
the fact, e.g., that the single-nucleon potential in SNM at large energies
saturates. Consequently, the above ``corrections'' in
$\Omega(\v{k},\v{k}')$ beyond the leading contribution ``1'' at
moderate momenta should be negative.
For example, we have $\Omega(\v{k},\v{k}')\approx 1-|\v{k}-\v{k}'|^2/\Lambda^2$ for the Yukawa potential and 
$\Omega(\v{k},\v{k}')\approx1-\mu^2|\v{k}-\v{k}'|^2/4$ for the Gogny force.
In fact, both of them can be approximated as $\Omega(\v{k},\v{k}')\approx 1+c_1\v{k}^2/\Lambda^2+c_2\v{k}'^2/\Lambda^2+c_3\v{k}\cdot\v{k}'/\Lambda^2$, where the coefficients $c_i$'s depend on the detailed form of the interactions.
We will use this perturbative expansion as a hint to construct a reasonable surrogate $\Omega(\v{k},\v{k}')$, see (\ref{MFunc}).

\item[(c)] Based on the last point, a perturbation-effective version of the
function $\Omega(\v{k},\v{k}')$ can be constructed generally as
$\Omega(\v{k},\v{k}')\approx
1+af(|\v{k}-\v{k}'|)+bg(|\v{k}-\v{k}'|)+\cdots$, where the
phenomenological functions $f,g,\cdots$ are small at moderate
momentum scale. This effective $\Omega(\v{k},\v{k}')$ can only be
used up to a momentum scale around $\Lambda$, see the sketch illustrating this feature in FIG.\,\ref{fig_Omega}, where the $\Omega(\v{k},\v{k}')$ denoted by the
blue line is perturbatively effective up to around a certain high momentum
cutoff. Namely, the surrogate $\Omega(\v{k},\v{k}')$ constructed in this manner grasps the main features of the original $\Omega(\v{k},\v{k}')$ up to the high momentum cutoff.
\end{enumerate}
Considering the SRC-induced HMT in the single-nucleon momentum distribution and the kernel $\Omega(\v{k},\v{k}')$ for the momentum dependent single-nucleon potential either with the Yukawa or Gogny force, their generally complicated form and nature hinder us from integrating out analytically the relevant integrations over momentum in the expressions of nucleon specific energy and single-nucleon potential. While these integrations are necessary for fixing all the EOS parameters using empirical properties of nuclear matter, symmetry energy, nucleon effective mass, and nucleon optical potentials at $\rho_0$, etc. 

Making use of the main features of $\Omega$ outlined above, we adopt in this work the following parametrization as a surrogate of $\Omega(\v{k},\v{k}')$ 
\begin{equation}\label{MFunc}
\Omega(\v{k},\v{k}')=1+{a}\left[\left(\frac{\v{k}\cdot\v{k}'}{\Lambda^2}\right)^2\right]^{1/4}
+{b}\left[\left(\frac{\v{k}\cdot\v{k}'}{\Lambda^2}\right)^2\right]^{1/6},
\end{equation}
where $a$ and $b$ are two phenomenological parameters, and
$\Lambda\approx1\,\rm{GeV}$ is an effective momentum scale. The
parameter $a$ should be negative according to the above discussion (see point (b)).
As pointed out in Ref.\,\cite{CLC18}, the advantages of using this surrogate is twofold: (1) the final 1/2 and 1/3 power of $\v{k}\cdot\v{k}'/\Lambda^2$ is relevant for describing properly the kinetic energy dependence of nucleon optical potential, as we shall show in FIG.\,\ref{fig_ab_U0}; (2) it enables analytical expressions for both the EOS of ANM and $U_J(\rho,\delta, |\v{k}|)$. 
Moreover,  the $\Omega$ function is only perturbatively effective at momenta smaller than the momentum scale $\Lambda$, indicating that the EDF thus constructed can only be used to a restricted range of momentum and/or density.  For example,  the cutoff of the HMT about $2.4k_{\rm{F}}\approx631\,\rm{MeV}$ or the density $\rho\approx7\rho_0\mbox{$\sim$}8\rho_0$ corresponding to $k_{\rm{F}}\approx503\,\rm{MeV}\mbox{$\sim$}526\,\rm{MeV}$ should be safely smaller than the $\Lambda$ parameter $\approx1\,\rm{GeV}$.
In this sense, the parameterized $\Omega$ (\ref{MFunc}) is a reliable surrogate/effective model\,\cite{SURR} for studying properties of neutron stars and intermediate-relativistic energy heavy-ion reactions where the maximum density reached is less than the above estimates.

\setcounter{equation}{0}
\section{Analytical expressions of the SNM EOS, nucleon effective mass, symmetry energy, isoscalar and isovector potentials}\label{sec4}
In this section, we present all the analytical expressions of quantities relevant for describing the ANM EOS using the surrogate function (\ref{MFunc}) for $\Omega(\v{k},\v{k}')$. Detailed derivations are given in the APPENDIX \ref{app1}. More specifically, in subsection \ref{sb_EOS_SNM} we give the SNM EOS together with the isoscalar nucleon effective mass $M_0^{\ast}(\rho)$, the coefficient of incompressibility $K_0(\rho)$ and the single-nucleon potential $U_0(\rho)$ in SNM. In subsection \ref{sb_Esym}, the expression for the symmetry energy is derived. For general expressions of the ANM EOS $E(\rho,\delta)$, see Eq.\,(\ref{HMT-Erd}) (together with Eqs.\,(\ref{HMT_Erd1}), (\ref{HMT_Erd2}) and (\ref{HMT_Erd3})) and Eq.\,(\ref{FFG-Erd}).\\

\subsection{EOS of SNM}\label{sb_EOS_SNM}

 For the kinetic EOS of SNM, we
have\,\cite{Cai15}
\begin{equation}\label{E0kin-HMT}
E^{\rm{kin}}_0(\rho)=\frac{3}{5}E_{\rm{F}}(\rho)\left[
1+{C}_0\left(5\phi_0+\frac{3}{\phi_0}-8\right)\right],
\end{equation}
where $E_{\rm{F}}=k_{\rm{F}}^2/2M$ is the Fermi kinetic energy. The
high momentum cutoff $\phi_0$ is generally larger than unity,
thus $\Upsilon_0=1+C_0(5\phi_0+3/\phi_0-8)\geq1$, indicating
that the $E^{\rm{kin}}_0(\rho)$ in the presence of HMT is enhanced
compared to the prediction of using the FFG nucleon momentum distribution\,\cite{Cai15}. This conclusion has
important consequences on the total EOS of SNM once the potential
part is taken into account, i.e., as the kinetic part is enhanced,
the potential part should be correspondingly reduced, leading to a
reduction of the single-nucleon potential. Furthermore, the
density-dependent terms in the EOS is the same as those in the MDI
model\,\cite{Das2003,LCK08,Che05,XuJ10,XuJ15}. The total EOS of SNM in
the HMT model is thus,
\begin{widetext}
\begin{align}\label{HMT-E0-mm}
E_0(\rho)=&\frac{3k_{\rm{F}}^2}{10M}\left(5\phi_0+\frac{3}{\phi_0}-8\right)
+\frac{1}{4}A_{\rm{tot}}\left(\frac{\rho}{\rho_0}\right)+\frac{B}{\sigma+1}\left(\frac{\rho}{\rho_0}\right)^{\sigma}\notag\\
&+\frac{1}{2}C_{\rm{tot}}\left(\frac{\rho}{\rho_0}\right)
\left[1+\frac{24a}{49}\left[\Delta_0+7C_0\left(1-\frac{1}{\sqrt{\phi_0}}\right)\right]^2\left(\frac{k_{\rm{F}}}{\Lambda}\right)+
\frac{243b}{400}\left[\Delta_0+5C_0\left(1-\frac{1}{\phi_0^{2/3}}\right)\right]^2\left(\frac{k_{\rm{F}}}{\Lambda}\right)^{2/3}\right],
\end{align}
\end{widetext}
see also (\ref{HMT-E0}), where
$A_{\rm{tot}}=A_\ell+A_{\rm{u}}=A_\ell^0+A_{\rm{u}}^0$ and
$C_{\rm{tot}}=C_\ell+C_{\rm{u}}$. The two factors $\Pi_1$ and $\Pi_2$ characterizing the HMT
in Eq.\,(\ref{HMT-E0-mm}) are given by
\begin{align}
\Pi_1=&\Delta_0+7C_0\left(1-\frac{1}{\sqrt{\phi_0}}\right),\label{def_Pi1}\\
\Pi_2=&\Delta_0+5C_0\left(1-\frac{1}{\phi_0^{2/3}}\right).
\end{align}

In the FFG model (as a reminder, the ``FFG'' here means that the $n_{\v{k}}^J$ takes the step function while the potential part is also included in calculating the EOS and related quantities), the high momentum cutoff $\phi_0$ becomes $\phi_0=1$ and $\phi_1=0$, we have $
\Delta_0=1-3C_0(1-\phi_0^{-1})\to1$,  and the two factors $\Pi_1$ and $\Pi_2$ become $\Pi_1\to1,\Pi_2\to1$. In the same limit, the Eq.\,(\ref{HMT-E0-mm}) is then reduced to
\begin{align}\label{FFG-E0-mm}
E_0(\rho)=&\frac{3k_{\rm{F}}^2}{10M}+\frac{1}{4}A_{\rm{tot}}\left(\frac{\rho}{\rho_0}\right)+\frac{B}{\sigma+1}\left(\frac{\rho}{\rho_0}\right)^{\sigma}\notag\\
&+\frac{1}{2}C_{\rm{tot}}\left(\frac{\rho}{\rho_0}\right)\left[1+\frac{24a}{49}\frac{k_{\rm{F}}}{\Lambda}
+\frac{243b}{400}\left(\frac{k_{\rm{F}}}{\Lambda}\right)^{2/3}\right],
\end{align}
which is derived (independently) in the FFG model, see derivations leading to Eq.\,(\ref{FFG-E0}).

The density structure of both the $E_0(\rho)$ in the
FFG and the HMT model is the same, i.e.,
\begin{equation}\label{str}
E_0(\rho)=\nu_1\rho^{2/3} +\nu_2\rho+\nu_3\rho^{\sigma}+
\nu_4\rho^{4/3}+\nu_5\rho^{11/9},
\end{equation}
with different coefficient $\nu_j$'s in the FFG/HMT model.
Thus, $E_0(\rho)$ is a generalized polynomial of density $\rho$ (or equivalently a polynomial of Fermi momentum $k_{\rm{F}}$),
indicating that it is only effectively applicable to low densities.
Consequently, the density structure of the pressure of SNM and the incompressibility can be worked out as
\begin{align}
P_0(\rho)/\rho=&\frac{2}{3}\nu_1\rho^{2/3} +\nu_2\rho
+\sigma\nu_3\rho^{\sigma}\notag\\
&+\frac{4}{3}\nu_4\rho^{4/3}
+\frac{11}{9}\nu_5\rho^{11/9},\\
K_0(\rho)=&-2\nu_1\rho^{2/3}+9\sigma(\sigma-1)\nu_3\rho^{\sigma}\notag\\
&+4\nu_4\rho^{4/3}+\frac{22}{9}\nu_5\rho^{11/9}.
\end{align}
Specifically, the expressions for the pressure $P_0(\rho)$ and incompressibility $K_0(\rho)$ in the FFG model are
given in Eqs.\,(\ref{FFG-p0}) and (\ref{FFG-K0}), while those in the
HMT model are given in Eqs.\,(\ref{HMT-p0}) and (\ref{HMT-K0}),
respectively.

Besides the SNM EOS $E_0(\rho)$ in the HMT model, the single-nucleon potential as a function of density $\rho$ and momentum $\v{k}$ in SNM could also be given,
\begin{align}\label{HMT-U0-mm}
&U_0(\rho,|\v{k}|)=\frac{1}{2}A_{\rm{tot}}\left(\frac{\rho}{\rho_0}\right)+B\left(\frac{\rho}{\rho_0}\right)^{\sigma}
+C_{\rm{tot}}\left(\frac{\rho}{\rho_0}\right)\notag\\
&\times\left[1+\frac{4a}{7}\Pi_1\left(\frac{|\v{k}|k_{\rm{F}}}{\Lambda^2}\right)^{1/2}
+\frac{27b}{40}\Pi_2\left(\frac{|\v{k}|k_{\rm{F}}}{\Lambda^2}\right)^{1/3}\right],
\end{align}
see Eq.\,(\ref{HMT-U0}) for a detailed derivation. Again the above expression can be reduced to the FFG one, see Eq.\,(\ref{FFG-U0}). The
momentum structure of the $U_0$ could be written explicitly as (due to the function $\Omega$),
\begin{equation}
U_0(\rho,|\v{k}|)=\zeta_0(\rho)+\zeta_1(\rho)|\v{k}|^{1/2}+\zeta_2(\rho)|\v{k}|^{1/3},
\end{equation}
with
$\zeta_0(\rho)=(A_{\rm{tot}}/2+C_{\rm{tot}})(\rho/\rho_0)+B(\rho/\rho_0)^{\sigma}$. The latter is unchanged in the presence of HMT, while the form of $\zeta_1(\rho)$ and $\zeta_2(\rho)$ should change correspondingly, compare (\ref{FFG-U0}) and (\ref{HMT-U0}).
The concavity of the $U_0$ at the saturation density\,\cite{Ham90} indicates that $C_{\rm{tot}}$ is negative according to $
\zeta_1(\rho)=4C_{\rm{tot}}(\rho/\rho_0)a\Pi_1k_{\rm{F}}^{1/2}/7\Lambda>0$,
where $a$ is a negative parameter as expected by general discussions (given in the last section) and
$\Pi_1$ is positive (see the expression (\ref{def_Pi1})). The momentum-dependence of the $U_0$ is characterized by introducing the nucleon effective k-mass\,\cite{LiBA2018PPNP,LiBA15,LiBA16}, i.e.,
\begin{equation}
M_0^{\ast}(\rho,|\v{k}|)/M=\left[1+\frac{M}{|\v{k}|}\frac{\partial
U_0}{\partial|\v{k}|}\right]^{-1}.
\end{equation}
Specifically, we obtain for the current EDF the following expression 
\begin{align}
\frac{M_0^{\ast}(\rho,|\v{k}|)}{M}=\Bigg[&1+C_{\rm{tot}}M\left(\frac{\rho}{\rho_0}\right)
\Bigg[\frac{2a\Pi_1}{7}\left(\frac{k_{\rm{F}}}{\Lambda^2}\right)^{1/2}|\v{k}|^{-3/2}\notag\\
&+\frac{9b\Pi_2}{40}\left(\frac{k_{\rm{F}}}{\Lambda^2}\right)^{1/3}|\v{k}|^{-5/3}\Bigg]\Bigg]^{-1},
\end{align}
which is further evaluated at the Fermi surface $|\v{k}|=k_{\rm{F}}$ to give the density dependence of the $M_0^{\ast}(\rho)/M$ as,
\begin{align}\label{FFG-M0-mm}
\frac{M_0^{\ast}(\rho)}{M}
=\Bigg[&1+C_{\rm{tot}}M\left(\frac{\rho}{\rho_0}\right)
\Bigg[\frac{2a\Pi_1}{7}\frac{1}{k_{\rm{F}}\Lambda}\notag\\
&+\frac{9b\Pi_2}{40}\left(\frac{1}{\Lambda^2k_{\rm{F}}^4}\right)^{1/3}\Bigg]\Bigg]^{-1}.
\end{align}
In the high density limit, we have
\begin{equation}\label{def_11}
\lim_{\rm{large }\rho}\frac{M_0^{\ast}(\rho)}{M}
\approx\left[1+\frac{2M\Pi_1aC_{\rm{tot}}}{7k_{\rm{F}}\Lambda}\left(\frac{\rho}{\rho_0}\right)+\mathcal{O}\left(k_{\rm{F}}^{-1/3}\right)\right]^{-1},
\end{equation}
which approaches zero in the limit $\rho\to\infty$. In the FFG
model, the density dependence of the effective mass is given in Eq.\,(\ref{FFG-M0}), i.e., by taking $\Pi_1=1$ in (\ref{def_11}).

\subsection{Symmetry Energy}\label{sb_Esym}

We now discuss the symmetry energy, its slope parameter, and the isovector (symmetry) potential $U_{\rm{sym}}(\rho,|\v{k}|)$. As derived in
detailed in APPENDIX \ref{app1}, the symmetry energy for the general HMT model is given by,
\begin{widetext}
\begin{align}\label{HMT-Esym-mm}
E_{\rm{sym}}(\rho)=&\frac{k_{\rm{F}}^2}{6M}\left[1+C_0(1+3C_1)\left(5\phi_0+\frac{3}{\phi_0}-8\right)
+3C_0\phi_1\left(1+\frac{5}{3}C_1\right)\left(5\phi_0-\frac{3}{\phi_0}\right)+\frac{27C_0\phi_1^2}{5\phi_0}\right]
\notag\\
&+\frac{1}{4}C_\ell\Delta_0^2\left(\frac{\rho}{\rho_0}\right)\left[
Y_{10}+Y_{11}a\frac{k_{\rm{F}}}{\Lambda}+Y_{12}b\left(\frac{k_{\rm{F}}}{\Lambda}\right)^{2/3}
\right]
+\frac{1}{4}C_{\rm{u}}\Delta_0^2\left(\frac{\rho}{\rho_0}\right)
\left[
Z_{10}+Z_{11}a\frac{k_{\rm{F}}}{\Lambda}+Z_{12}b\left(\frac{k_{\rm{F}}}{\Lambda}\right)^{2/3}
\right]\notag\\
&+\frac{3}{2}C_\ell\Delta_0C_0\left(\frac{\rho}{\rho_0}\right)\left[
Y_{20}+Y_{21}a\frac{k_{\rm{F}}}{\Lambda}+Y_{22}b\left(\frac{k_{\rm{F}}}{\Lambda}\right)^{2/3}
\right]
+\frac{3}{2}C_{\rm{u}}\Delta_0C_0\left(\frac{\rho}{\rho_0}\right)\left[
Z_{20}+Z_{21}a\frac{k_{\rm{F}}}{\Lambda}+Z_{22}b\left(\frac{k_{\rm{F}}}{\Lambda}\right)^{2/3}
\right]\notag\\
&+\frac{9}{4}C_\ell C_0^2\left(\frac{\rho}{\rho_0}\right)\left[
Y_{30}+Y_{31}a\frac{k_{\rm{F}}}{\Lambda}+Y_{32}b\left(\frac{k_{\rm{F}}}{\Lambda}\right)^{2/3}
\right]+\frac{9}{4} C_{\rm{u}} C_0^2\left(\frac{\rho}{\rho_0}\right)\left[
Z_{30}+Z_{31}a\frac{k_{\rm{F}}}{\Lambda}+Z_{32}b\left(\frac{k_{\rm{F}}}{\Lambda}\right)^{2/3}
\right]\notag\\
&+\frac{1}{4}A_{\rm{d}}\left(\frac{\rho}{\rho_0}\right)-\frac{Bx}{\sigma+1}\left(\frac{\rho}{\rho_0}\right)^{\sigma},\end{align}
\end{widetext}
with the expressions for $Y_{ij}$ and $Z_{ij}$ given in Eqs.\,(\ref{def_Y1}), (\ref{def_Y20}), (\ref{def_Y21}), (\ref{def_Y22}),
(\ref{def_Y30}), (\ref{def_Y31}), (\ref{def_Y32}),
 (\ref{def_Z1}), (\ref{def_Z20}), (\ref{def_Z21}), 
(\ref{def_Z22}), (\ref{def_Z30}), (\ref{def_Z31}), and (\ref{def_Z32}), respectively.
The second line of Eq.\,(\ref{HMT-Esym-mm}) represents the
contribution from the depletion in the $n_{\v{k}}^J$ characterized by $\Delta_0^2$, while the fourth line has the HMT contribution indicated by the
$C_0^2$. The third line of Eq.\,(\ref{HMT-Esym-mm}) is the mixing of
the depletion and the HMT characterized by $\Delta_0C_0$. In the FFG
model, only the first two lines of Eq.\,(\ref{HMT-Esym-mm}) survive.
Moreover, $A_{\rm{d}}$ and $C_{\rm{d}}$ are the
difference between the like and unlike terms,
\begin{equation}\label{ad}
A_{\rm{d}}=A_{\rm{d}}^0+\frac{4xB}{1+\sigma},~~C_{\rm{d}}=C_\ell-C_{\rm{u}},
\end{equation}
with $
A_{\rm{d}}^0=A_\ell^0-A_{\rm{u}}^0$. The density structure of
the symmetry energy is the same as that of the SNM EOS, see Eq.\,(\ref{str}), thus it is straightforward to obtain the slope
parameter $L(\rho)$, see Eq.\,(\ref{HMT-L}).

Similarly, the isovector (symmetry) potential $U_{\rm{sym}}(\rho,|\v{k}|)$ is defined through the following Lane potential \,\cite{Lan62} 
\begin{equation}
U_J(\rho,\delta,|\v{k}|)\approx
U_0(\rho,|\v{k}|)+U_{\rm{sym}}(\rho,|\v{k}|)\tau_3^J\delta+\mathcal{O}(\delta^2).
\end{equation}
Such a decomposition of single-nucleon potential in ANM has been verified by various phenomenological model analyses of nucleon-nucleus scattering data and predictions of microscopic nuclear many-body theories, see, e.g., Refs. \cite{mu04,Li04,Ron06,Dal1,zuo05,Beh11,LiX13,LiX15}. In the HMT model, it is given by
\begin{align}\label{HMT-Usym-mm}
&U_{\rm{sym}}(\rho,|\v{k}|)=\frac{1}{2}A_{\rm{d}}\left(\frac{\rho}{\rho_0}\right)-\frac{2Bx}{\sigma+1}\left(\frac{\rho}{\rho_0}\right)^{\sigma}\notag\\
&+
C_{\rm{d}}\left(\frac{\rho}{\rho_0}\right)\left[\gamma_0+\gamma_1a\left(\frac{|\v{k}|k_{\rm{F}}}{\Lambda^2}\right)^{1/2}
+\gamma_2b\left(\frac{|\v{k}|k_{\rm{F}}}{\Lambda^2}\right)^{1/3}\right].
\end{align}
The expressions for $\gamma_0,~\gamma_1$ and $\gamma_2$ are given in
Eqs.\,(\ref{def_gamma0}), (\ref{def_gamma1}) and (\ref{def_gamma2}), respectively.
Notice that in the FFG model, besides $
\Delta_0\to1$, we also have
\begin{align}
\Delta_0\Delta_1=&-{3C_0[C_1(\phi_0-1)+\phi_1]}/{\phi_0}\to0,\\
\Delta_0\Delta_2=&-{3C_0\phi_1(C_1-\phi_1)}/{\phi_0}\to0,
\end{align}
see (\ref{Def_Delta012}), thus,
\begin{align}
&Y_{10}\to2,~~Y_{11}\to{32}/{21},~~Y_{12}\to{33}/{20},\\
&Z_{10}\to-2,~~Z_{11}\to-{8}/{7},~~Z_{12}\to-{27}/{20},
\end{align}
together with $Y_{2j}\to0,~Z_{2j}\to0,~Y_{3j}\to0,~Z_{3j}\to0$ with $j=0,1,2$, see
the expressions (\ref{def_Y20}), (\ref{def_Y21}), (\ref{def_Y22}),
(\ref{def_Z20}), (\ref{def_Z21}), (\ref{def_Z22}), (\ref{def_Y30}),
(\ref{def_Y31}), (\ref{def_Y32}), (\ref{def_Z30}),
(\ref{def_Z31}) and (\ref{def_Z32}) and the consequent result is the
symmetry energy in the FFG model, see Eq.\,(\ref{FFG-Esym}).
Moreover, the expressions for the slope parameter $L(\rho)$ and the symmetry potential
$U_{\rm{sym}}(\rho,|\v{k}|)$ in the FFG model are given in Eq.\,(\ref{FFG-L}) and Eq.\,(\ref{FFG-Usym}), respectively.

Considering the expression for the symmetry energy, the factor characterizing the effects of the HMT, i.e.,
\begin{align}\label{def_Upsilon}
&\Upsilon_{\rm{sym}}=1+C_0(1+3C_1)\left(5\phi_0+\frac{3}{\phi_0}-8\right)\notag\\
&+3C_0\phi_1\left(1+\frac{5}{3}C_1\right)\left(5\phi_0-\frac{3}{\phi_0}\right)
+\frac{27C_0\phi_1^2}{5\phi_0},
\end{align}
is generally smaller than unity, leading to a reduction of the
kinetic symmetry energy\,\cite{Cai15}. According to the single-nucleon potential decomposition of the symmetry
energy\,\cite{XuC10,XuC11,CheR12} based on the  Hugenholtz-Van Hove (HVH) theorem\,\cite{Hug58}, a contribution of
$2^{-1}U_{\rm{sym}}(\rho,|\v{k}|=k_{\rm{F}})$ to the
symmetry energy is obtained. Thus a reduction of the kinetic symmetry energy
will induce an enhancement of the symmetry potential, i.e., the
symmetry energy encapsulating the SRC effects will intrinsically
have impacts on the symmetry potential.
In addition, by using the expression of $A_{\rm{d}}$ in Eq.\,({\ref{ad}) the last two terms in Eq.\,(\ref{HMT-Esym-mm}) can be rewritten as
\begin{equation}
\frac{1}{4}A_{\rm{d}}^0\left(\frac{\rho}{\rho_0}\right)+\frac{xB}{1+\sigma}\left(\frac{\rho}{\rho_0}\right)
\left[1-\left(\frac{\rho}{\rho_0}\right)^{\sigma-1}\right],
\end{equation}
indicating that the $x$ does not affect the symmetry energy at
$\rho_0$\,\cite{Che05,XuJ10,XuJ15}. On the other hand, the corresponding contribution to the slope parameter of symmetry energy
$L(\rho)$ is $
3A_{\rm{d}}^0/4+3xB({1-\sigma})/({1+\sigma})$ at $\rho=\rho_0$. Thus
a larger (positive) $x$ corresponds to a softer symmetry energy if
$B(1-\sigma)<0$, once the other phenomenological parameters are
fixed\,\cite{Che05}, see TABLE \ref{tab_para}.


\setcounter{equation}{0}
\section{Numerical Demonstrations}\label{sec5}

\subsection{Schemes for Determining the Coupling Constants}\label{sb_parameter}

In this subsection, we discuss the scheme to determine the model couplings, i.e., $A_{\rm{tot}},~B,~C_{\rm{tot}},~A_{\rm{d}},~C_{\rm{d}},~\sigma,~a,~b,~x$ and $\Lambda$.
As pointed out in Ref.\,\cite{CLC18} that the function $\Omega(\v{k},\v{k}')$ is invariant under the transformations,
\begin{equation}
a\to a/\xi^{3/2},~~b\to\xi b,~~\Lambda\to\Lambda/\xi^{3/2},
\end{equation}
with $\xi>0$ being any scaling factor, i.e., we have the freedom to first fix one of them without affecting the physical results. We set $b = 2$ in the following and then determine other parameters using known empirical constraints.

\begin{table*}[t!]
\caption{Coupling constants used in the two models (right side) and
some empirical properties of ANM used to fix
them (left side), $b=2$ and $\Lambda=1.6\,\rm{GeV}$ are fixed,
$K_0\equiv K_0(\rho_0),~M_0^{\ast}\equiv M_0^{\ast}(\rho_0),~L\equiv
L(\rho_0)$.
See also Ref.\,\cite{CLC18}.
}\label{tab_para} {\normalsize\begin{tabular}{lr||cccc}
\hline\hline Quantity& Value & Coupling & FFG&HMT-SCGF&HMT-exp\\
\hline $\rho_0$ (fm$^{-3}$) & $0.16$ & $A_\ell^0$ $(\rm{MeV})$ &
$-266.2934$&$520.3611$&$146.6085$\\
\hline $E_0(\rho_0)$
$(\rm{MeV})$&$-16.0$&$A_{\rm{u}}^0$ $(\rm{MeV})$&$-86.8331$&$805.3082$&$1216.2500$\\
\hline $M_0^{\ast}/M$ &$0.58$&$B$ $(\rm{MeV})$&$517.5297$&$-256.9850$&$-64.5669$\\
\hline $K_0$ ($\rm{MeV}$) &$230.0$&$C_\ell$ $(\rm{MeV})$&$-155.6406$&$-154.7508$&$-37.3249$\\
\hline $U_0(\rho_0,0)$ ($\rm{MeV}$)
&$-100.0$&$C_{\rm{u}}$ $(\rm{MeV})$&$-285.3256$&$-351.0989$&$-679.5379$\\
\hline $E_{\rm{sym}}(\rho_0)$ ($\rm{MeV}$)&$31.6$&$\sigma$&$1.0353$&$0.9273$&$0.6694$\\
\hline $L$ ($\rm{MeV}$)&$58.9$&$a$&$-5.4511$&$-5.0144$&$-4.1835$\\
\hline $U_{\rm{sym}}(\rho_0,1\,\rm{GeV})$ ($\rm{MeV}$)&$-20.0$&$x$&$0.6144$&$0.3774$&$-0.2355$\\
\hline\hline
\end{tabular}}
\end{table*}

In our scheme, the values of $E_0(\rho_0),~ \rho_0, ~K_0\equiv
K_0(\rho_0), ~M_0^{\ast}(\rho_0)$ and $U_0(\rho_0,0\,\rm{MeV})$
for SNM, and those of $E_{\rm{sym}}(\rho_0), ~L\equiv L(\rho_0)$ and
$U_{\rm{sym}}(\rho_0,1\,\rm{GeV})$ for the symmetry energy, its slope and the
symmetry potential are fixed at their currently known most probable empirical values. For example, the value of the incompressibility $K_0=230\pm20\,\rm{MeV}$ was
determined from analyzing nuclear giant resonances
(GMR)\,\cite{You99,Shl06,Pie10,Che12,Col14,Stone2014PRC,Garg2018PPNP,LiBA2021PRC,XuJ2021PRC,ZhangZ2021CPC}, and the nucleon effective mass at $\rho_0$ is fixed at $0.58M$\,\cite{CLC18,LiBA2018PPNP}.
For the $E_\text{sym}(\rho_0)$ and $L $, all existing constraints extracted
so far from both terrestrial laboratory measurements and
astrophysical observations are found to be essentially consistent
with the global averages of $E_{\text{sym}}({\rho _{0}}) \approx
31.6\pm2.66$ MeV and $L \approx 58.9 \pm 16$ MeV\,\cite{LiBA13}, see also Ref.\,\cite{Chen17,LiBA21}. Consequently, we can fix for the SNM five phenomenological parameters, i.e.,
$A_{\rm{tot}},~B,~C_{\rm{tot}},~\sigma$ and $a$, while for the symmetry
energy and the symmetry potential, the remaining three parameters,
i.e., $A_{\rm{d}},~C_{\rm{d}}$ and $x$ can be fixed. Equivalently,
the totally eight parameters, i.e.,
$A_\ell^0,~A_{\rm{u}}^0,~B,~C_\ell,~C_{\rm{u}},~\sigma,~a$ and $x$ can be
determined.

Next, we need to select the cutoff $\Lambda$.
At first glance, the $\Lambda$ could be determined by adopting another empirical constraint, e.g., the symmetry energy at two times the saturation density $E_{\rm{sym}}(2\rho_0)$\,\cite{LiBA21}.
But it does not work in this manner.
In particular, one can demonstrate that once we fix the above eight quantities, i.e., $E_0(\rho_0),~\rho_0,~K_0,~M_0^{\ast}(\rho_0),~U_0(\rho_0,0\,\rm{MeV}),~E_{\rm{sym}}(\rho_0),~L$ and
$U_{\rm{sym}}(\rho_0,1\,\rm{GeV})$, these quantities at other densities or other energy scales do not further depend on the cutoff $\Lambda$, see APPENDIX \ref{app2} for more discussions.
The role of the $\Lambda$ is different from the other model parameters.
However, the $\Lambda$ parameter may affect quantities which are not used in the fixing scheme, e.g., $E_{\rm{sym},4}(\rho)$.
In the following we select the $\Lambda$ parameter based on the empirical fact that the fourth-order symmetry energy at $\rho_0$ in the FFG model is smaller than about 3\,MeV\,\cite{Cai12,Sei14,Gon17,PuJ17,CWZC2022}.

We construct three models of EOS of ANM using
different HMT input parameters described in section \ref{sec2},
i.e., the FFG model as well as the HMT-SCGF and HMT-exp models. Again, we emphasize that the ``FFG'' or the ``HMT'' is only used to label the $n_{\v{k}}^J(\rho,\delta)$ and the potential EOS is included in all of these models.
See TABLE \ref{tab_para} for the values of the model parameters, here the cutoff $\Lambda$ is set to be 1.6\,GeV to fulfill the aforementioned constraint (specifically $1.40\,\rm{GeV}\lesssim\Lambda\lesssim1.65\,\rm{GeV}$).
It is necessary to point out that since we fixed the parameters of
the HMT by using experimental data and/or microscopic-theory calculations at the saturation density, the possible density dependence of those
parameters, i.e., $C_0,~C_1,~\phi_0$ and $\phi_1$ is not
explored here (also see discussions given in Ref.\,\cite{Cai15}). The density dependence of the various terms
in the kinetic EOS is thus only due to that of the nucleon Fermi momentum $k_{\rm{F}}$.
From the table, one can clearly find that the parameter $a$ is negative, i.e., the moving nucleons feel weaker interaction compared to the static ones,
see the discussions given after the formula (\ref{MFunc}).
The EDF constructed is abbreviated as the abMDI\,\cite{CLC18}.

\begin{figure}[h!]
\centering
  \includegraphics[height=2.8cm]{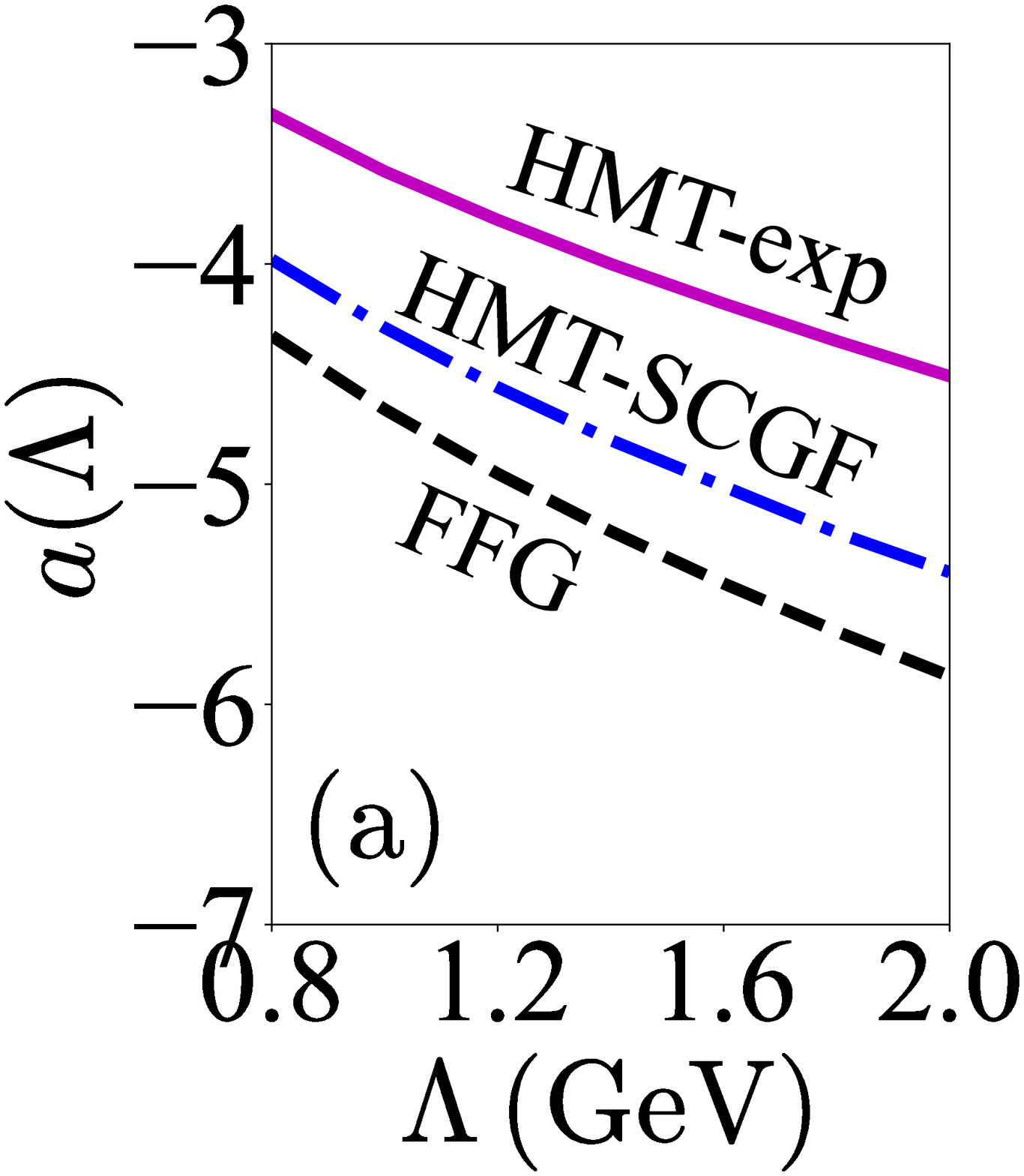}\quad
  \includegraphics[height=2.8cm]{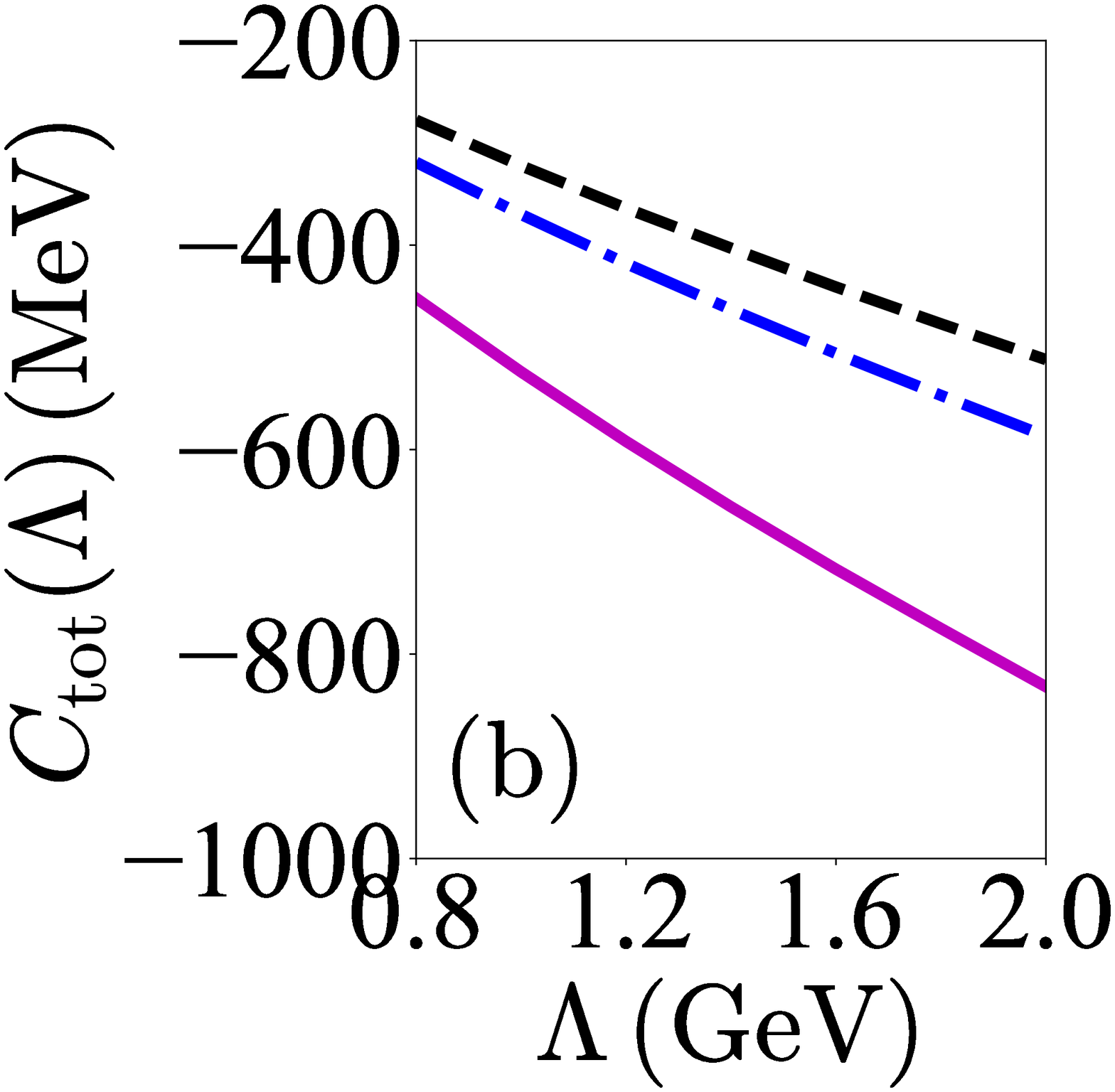}\quad
  \includegraphics[height=2.8cm]{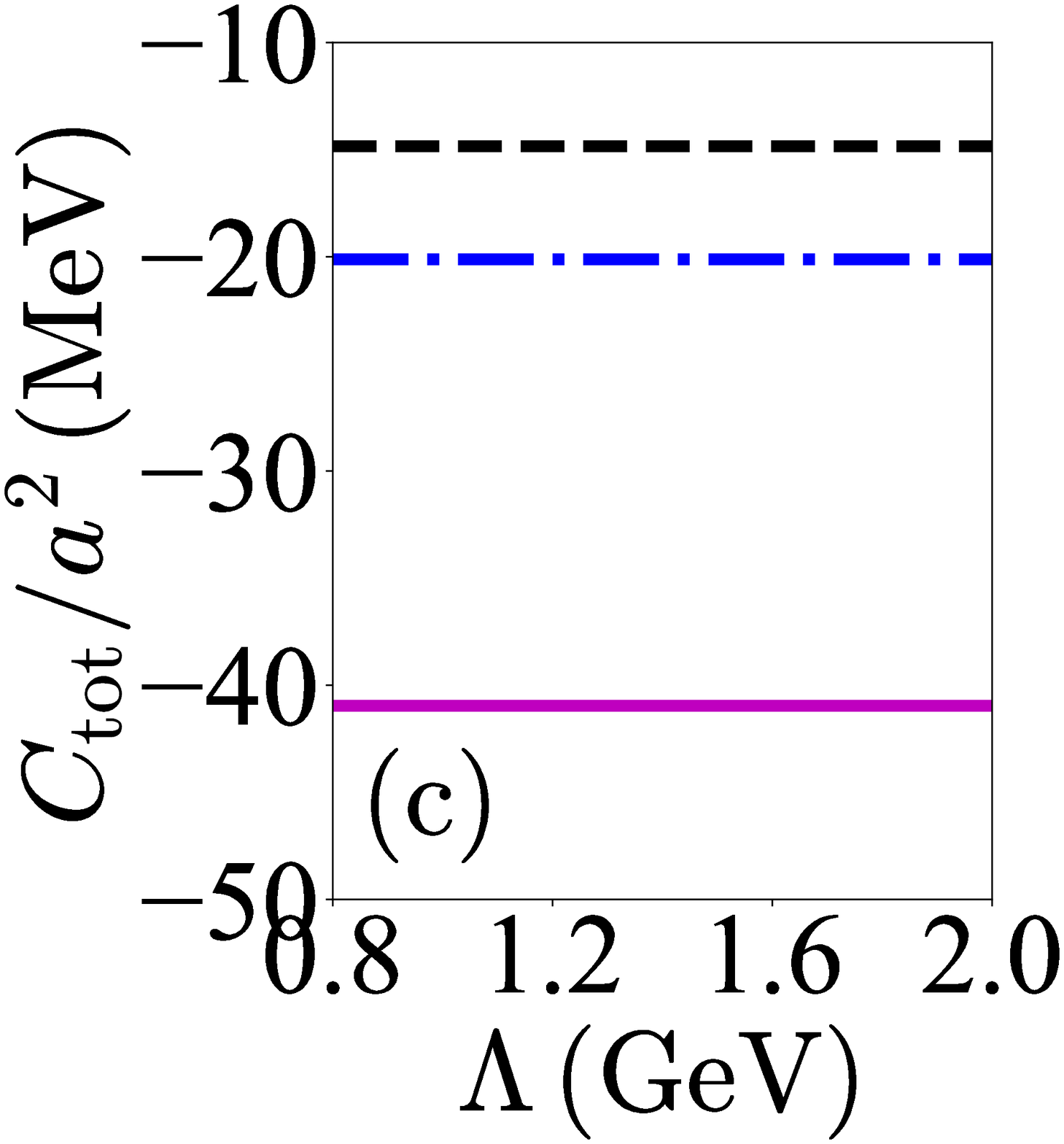}
  \caption{The $\Lambda$-dependence of the coupling constants $a$ (left) and $C_{\rm{tot}}$ (middle) and the ratio $C_{\rm{tot}}/a^2$ (right) in the three models.}
  \label{fig_CC-Lambda}
\end{figure}

We show in FIG.\,\ref{fig_CC-Lambda} the $\Lambda$-dependence of the coupling constants $a$ (left panel) and $C_{\rm{tot}}$ (middle panel) as well as the ratio $C_{\rm{tot}}^2/a$ (right panel) in the three models. The parameters $a$ and $C_{\rm{tot}}$ evolve according to $a=a^0(\Lambda/\Lambda_0)^{1/3}$ and $C_{\rm{tot}}=C_{\rm{tot}}^0(\Lambda/\Lambda_0)^{2/3}$, respectively, see (\ref{CC-aa}), and thus $C_{\rm{tot}}/a^2=C_{\rm{tot}}^0/(a^{0})^2=\rm{const.}$, with the constant depending on the fitting scheme, i.e., it depends on the values of $\rho_0,~E_0(\rho_0),~K_0,~M_0^{\ast}(\rho_0)$ and $U_0(\rho_0,0)$ via the specific model.
See the right panel of FIG.\,\ref{fig_CC-Lambda}, e.g., the ratio $C_{\rm{tot}}/a^2$ in the HMT-exp model is about 
$-40.95\,\rm{MeV}$ (see TABLE \ref{tab_para}).
One can also show numerically that as $\Lambda\to0$, both $a$ and $C_{\rm{tot}}$ approach zero, but their ratio keeps a constant.
The $\Lambda$-dependence of other coupling constants ($A_{\ell}^0,~A_{\rm{u}}^0,~C_{\ell},~C_{\rm{u}}$) could be obtained similarly, and would not be analyzed here further (see APPENDIX B for more discussions).

\subsection{SNM EOS $E_0(\rho)$ and Single-nucleon Potential $U_0(\rho,|\v{k}|)$}

In FIG.\,\ref{fig_ab_E0}, we show the SNM EOSs in the three models.
It is first interesting to see that the HMT models predict a
slightly harder SNM EOS at supra-saturation densities than the
FFG model, while by design they all have the same values of
$M_0^{\ast}$, $\rho_0$, $E_0(\rho_0)$ and $K_0$. Physically, this is
easy to understand because of the large contribution to the kinetic EOS by the
high momentum nucleons in the HMT models\,\cite{Cai16b}. For
example, the skewness of the SNM characterizing the high density
behavior of the EOS in the FFG model is found to be
$J_0(\rm{FFG})\approx-381\,\rm{MeV}$, while that in the two HMT
models is about $J_0(\rm{HMT-SCGF})\approx-376\,\rm{MeV}$ and
$J_0(\rm{HMT-exp})\approx-329\,\rm{MeV}$, respectively. The small
change in the skewness from the FFG model to the HMT-SCGF model can
be traced back to the enhancement factor
$\Upsilon_0=1+C_0(5\phi_0+3/\phi_0-8)$, which is unity in the FFG
model and becomes about 1.18/1.83 in the HMT-SCGF/HMT-exp model. It
is thus not surprising that the predictions on the pressure in SNM
is also very similar, and can all pass through the constraints from analyzing 
the collective flow date from relativistic heavy-ion collisions\,\cite{Dan02},
see FIG.\,\ref{fig_ab_Flow}.
It is necessary to point out that since the high momentum nucleon fraction in the HMT-SCGF is relatively lower than that in the HMT-exp model, the former is much closer to the FFG model (see the black dash line and the blue dash-dotted line shown in the inset of FIG.\,\ref{fig_ab_E0}).

\begin{figure}[h!]
\centering
  \includegraphics[width=6.5cm]{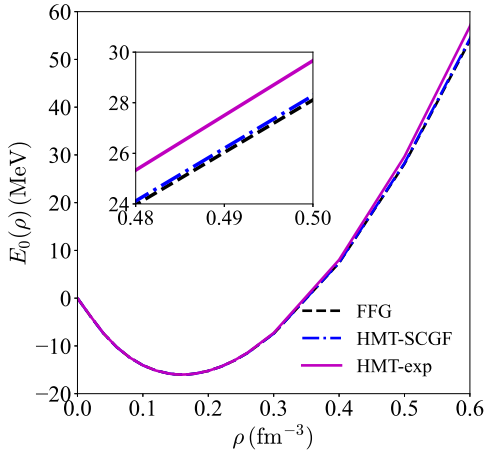}
  \caption{EOS of SNM in the FFG model and the HMT-SCGF as well as the HMT-exp models.}
  \label{fig_ab_E0}
\end{figure}

\begin{figure}[h!]
\centering
  \includegraphics[width=6.5cm]{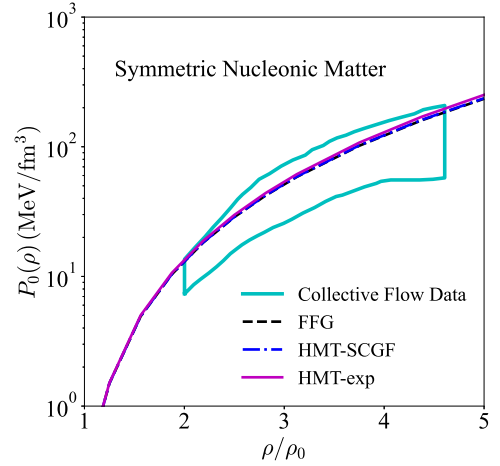}
  \caption{Comparison between the pressure $P_0$ of SNM with the experimental constraints from analyzing nuclear collective flows in heavy ion collisions\,\cite{Dan02}.}
  \label{fig_ab_Flow}
\end{figure}

In FIG.\,\ref{fig_ab_U0}, we show the single-nucleon potential in SNM
$U_0$ as a function of kinetic energy $E_{\rm{kin}}=[|\v{k}|^2+M^2]^{1/2}+U_0(\rho_0,|\v{k}|)-M$
in the three models. As discussed earlier, the SRC-induced HMT (slightly) enhances
the kinetic EOS of SNM, and in order to maintain the total EOS of SNM
at the saturation density to be consistent with empirical constraints the
potential part should be correspondingly reduced, which is reflected
in FIG.\,\ref{fig_ab_U0}. As the difference of the fraction of high
momentum nucleons in SNM and in PNM becomes larger, the kinetic EOS
of SNM is much more enhanced, leading to a much softer potential
part, i.e., the $U_0$ in the HMT-exp model
is smaller than that in the HMT-SCGF model. Also shown in FIG.\,\ref{fig_ab_U0} are the predictions on the $U_0(\rho_0,|\v{k}|)$
from other approaches, including the global relativistic fitting of
electron-scattering data up to about 1\,GeV\,\cite{Ham90} (blue ``+''); the
predictions by the neutron optical model up to about
200\,MeV\,\cite{LiX15} (dashed black band), and those from the chiral effective field
theories\,\cite{Hol16} (cyan band). It is obvious that the predictions on $U_0$
both in the HMT-SCGF and the HMT-exp models are consistent with these approaches, at kinetic energies $\lesssim600\,\rm{MeV}$.
In this sense, the momentum-dependence in the function (\ref{MFunc})
gives a reasonable description of the kinetic-energy-dependence of the single-nucleon isoscalar
potential. As the kinetic energy $E_{\rm{kin}}$ becomes larger, the
deviation between the predictions on the $U_0$ in the three models and
the global relativistic fitting also becomes larger, indicating the breakdown of the perturbative construction for $\Omega(\v{k},\v{k}')$.

\begin{figure}[h!]
\centering
  \includegraphics[width=6.5cm]{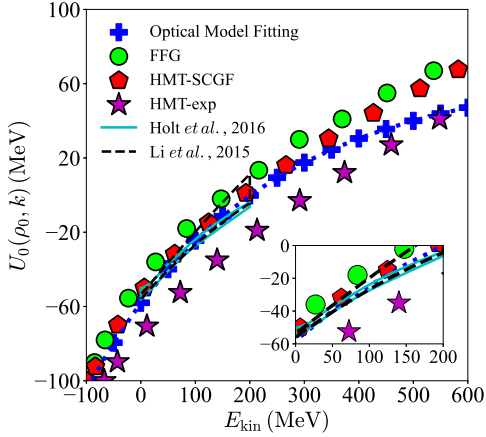}
  \caption{single-nucleon potential in SNM as a function of kinetic energy in the FFG model and
  HMT-SCGF/HMT-exp model. Constraints from other approaches are shown for comparison.}
  \label{fig_ab_U0}
\end{figure}

\begin{figure}[h!]
\centering
  \includegraphics[width=6.5cm]{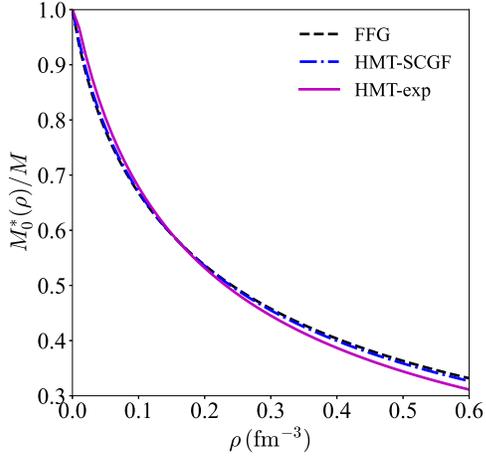}
  \caption{Nucleon effective k-mass as a function of density in the FFG and HMT-SCGF/HMT-exp models.}
  \label{fig_ab_M0}
\end{figure}

In FIG.\,\ref{fig_ab_M0}, the density dependence of the effective k-mass is
shown. It is straightforward to understand that the three models give very
similar results. On one hand, three points of the effective mass are
fixed, i.e., $M_0^{\ast}(0)/M=1$, $M_0^{\ast}(\rho_0)/M=0.58$ and
$M_0^{\ast}(\infty)/M=0$ (see Eq.\,(\ref{def_11}) since both $a$ and $C_{\rm{tot}}$ are negative, leading to $aC_{\rm{tot}}>0$). Moreover, the
$M_0^{\ast}/M$ monotonically decreases in the whole density range
and is concave, e.g., for large density $\rho$, one has from (\ref{def_11}) that $
M_0^{\ast}/M\sim\rho^{-2/3}$. Meeting all of these common constraints the
$M_0^{\ast}(\rho)/M$ in the three models behaves thus very similarly (see similar discussions given in Ref.\,\cite{Cai16b} for the nonlinear RMF model counterpart).

\subsection{Reduction of the Kinetic Symmetry Energy $E_{\textmd{sym}}^{\textmd{kin}}(\rho)$ and Enhancement of the Symmetry Potential $U_{\textmd{sym}}(\rho,|\v{k}|)$}

We now discuss effects of the SRC/HMT on the symmetry energy and isovector (symmetry) potential.
In FIG.\,\ref{fig_ab_Esym}, we show the density dependence of the nuclear
symmetry energy in the three models\,\cite{CLC18}.
Also shown include the constraints on the $E_{\rm{sym}}(\rho)$
around the saturation density from analyses of heavy-ion collisions\,\cite{Tsa12} and isobaric analog state studies\,\cite{Dan14}. Although the symmetry energy from the three models can pass through these constraints, they have very different behaviors at super-saturation
densities\,\cite{CLC18}. Specifically, as discussed in the above sections, the
reduction of the kinetic symmetry energy should be compensated by
the potential part to keep the total symmetry energy around $\rho_0$ consistent with certain empirical constraints\,\cite{LiBA13,Chen17,LiBA21}.
More quantitatively, the factor $\Upsilon_{\rm{sym}}$ (defined in Eq.\,(\ref{def_Upsilon})) in the HMT-SCGF model is found to be about
$\Upsilon_{\rm{sym}}\approx0.71$, which is not far different from
that in the FFG model (which is unity), while the
$\Upsilon_{\rm{sym}}$ factor in the HMT-exp model is about
$\Upsilon_{\rm{sym}}\approx-1.12$\,\cite{Cai15}, totally different from the FFG
prediction. Thus it is not surprising that the effects of HMT in the
HMT-exp model is much more apparent. 

In addition, there are two critical densities $\rho_1$ and $\rho_2$ corresponding to the point above which the symmetry energy starts decreasing and its zero-point when the symmetry energy vanishes. They are defined as $L(\rho_1)=0$ and $E_{\rm{sym}}(\rho_2)=0$, respectively. The latter case indicates the onset of the so-called isospin separation instability, namely it is energetically more favorable to split SNM into pure neutron matter and proton matter when the  symmetry energy becomes negative\,\cite{Kut93,Kut94,LiBA02}. Possible appearances and ramifications of such instability in heavy-ion reactions and neutron stars have been studied in several works, see., e.g., Refs.\,\cite{Kut93,Kut94,LiBA02,Szm06,wen,Kho96,Bas07,Ban00,Kubis1}.
Without considering the SRC-induced HMT, the FFG model predicts $\rho_1^{\rm{FFG}}\approx4.4\rho_0$ and $\rho_2^{\rm{FFG}}\approx11.2\rho_0$, while the HMT reduces the $\rho_1$ to $3.5\rho_0$ ($1.9\rho_0$) and the $\rho_2$ to $8.5\rho_0$ ($4.1\rho_0$) in the HMT-SCGF (HMT-exp) model.
We see that the experimental HMT constraints largely reduce the zero-point density $\rho_2$, with an effect of about 63\%.
Since the $\rho_2^{\rm{HMT}\mbox{-}\rm{exp}}\approx4.1\rho_0$ is quite low considering neutron star issues, it is expected that the reduction of the symmetry energy at supra-saturation densities in the HMT-exp model may have sizable effects on properties of neutron stars, e.g., the mass-radius relation.
On the other hand, the reduction of the $E_{\rm{sym}}(\rho)$ at sub-saturation densities (see the inset of FIG.\,\ref{fig_ab_Esym}) may also induce changes on, e.g., the core-crust transition densities in neutron stars.
These issues are investigated in some details in subsection \ref{sb_NS}.

Furthermore, we also show in FIG.\,\ref{fig_ab_Esym} two constraints on the $E_{\rm{sym}}(2\rho_0)$. The pink solid circle at $E_{\rm{sym}}(2\rho_0)\approx45\pm3\,\rm{MeV}$ is from chiral perturbation theories with consistent nucleon-nucleon and three-nucleon interactions up to the fourth-order\,\cite{Dri20}. The grey diamond at $E_{\rm{sym}}(2\rho_0)\approx51\pm13\,\rm{MeV}$ is the fiducial value from surveying analyses of both relativistic heavy-ion reactions and neutron star properties since GW170817\,\cite{LiBA21}. It is clearly seen that the calculations with the SRC/HMT-induced reduction of $E_{\rm{sym}}(2\rho_0)$ are consistent with these constraints\,\cite{LiBA21,Dri20,CL2021}.

\begin{figure}[h!]
\centering
  \includegraphics[width=6.5cm]{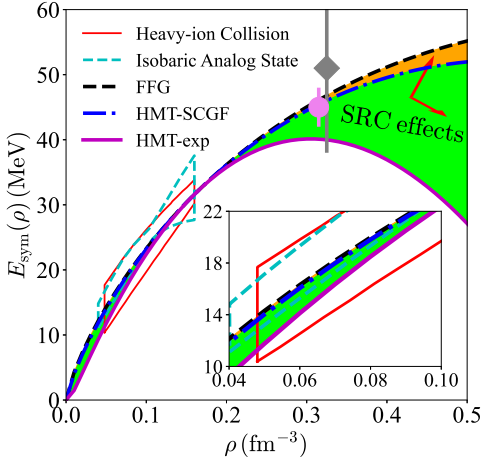}
  \caption{Density dependence of the symmetry energy in the FFG model and HMT-SCGF/HMT-exp model. Constraints from heavy-ion collisions\,\cite{Tsa12} and the isobaric analog state studies\,\cite{Dan14} are shown for comparison.
See also Ref.\,\cite{CLC18} and the text for details on the pink solid circle and the grey diamond.}
  \label{fig_ab_Esym}
\end{figure}

The reduction of the
symmetry energy found here is qualitatively consistent with the one from the nonlinear RMF
models\,\cite{Cai16b}. 
However, they give quantitatively different results. 
In particular, due to the specific structure of the nonlinear RMF model, the symmetry energy should never becomes negative.
On the other hand, the EOS in a non-relativistic EDF could be treated as an effective expansion in density. 
In order to investigate the probable origin of this quantitative difference, it is enough for our purpose to assume that the symmetry energy takes the form $E_{\rm{sym}}(\rho)=E_{\rm{sym}}^{\rm{kin}}(\rho_0)(\rho/\rho_0)^{2/3}+s_1(\rho/\rho_0)^{\alpha}+s_2(\rho/\rho_0)^{\beta}$, where $\beta>\alpha>2/3$, and $E_{\rm{sym}}^{\rm{kin}}(\rm{FFG})\approx12.3\,\rm{MeV}$.
The coefficients $s_1$ and $s_2$ can be determined via the values of $E_{\rm{sym}}(\rho_0)$ and $L\equiv L(\rho_0)$. Specifically, the expression for $s_2$ is given by,
\begin{equation}
s_2=\frac{1}{\beta-\alpha}\left(\alpha-\frac{2}{3}\right)E_{\rm{sym}}^{\rm{kin}}(\rho_0)+\frac{L-3\alpha E_{\rm{sym}}(\rho_0)}{\beta-\alpha}.
\end{equation}
The second term here is negative since $\alpha>2/3$ and thus $L-3\alpha E_{\rm{sym}}(\rho_0)<0$ while $\beta-\alpha>0$. 
Similarly, the coefficient of the first term of $E_{\rm{sym}}^{\rm{kin}}(\rho_0)$ is positive (due to the same consideration), thus:
\begin{enumerate}
\item[(a)]One has $s_2^{\rm{FFG}}>s_2^{\rm{HMT}}$ as $E_{\rm{sym}}^{\rm{kin}}(\rm{FFG})>E_{\rm{sym}}^{\rm{kin}}(\rm{HMT})$. 
\item[(b)]In addition, $s_2<0$ for the HMT-exp model, as the $E_{\rm{sym}}^{\rm{kin}}(\rho_0)\approx-13.8\,\rm{MeV}$ is negative. 
\end{enumerate}
The point (a) explains why the $E_{\rm{sym}}(\rho)$ considering the HMT should be reduced while the point (b) indicates that in the HMT-exp model (with a negative $E_{\rm{sym}}^{\rm{kin}}(\rho_0)$) the $E_{\rm{sym}}(\rho)$ must start decreasing above some critical density, and become negative at even higher densities. In our EDF, $\beta=4/3$ and one can find actually that the $s_2$'s in all three models are negative, see TABLE \ref{tab_para1}.
This explains why the reduction of symmetry energy using the non-relativistic EDF should be much stronger than the one in the nonlinear RMF models.

The reduction of the symmetry energy at supra-saturation densities also generates corresponding reduction of its curvature coefficient $K_{\rm{sym}}$. Quantitatively, its value changes from $-109$\,MeV in the FFG model to about
$-121\,\rm{MeV}$ ($-223$\,MeV) in the HMT-SCGF (HMT-exp) model.
While the $K_{\rm{sym}}\approx-223\,\rm{MeV}$ in the HMT-exp model is somewhat smaller than its current fiducial value of about $-107\pm 88$ MeV (see FIG.\,2 of Ref.\,\cite{LiBA21}), those from the FFG and the HMT-SCGF model are very consistent with the fiducial value. Thus, the general tendency is that the SRC/HMT reduces the symmetry energy curvature coefficient $K_{\rm{sym}}$. This may have some other consequences. For example, the pressure of neutron-rich nucleonic matter is given by
\begin{align}
P/3\rho\approx&L(\rho)\delta^2+3\rho\frac{\d E_0(\rho)}{\d\rho}\notag\\
\approx&
L\delta^2\left(1+3\chi+\frac{9}{2}\chi^2\right)
+K_0\chi(1+3\chi)+\frac{1}{2}\chi^2J_0\notag\\
&+K_{\rm{sym}}\delta^2\chi(1+3\chi)\left[1+\frac{J_{\rm{sym}}}{K_{\rm{sym}}}
\frac{\chi}{2(1+3\chi)}\right]\notag\\
\approx&L\delta^2\left(1+3\chi+\frac{9}{2}\chi^2\right)
+K_0\chi(1+3\chi)\notag\\
&+\frac{1}{2}\chi^2J_0+K_{\rm{sym}}\delta^2\chi(1+3\chi),
\end{align}
where the skewness coefficient $J_{\rm{sym}}$ is neglected in the last step.
Since the $K_0$ and $L$ are fixed in our study,  and as discussed earlier the $J_0$ is slightly increased by the SRC/HMT, the reduction of $K_{\rm{sym}}$ may reduce the pressure $P$ at densities below about $3\rho_0$ above which the $J_{\rm{sym}}$ becomes important (the $\chi$ is 2/3 for $\rho=3\rho_0$ and thus $\chi/[2(1+3\chi)]=1/9$). Moreover,  the reduction of $K_{\rm{sym}}$ may also modify the saturation line (on which the pressure is zero) of ANM. More quantitatively, the incompressibility coefficient of ANM along its saturation line is $K(\delta)=K_0+K_{\rm{sat},2}\delta^2$\,\cite{LWChen2009} where the $K_{\rm{sat},2}$ is
 \begin{equation}
K_{\rm{sat,2}}=K_{\rm{sym}}-6L-\frac{J_0L}{K_0}.
\end{equation}
It changes from $-365\,\rm{MeV}$ in the FFG model to $-378\,\rm{MeV}$ ($-492\,\rm{MeV}$) in the HMT-SCGF (HMT-exp) model.  
It is seen that the maximum reduction is about 127\,\rm{MeV}.

\begin{table*}[t!]
\caption{Parameterizations of the EOS of SNM and the symmetry energy as
well as the nucleon potential. Momentum $|\v{k}|$ is measured in MeV and
the density $\rho$ in $\rm{fm}^{-3}$,
$E_0(\rho),E_{\rm{sym}}(\rho),U_0(\rho,|\v{k}|)$ and
$U_{\rm{sym}}(\rho,|\v{k}|)$ are in MeV.}\label{tab_para1}
{\normalsize
\begin{tabular}{c||c}
\hline\hline Model for $n_{\v{k}}^J$& Parametrization of the Physical Quantity\\\hline 
\hline FFG  &$x_{\rm{SNM}}^{\rm{HMT}}=0\%,x_{\rm{PNM}}^{\rm{HMT}}=0\%$
\\
\hline $E_0(\rho)$ &
$E_0(\rho)\approx75.00\rho^{2/3}+1695.42\rho^{1.04}-551.76\rho-1378.02\rho[1-0.81\rho^{1/3}+0.55\rho^{2/9}]$\\
\hline $U_0(\rho,|\v{k}|)$ &$U_0(\rho,|\v{k}|)\approx3450.68\rho^{1.04}-1103.52\rho-2756.04\rho[1-0.04|\v{k}|^{1/2}\rho^{1/6}+0.08|\v{k}|^{1/3}\rho^{1/9}]$\\
\hline $E_{\rm{sym}}(\rho)$ & $E_{\rm{sym}}(\rho)\approx41.67\rho^{2/3}+1101.25\rho-1041.63\rho^{1.04}-229.34\rho^{4/3}+180.92\rho^{11/9}$\\
\hline $U_{\rm{sym}}(\rho,|\v{k}|)$ &
$U_{\rm{sym}}(\rho,|\v{k}|)\approx-2083.27\rho^{1.04}+\rho[2202.51-40.52|\v{k}|^{1/2}\rho^{1/6}+69.80|\v{k}|^{1/3}\rho^{1/9}]$\\\hline
\hline HMT-SCGF  &$x_{\rm{SNM}}^{\rm{HMT}}=12\%,x_{\rm{PNM}}^{\rm{HMT}}=4\%$
\\
\hline $E_0(\rho)$   & $E_0(\rho)\approx88.28\rho^{2/3}-729.40\rho^{0.93}+2071.36\rho-1580.78\rho[1-0.79\rho^{1/3}+0.57\rho^{2/9}]$\\
\hline $U_0(\rho,|\v{k}|)$  &$U_0(\rho,|\v{k}|)\approx-1405.76\rho^{0.93}+4142.72\rho-3161.56\rho[1-0.04|\v{k}|^{1/2}\rho^{1/6}+0.08|\v{k}|^{1/3}\rho^{1/9}]$\\
\hline $E_{\rm{sym}}(\rho)$ & $E_{\rm{sym}}(\rho)\approx29.53\rho^{2/3}-146.18\rho+275.30\rho^{0.93}-467.07\rho^{4/3}+343.07\rho^{11/9}$\\
\hline $U_{\rm{sym}}(\rho,|\v{k}|)$ &
$U_{\rm{sym}}(\rho,|\v{k}|)\approx550.60\rho^{0.93}+\rho[-292.37-57.07|\v{k}|^{1/2}\rho^{1/6}+106.42|\v{k}|^{1/3}\rho^{1/9}]$\\\hline
\hline HMT-exp  &$x_{\rm{SNM}}^{\rm{HMT}}=28\%,x_{\rm{PNM}}^{\rm{HMT}}=1.5\%$
\\
\hline $E_0(\rho)$  & $E_0(\rho)\approx137.31\rho^{2/3}-131.89\rho^{0.67}+2129.47\rho-2240.19\rho[1-0.77\rho^{1/3}+0.63\rho^{2/9}]$\\
\hline $U_0(\rho,|\v{k}|)$ &$U_0(\rho,|\v{k}|)\approx-220.18\rho^{0.67}+4258.93\rho-4480.39\rho[1-0.04|\v{k}|^{1/2}\rho^{1/6}+0.08|\v{k}|^{1/3}\rho^{1/9}]$\\
\hline $E_{\rm{sym}}(\rho)$ & $E_{\rm{sym}}(\rho)\approx-46.83\rho^{2/3}+392.52\rho-31.06\rho^{0.67}-1837.48\rho^{4/3}+1421.05\rho^{11/9}$\\
\hline $U_{\rm{sym}}(\rho,|\v{k}|)$ &
$U_{\rm{sym}}(\rho,|\v{k}|)\approx-62.11\rho^{0.67}+\rho[785.04-156.10|\v{k}|^{1/2}\rho^{1/6}+348.23|\v{k}|^{1/3}\rho^{1/9}]$\\
\hline\hline
\end{tabular}}
\end{table*}

\begin{figure}[h!]
\centering
  \includegraphics[height=3.8cm]{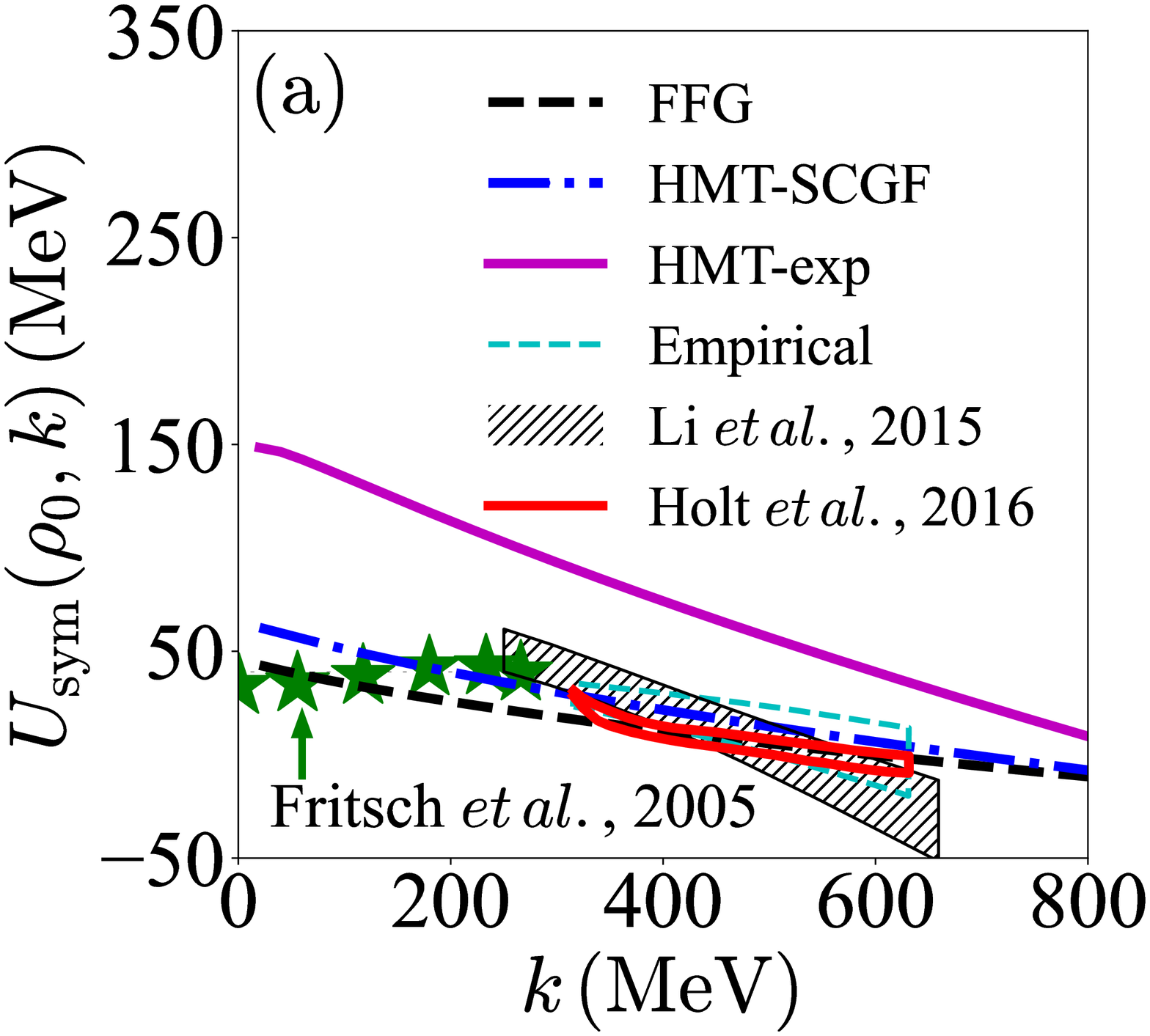}
  \includegraphics[height=3.8cm]{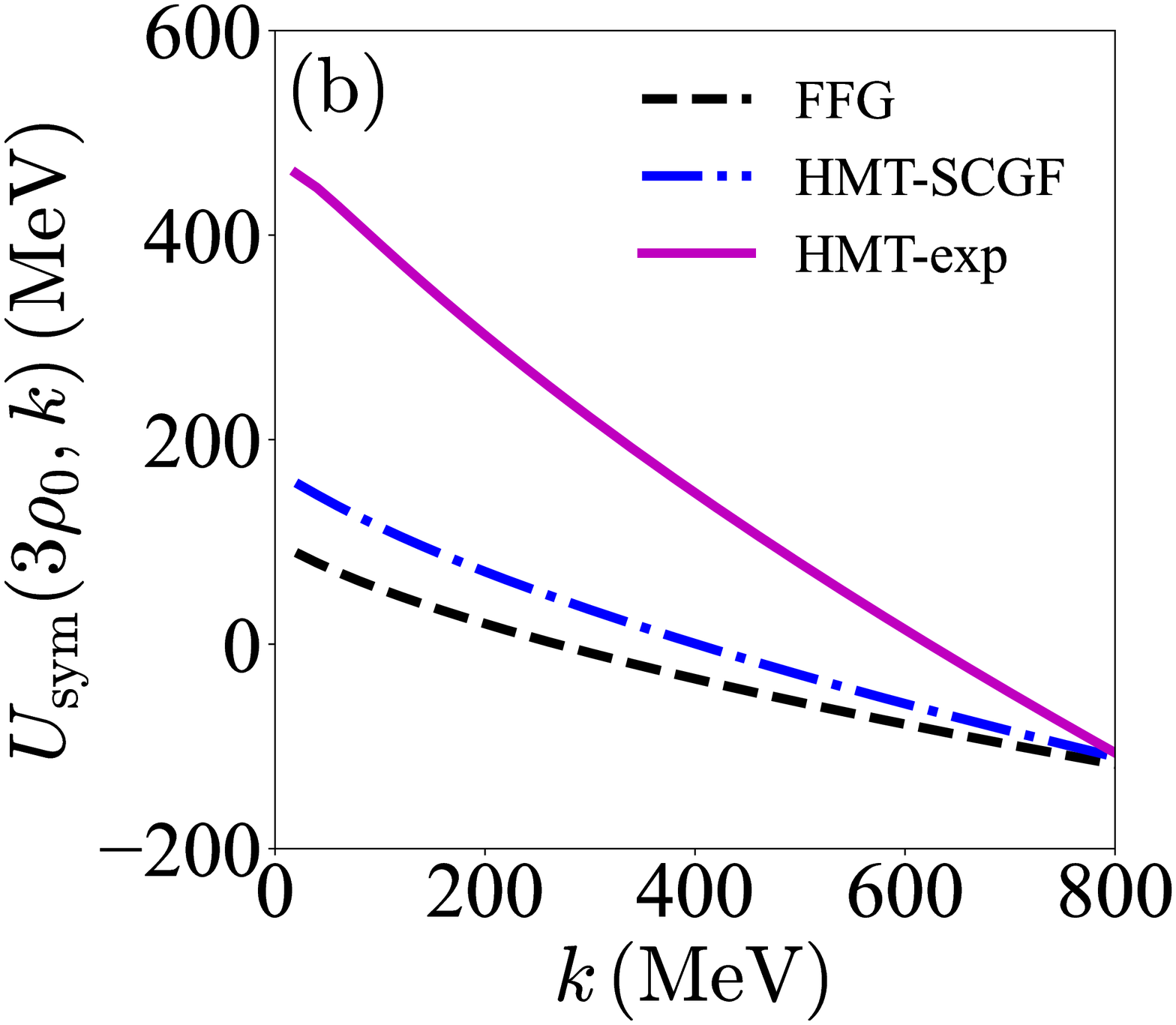}
  \caption{Symmetry potential in the FFG model and the HMT-SCGF/HMT-exp model at $\rho=\rho_0$ (left) and
  at $\rho=3\rho_0$ (right).
Constraints on the symmetry potential at $\rho=\rho_0$ from other approaches are shown for
comparisons.}
  \label{fig_ab_Usym}
\end{figure}

In FIG.\,\ref{fig_ab_Usym}, we show the momentum dependence of the
symmetry potential $U_{\rm{sym}}(\rho,|\v{k}|)$ at $\rho_0$ (left panel) and $3\rho_0$ (right panel), respectively. The predictions on
$U_{\rm{sym}}(\rho_0,|\v{k}|)$ from other microscopic
approaches\,\cite{LiX15,Hol16,Fri05} are also shown for comparisons. The empirical band is parametrized as $U_{\rm{sym}}(E)\approx
(28\pm6)\,\rm{MeV}-(0.15\pm0.05)E$\,\cite{Hol16}. It can be found 
that the uncertainties on the symmetry potential are larger than
those of the isoscalar potential $U_0$. This is mainly due to the poorly-known isospin dependence of nuclear interactions\,\cite{LCK08,LiBA2018PPNP}.
While the prediction by the HMT-exp on the
$U_{\rm{sym}}(\rho_0,|\v{k}|)$ has certain deviation from the microscopic model predictions, the FFG and HMT-SCGF model
predictions are quite consistent with the optical model fitting\,\cite{LiX15} as well as the chiral effective theory prediction\,\cite{Hol16}.
Since there exists no direct constraint on the symmetry potential at, e.g., a large density $\rho=3\rho_0$, and it is beyond the current application domain of microscopic nuclear many-body theories, it will be interesting to test the predictions shown in the right window of  FIG.\,\ref{fig_ab_Usym}, thus the SRC/HMT effects in dense neutron-rich matter, using nuclear reactions with high-energy radioactive beams. Of course, for this purpose it is first necessary to investigate experimental observables sensitive to the high-density symmetry potential. The results presented here especially the various analytical expressions and parameterizations will facilitate the explorations of this topic using transport models for nuclear reactions with radioactive beams. 
Listed in TABLE \ref{tab_para1} are parametrizations of the SNM EOS $E_0(\rho)$, nucleon isoscalar potential $U_0(\rho,|\v{k}|)$, symmetry energy $E_{\rm{sym}}(\rho)$ and isovector potential $U_{\rm{sym}}(\rho,|\v{k}|)$ corresponding to the three models (with their parameters given in TABLE \ref{tab_para}), for future applications.
\begin{figure}[h!]
\centering
  \includegraphics[height=3.3cm]{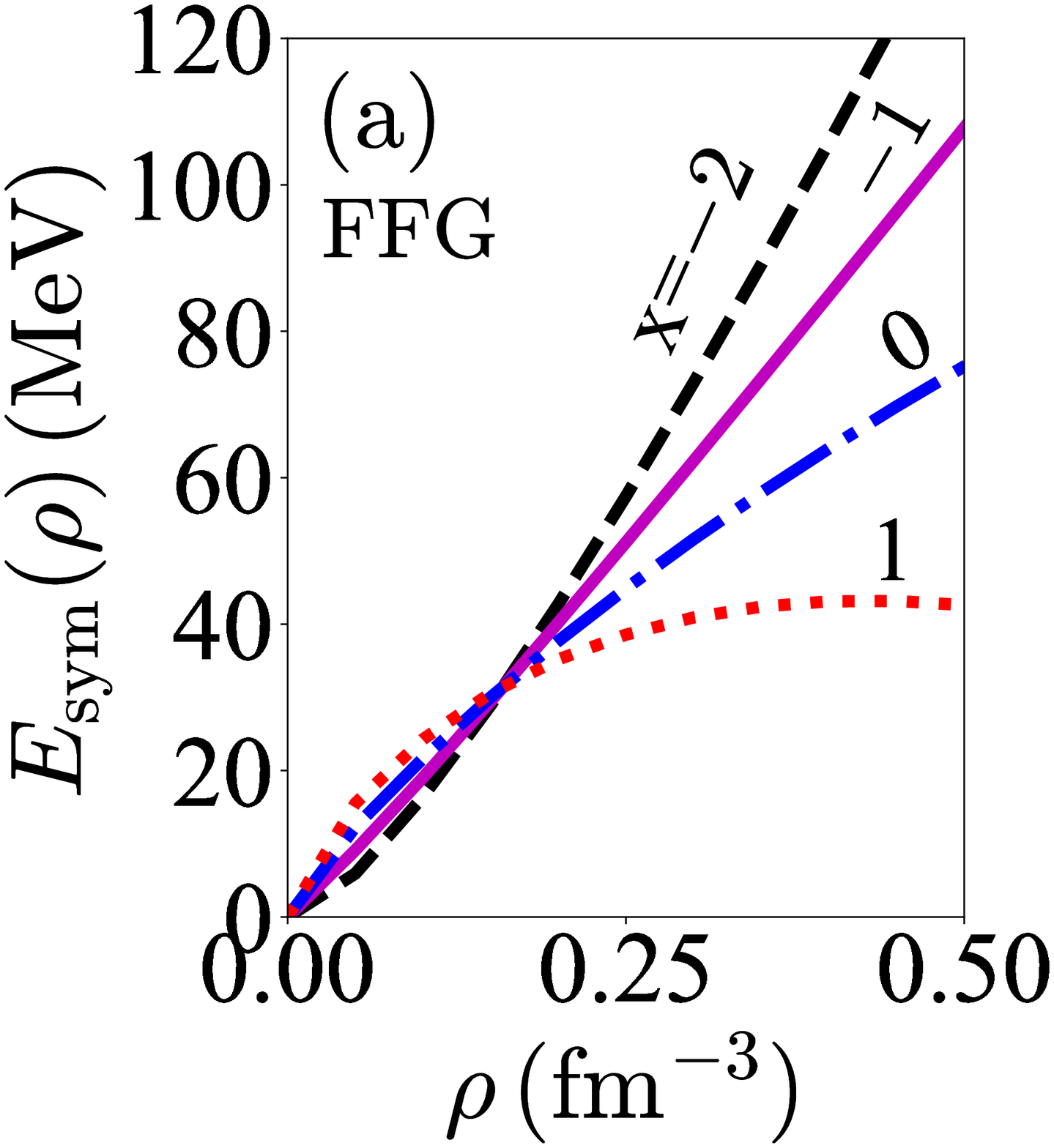}
  \includegraphics[height=3.3cm]{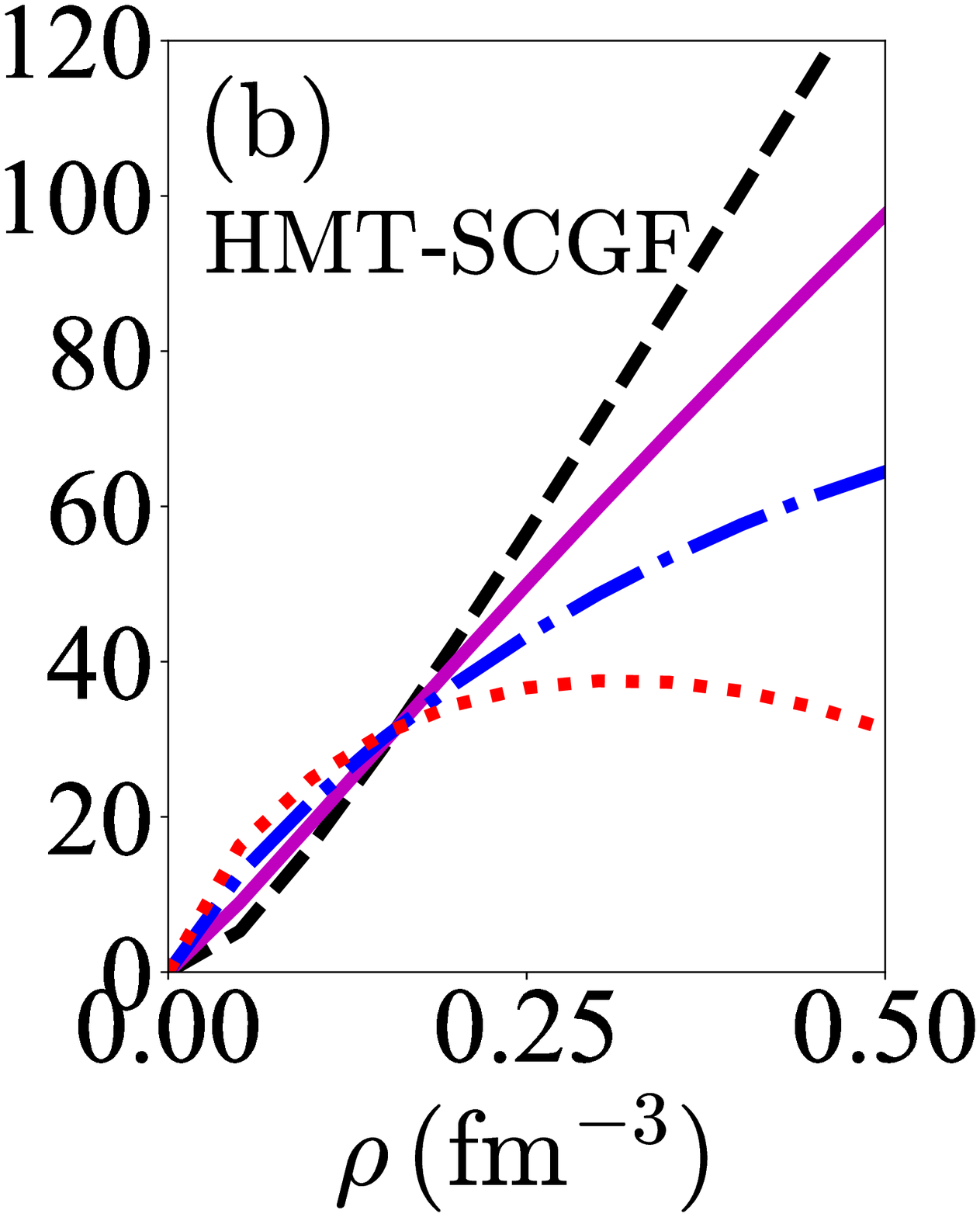}
  \includegraphics[height=3.3cm]{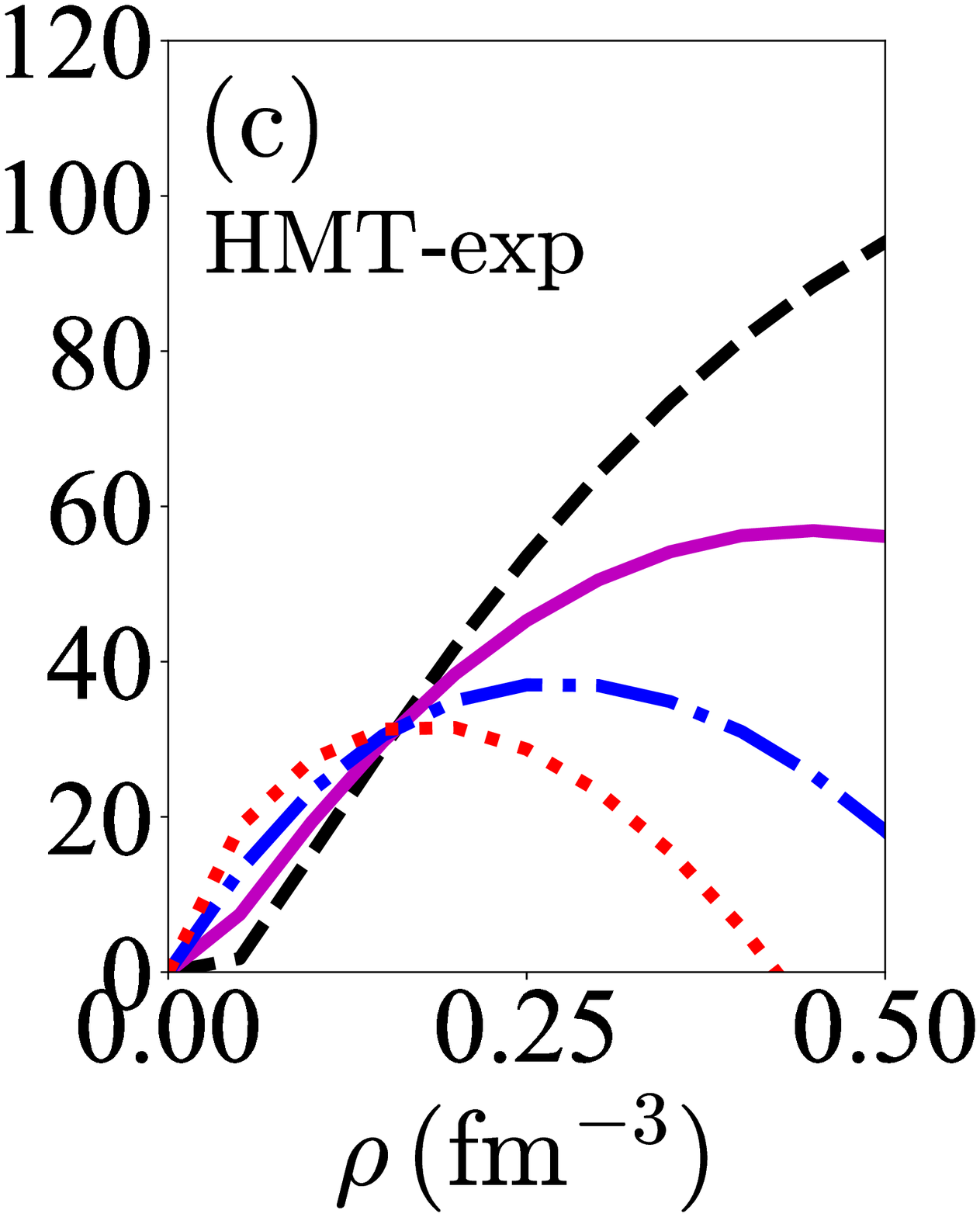}
  \caption{Symmetry energy with different effective three-nucleon force parameter $x$ in the three models indicated. }
  \label{fig_ab_Esym_x}
\end{figure}

In our calculations up to now, the effective three-nucleon force parameter $x$ is self-consistently determined by fixing the $L(\rho_0)$, see TABLE \ref{tab_para}.
Since the $x$ parameter only affects the slope parameter of the symmetry energy instead of its magnitude at $\rho_0$ itself, we can adjust $x$ to investigate how the density dependence of $E_{\rm{sym}}(\rho)$ changes when the SRC-induced HMT is taken into account. In FIG.\,\ref{fig_ab_Esym_x}, the symmetry energy obtained within the three
models indicated by only changing the $x$ parameter (corresponding to changing only the $L$ coefficient) while fixing at the same time all the other seven parameters listed
in TABLE \ref{tab_para}, is shown. For the FFG model, the $x$-dependence of $E_{\rm{sym}}(\rho)$ is the same as in the conventional MDI
model\,\cite{Das2003,Che05,LCK08,XuJ10,XuJ15}. In the presence of the SRC/HMT, the large
uncertainties of $E_{\rm{sym}}(\rho)$ at high densities can come from the poorly known physics of either the three-nucleon forces and/or the SRC/HMT.

\begin{figure}[h!]
\centering
  \includegraphics[height=3.7cm]{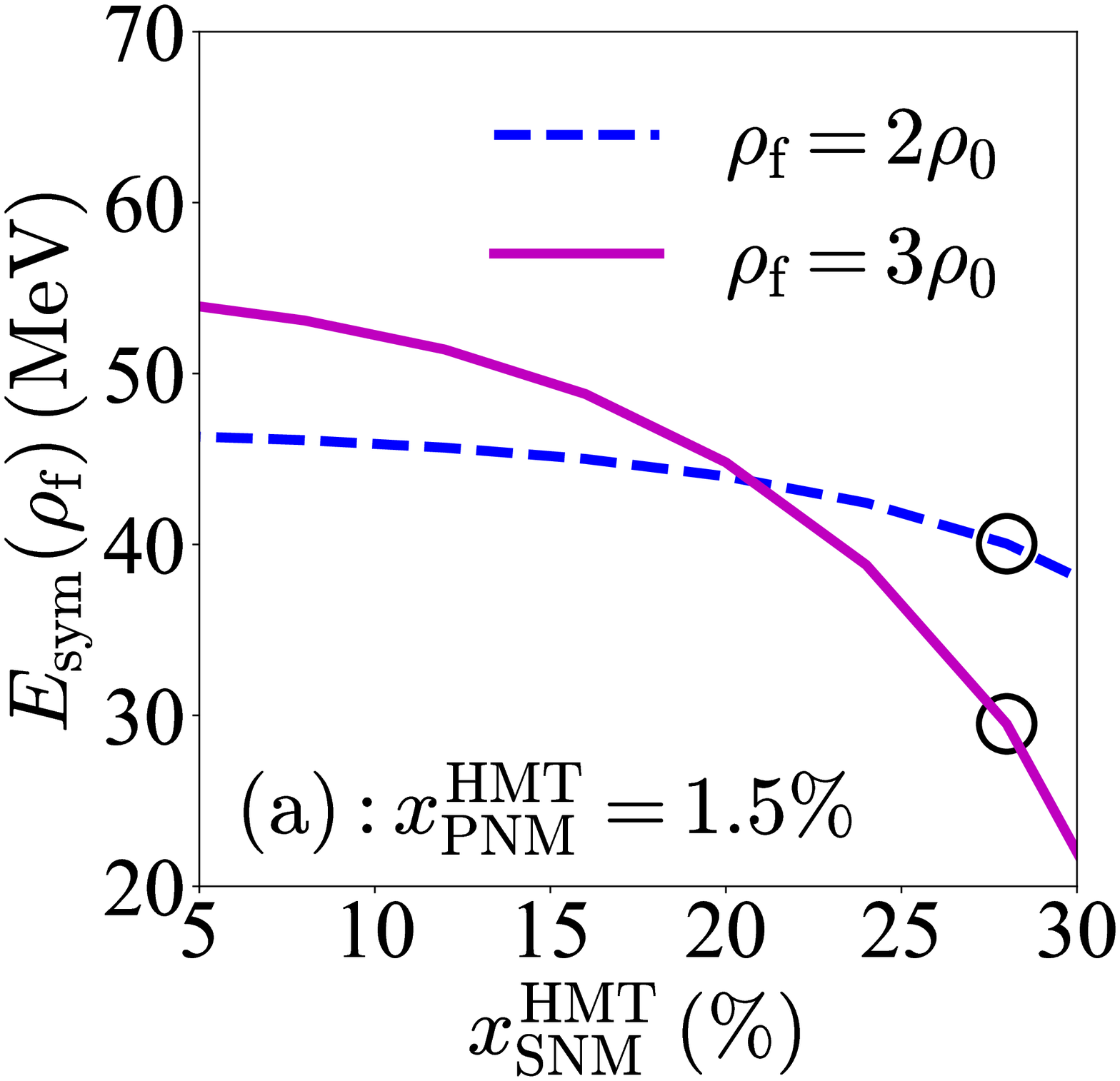}\quad
  \includegraphics[height=3.7cm]{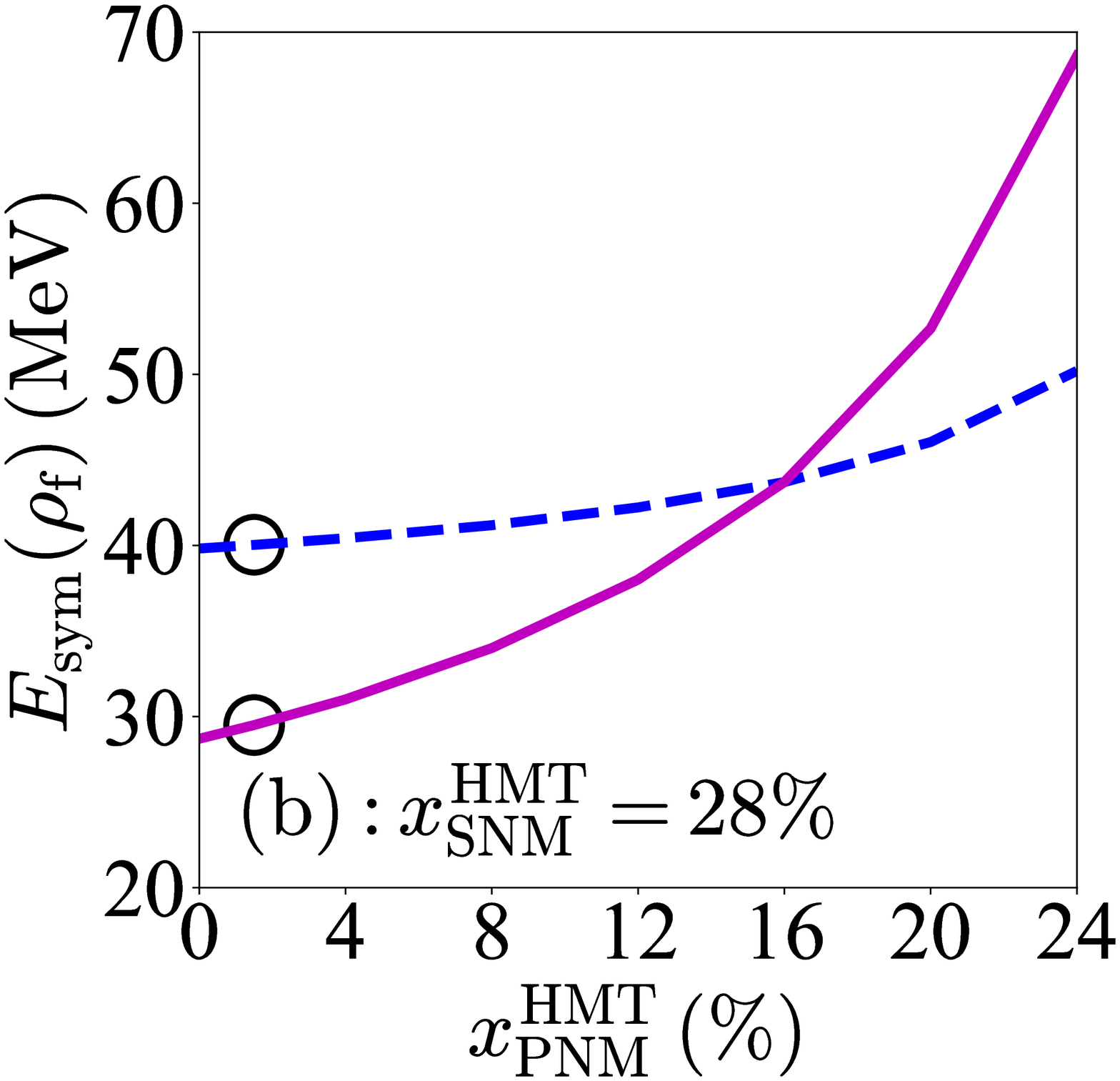}
  \caption{The symmetry energy at $2\rho_0$ and $3\rho_0$ with different fractions of high momentum nucleons in SNM and/or in PNM, respectively. The black circles are for the default set on $x_{\rm{SNM}}^{\rm{HMT}}$ and $x_{\rm{PNM}}^{\rm{HMT}}$, i.e., $x_{\rm{SNM}}^{\rm{HMT}}=28\%$ and $x_{\rm{PNM}}^{\rm{HMT}}=1.5\%$.}
  \label{fig_ab_EsymFrac}
\end{figure}

As discussed in the INTRODUCTION, there are still uncertainties about the fraction of high momentum nucleons in either SNM and/or PNM and thus its isospin dependence\,\cite{Hen14,Rio14}. Thus, it is interesting to study how the relevant quantities change when the HMT fractions in SNM and/or PNM are modified.
In FIG.\,\ref{fig_ab_EsymFrac}, we show the symmetry energy at two
reference densities, i.e., $\rho_{\rm{f}}=2\rho_0$ and
$\rho_{\rm{f}}=3\rho_0$, by continuously changing the high
momentum nucleon fraction $x_{\rm{SNM}}^{\rm{HMT}}$ in SNM when fixing the $x_{\rm{PNM}}^{\rm{HMT}}$ in PNM at the
default value of the HMT-exp model (left panel) (in PNM when fixing
that in SNM at the default value of the HMT-exp model (right panel)), respectively. 
The values of $C_0$ and $C_{\rm{n}}^{\rm{PNM}}=C_0(1+C_1)$ are fixed at 0.161 and 0.12, respectively, while the values of  $\phi_0$ and $\phi_1$ are readjusted correspondingly according to the given $x_{\rm{SNM}}^{\rm{HMT}}$ and $x_{\rm{PNM}}^{\rm{HMT}}$ values. The black circles are for the default parameter set of $x_{\rm{SNM}}^{\rm{HMT}}$ and $x_{\rm{PNM}}^{\rm{HMT}}$, i.e., $x_{\rm{SNM}}^{\rm{HMT}}=28\%$ and $x_{\rm{PNM}}^{\rm{HMT}}=1.5\%$.
As the difference between the $x_{\rm{SNM}}^{\rm{HMT}}$ and
$x_{\rm{PNM}}^{\rm{HMT}}$, i.e., $\delta x=x_{\rm{SNM}}^{\rm{HMT}}-x_{\rm{PNM}}^{\rm{HMT}}$, increases, the reduction of the
$E_{\rm{sym}}(\rho_{\rm{f}})$ becomes much more apparent. While the
symmetry energy at the saturation density is fixed (at $31.6\,\rm{MeV}$),
these effects on $E_{\rm{sym}}(2\rho_0)$ are relatively minor compared with those at $\rho=3\rho_0$.

\begin{figure}[h!]
\centering
  \includegraphics[height=3.6cm]{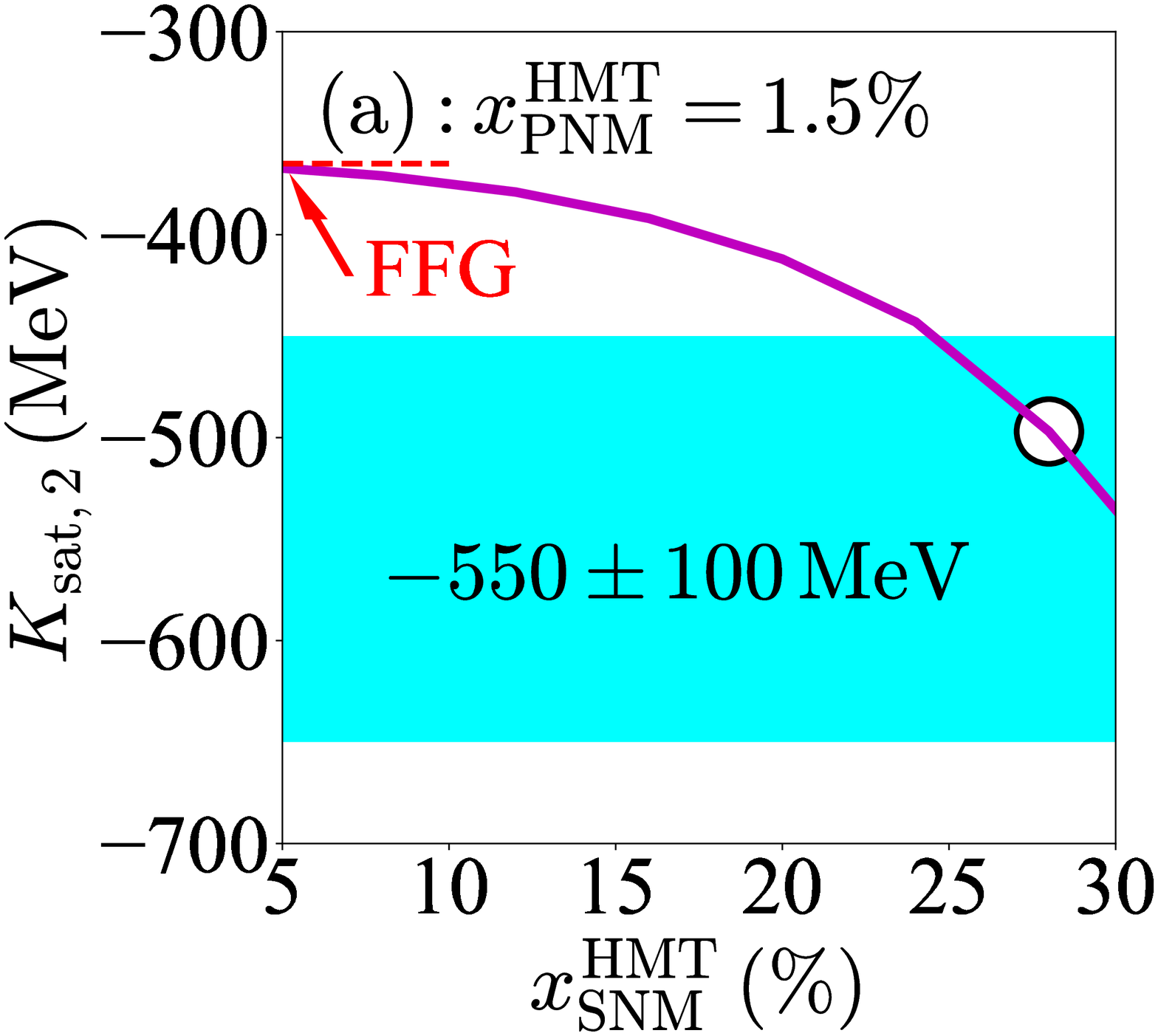}\quad
  \includegraphics[height=3.6cm]{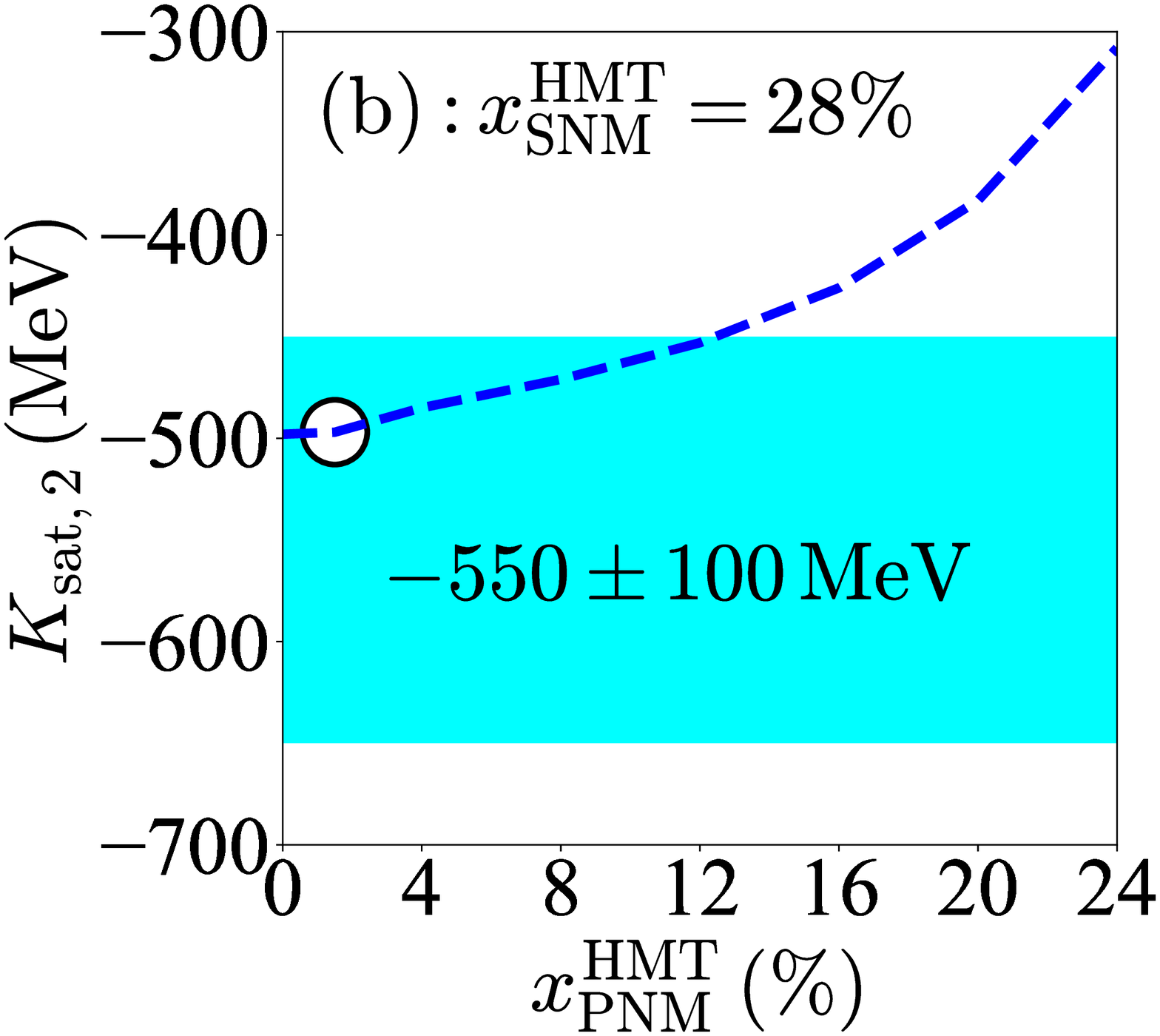}
  \caption{The same as FIG.\,\ref{fig_ab_EsymFrac} but for the isospin incompressibility coefficient $K_{\rm{sat},2}=K_{\rm{sym}}-6L-J_0L/K_0$. }
  \label{fig_Ksat2_frac}
\end{figure}

Similarly, we show in FIG.\,\ref{fig_Ksat2_frac} the dependence of the isospin incompressibility coefficient $K_{\rm{sat},2}$ on the high momentum nucleon fraction.
As the $L$ and $K_0$ are fixed in our calculations, the dependence of the coefficient $K_{\rm{sat},2}$ on $x_{\rm{SNM}}^{\rm{HMT}}$ and/or
$x_{\rm{PNM}}^{\rm{HMT}}$ mainly reflects the dependence of the curvature coefficient $K_{\rm{sym}}$ and the skewness parameter $J_0$ on them.
The $K_{\rm{sat},2}$ shown in FIG.\,\ref{fig_Ksat2_frac} is in good agreement with the estimate of $K_{\rm{sat},2}\approx-550\pm100\,\rm{MeV}$ from analyzing different types
of experimental data currently available\,\cite{Col14}, as shown using the cyan bands.
For example, if $x_{\rm{PNM}}^{\rm{HMT}}\approx1.5\%$ is fixed, then the constraint on $K_{\rm{sat},2}$ leads to about $x_{\rm{SNM}}^{\rm{HMT}}\gtrsim24\%$. Similarly, if $x_{\rm{SNM}}^{\rm{HMT}}\approx28\%$ is fixed, then we can roughly find the constraint that $x_{\rm{PNM}}^{\rm{HMT}}\lesssim12\%$.

\subsection{SRC/HMT Effects on the Proton Fraction $x_{\textmd{p}}$, Core-crust Transition Density $\rho_{\textmd{t}}$ and pressure $P_{\textmd{t}}$ as well as the Mass-radius Relation of Cold Neutron Stars at $\beta$-equilibrium}\label{sb_NS}

As an example of applying the Gogny-like EDFs encapsulating the SRC/HMT in astrophysical studies, we investigate here the density profile of proton fraction $x_{\textmd{p}}(\rho)=[1-\delta(\rho)]/2$, the core-crust transition density $\rho_{\rm{t}}$ and transition pressure $P_{\rm{t}}$ as well as the mass-radius correlation in cold neutron stars at $\beta$-equilibrium.
The transition density $\rho_{\rm{t}}$ is the nucleon number
density that separates the liquid core from the inner crust in
neutron stars. It plays an important role in determining many
properties of neutron
stars\,\cite{Hor01,Pro06,Duc08a,Duc08b,XuJ09}.
One simple and widely used approach to determine the core-crust transition density $\rho
_{\rm{t}}$ is the thermodynamical method. Normally, neutron star matter has to obey the intrinsic stability
condition\,\cite{Lat04,XuJ09,NBZ19,Kub07}
\begin{align}
\mathcal{U}_{\rm{ther}}(\rho)=&2\rho \frac{\partial E(\rho
,x_{\rm{p}})}{\partial \rho }+\rho ^{2}\frac{\partial
^{2}E(\rho ,x_{\rm{p}})}{\partial \rho ^{2}}\notag\\
& -\left. \left( \frac{\partial ^{2}E(\rho ,x_{\rm{p}})}{\partial
\rho
\partial x_{\rm{p}}}\rho \right) ^{2}\right/ \frac{\partial ^{2}E(\rho ,x_{\rm{p}})%
}{\partial x_{\rm{p}}^{2}}\notag\\
=&\rho^2\left(\frac{\partial^2E(\rho,x_{\rm{p}})}{\partial x_{\rm{p}}^2}\right)^{-1} 
\left[\frac{\partial\mu_{\rm{n}}}{\partial\rho_{\rm{n}}}\frac{\partial \mu_{\rm{p}}}{\partial\rho_{\rm{p}}}-\left(\frac{\partial\mu_{\rm{n}}}{\partial\rho_{\rm{p}}}\right)^2\right]>0. \label{Vther}
\end{align}
In the above, the density profile $x_{\rm{p}}=x_{\rm{p}}(\rho)$ is determined by the $\beta$-equilibrium and charge neutrality condition of neutron star matter consisting of neutrons, protons and electrons (the core-crust transition occurs around $\rho_0/3\mbox{$\sim$}\rho_0/2$ where muons generally do not emerge). It is uniquely determined by the density dependence of nuclear symmetry energy\,\cite{Lat04}.
Once the above stability condition is broken, the speed of sound in the uniform outer core becomes imaginary and small density oscillations will then grow exponentially, indicating the onset of a transition from the uniform liquid core to the crust.  
In (\ref{Vther}), $E(\rho ,x_{\rm{p}})$ is the energy per nucleon in
the $\beta $-equilibrated neutron star matter.  The corresponding pressure is obtained as $P=P_{\rm{N}}+P_{\rm{e}}$ with
$P_{\rm{N}}$ and $P_{\rm{e}}$ denoting contributions from nucleons and electrons, respectively.
The electron pressure $P_{\rm{e}}$ could be obtained through the FFG model\,\cite{XuJ09}, i.e., 
\begin{equation}\label{dd_e}
\varepsilon_{\rm{e}}(\rho,\delta)=\eta_{\rm{e}}\Phi_{\rm{e}}(t_{\rm{e}}),
\end{equation}
here \begin{equation}
\eta_{\rm{e}}={m_{\rm{e}}}/{8\pi^2\lambda_{\rm{e}}^3},~~
\lambda_{\rm{e}}={1}/{m_{\rm{e}}},~~t_{\rm{e}}=\lambda_{\rm{e}}\left(3\pi^2\rho_{\rm{e}}\right)^{1/3}
,\end{equation} and,
\begin{equation}\label{dd_e1}
\Phi_{\rm{e}}(t_{\rm{e}})=t_{\rm{e}}\left(1+2t_{\rm{e}}^2\right)\sqrt{1+t_{\rm{e}}^2}-\ln\left(t_{\rm{e}}+\sqrt{1+t_{\rm{e}}^2}\right)
,\end{equation} 
where $m_{\rm{e}}$ and $\rho_{\rm{e}}$ are the electron mass and density respectively.
Consequently, $P_{\rm{e}}(\rho,\delta)=\rho_{\rm{e}}\mu_{\rm{e}}-\varepsilon_{\rm{e}}(\rho,\delta)
$, with $
\mu_{\rm{e}}=[{k_{\rm{e}}^2+m_{\rm{e}}^2}]^{1/2}\approx k_{\rm{e}}$ and $k_{\rm{e}}=(3\pi^2\rho_{\rm{e}})^{1/3}$.

Shown in FIG.\,\ref{fig_ab_rhot} are our results for the core-crust transition density $\rho_{\rm{t}}$ (upper panel) and pressure $P_{\rm{t}}$ (lower panel) as functions of the slope parameter $L(\rho_{\rm{r}})$ at a sub-saturation density $\rho_{\rm{r}}=0.11\,\rm{fm}^{-3}$.}
It is clear that the SRC/HMT affects significantly the
transition density $\rho_{\rm{t}}$, e.g., a reduction as high as about
58\% with the HMT-exp parameter set compared to the FFG model. More quantitatively, the crust-core transition density in the FFG model is found to be about
$\rho_{\rm{t}}\approx0.086\,\rm{fm}^{-3}$, while that in the HMT
models is about $\rho_{\rm{t}}\approx0.079\,\rm{fm}^{-3}$
(HMT-SCGF) and $\rho_{\rm{t}}\approx0.036\,\rm{fm}^{-3}$ (HMT-exp), respectively.
Correspondingly, the transition pressure in the FFG model and the
HMT-SCGF (HMT-exp) model is about
$0.28\,\rm{MeV}/\rm{fm}^{3}$ and $0.26\,\rm{MeV}/\rm{fm}^{3}$
($0.17\,\rm{MeV}/\rm{fm}^{3}$), respectively. 

As discussed in the
above sections, the symmetry energy $E_{\rm{sym}}(\rho)$ at sub-saturation densities is reduced by the SRC/HMT while its magnitude and slope at $\rho_0$ are fixed at their currently known empirical values. Thus, the $E_{\rm{sym}}(\rho)$ is effectively hardened within the sub-saturation density region (see the inset of FIG.\,\ref{fig_ab_Esym}), leading to an enhancement of the slope
parameter $L(\rho)$ for $\rho\lesssim\rho_0$. It is known that a larger $L(\rho_{\rm{r}})$ with $\rho_{\rm{r}}\lesssim\rho_0$\,\cite{Zha13} leads to a
smaller core-crust transition density as studied carefully in, e.g.,
Refs.\,\cite{XuJ09}. The observed SRC/HMT effects on the crust-core transition density and pressure are qualitatively consistent with expectations based on previous studies. 
\begin{figure}[h!]
\centering
 \includegraphics[height=2.7cm]{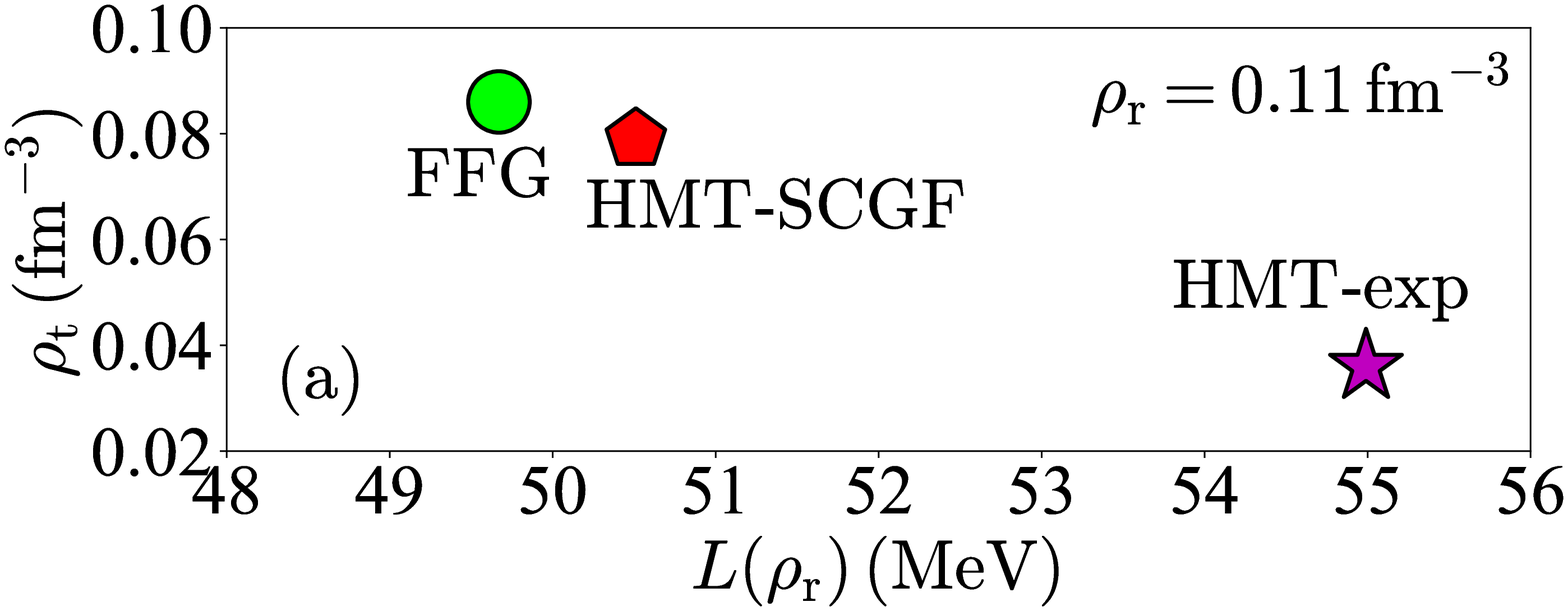}\\
 \hspace{0.cm}
 \includegraphics[height=2.7cm]{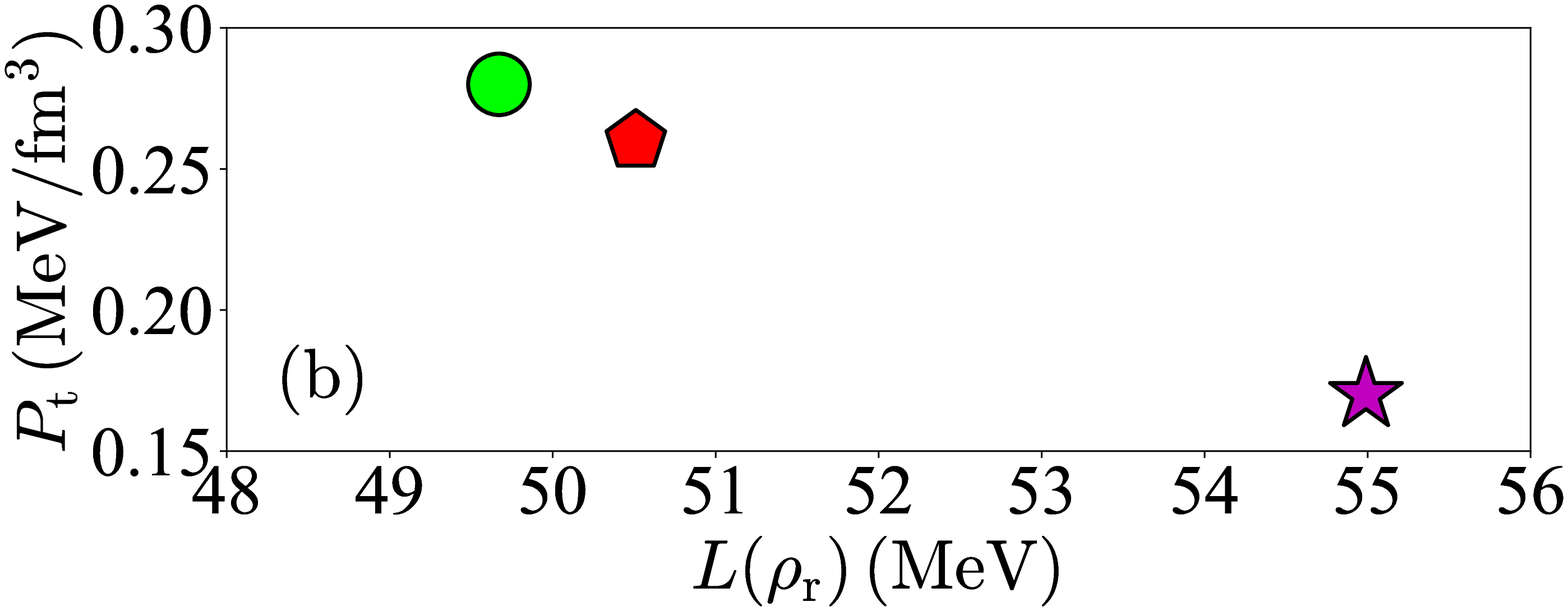}
  \caption{Correlation between the core-crust transition density (upper) and the transition pressure (lower) in $\beta$-stable neutron star matter
  and the slope parameter $L(\rho_{\rm{r}})$ at $\rho_{\rm{r}}=0.11\,\rm{fm}^{-3}$.}
  \label{fig_ab_rhot}
\end{figure}

As we discussed earlier, the fundamentally important quantity that is growing from the FFG to HMT-SCGF then HMT-exp is the strength of the underlying isospin dependence of SRC/HMT, namely, 
the difference $\delta x=x_{\rm{HMT}}^{\rm{SNM}}-x_{\rm{HMT}}^{\rm{PNM}}$ is increasing. Our results shown in FIG.\,\ref{fig_ab_rhot} indicates an anti-correlation between the $\delta x$  and the transition density $\rho_{\rm{t}}$.
It provides an important connection to the study of the thickness of inner crusts\,\cite{XuJ09} as well as the related observational phenomena\,\cite{Gear11,Wen12,Newton12,Newton14,Hooker15,Zhou21}, such as the crustal oscillations and glitches as well as the $r$-mode stability window of rapidly rotating neutron stars. Our findings here may also have important implications on the heat transport process in neutron stars\,\cite{Cac08}.

Within the minimal model of neutron stars assuming no hadron-quark phase transition as well as baryon resonance and hyperon production, the compositions of neutron stars are 
determined by the requirement of charge neutrality and $\beta$-equilibrium conditions. From the nucleon specific energy $E(\rho,\delta)$ we can calculate numerically the proton fraction $x_{\rm{p}}(\rho)$ for $\beta$-stable matter and examine its dependence on the strength of SRC/HMT. More specifically, for neutrino free $ \beta $-stable matter, the chemical
equilibrium for the reactions $
\rm{n}\rightarrow\rm{p}+\rm{e}^{-}+\overline{\nu}_{\rm{e}}$ and
$\rm{p}+\rm{e}^{-}\rightarrow \rm{n}+\nu _{\rm{e}}$ requires $ \mu
_{\rm{e}}=\mu_{\rm{n}}-\mu_{\rm{p}}\approx
4E_{\rm{sym}}(\rho)\delta+8E_{\rm{sym,4}}(\rho)\delta^3+12E_{\rm{sym},6}(\rho)\delta^5+\cdots$, here $E_{\rm{sym},6}(\rho)\equiv 120^{-1}\partial^6E(\rho,\delta)/\partial\delta^6|_{\delta=0}$ is the sixth-order symmetry energy.
For relativistically degenerate electrons, we have $ \mu _{\rm{e}}=[
m_{\rm{e}}^{2}+(3\pi ^{2}\rho x_{\rm{e}})^{2/3}] ^{1/2}\approx
( 3\pi ^{2}\rho x_{\rm{e}}) ^{1/3} $, where
$m_{\rm{e}}\approx0.511\,$MeV is the electron mass, and
$x_{\rm{p}}=x_{\rm{e}}$ because of charge neutrality. Above a certain density where $\mu _{\rm{e}}$ exceeds the muon
mass $m_{\mu }\approx105.7\,$MeV, the reactions
$\rm{e}^{-}\rightarrow \mu ^{-}+\nu
_{\rm{e}}+\overline{\nu}_{\mu }$, $\rm{p}+\mu ^{-}\rightarrow \rm{n}+\nu _{\mu }$ and $%
\rm{n}\rightarrow\rm{p}+\mu ^{-}+\overline{\nu}_{\mu }$ are
energetically allowed. Then, both electrons and muons are present
in $\beta $-stable matter. This alters the $\beta$-stability condition
to $\mu_{\rm{e}}=\mu _{\mu }=[ m_{\mu }^{2}+(3\pi ^{2}\rho x_{\mu
})^{2/3}] ^{1/2}$ with $x_{\rm{p}}=x_{\rm{e}}+x_{\mu }$. 
Generally, when the nucleon momentum distribution function $n_{\v{k}}^J(\rho,\delta)$ has a low- (high-) momentum depletion (tail), the nucleon chemical potential is not given by 
$k_{\rm{F}}^{\rm{n/p},2}/2M+U_{\rm{n/p}}(\rho,\delta,k_{\rm{F}}^{\rm{n/p}})$, i.e., it could not be obtained by simply letting the momentum $|\v{k}|$ to be the Fermi momentum in the single-nucleon potential. 
In fact, determining the relation between the nucleon chemical potential and the single-nucleon potential for a general $n_{\v{k}}^J(\rho,\delta)$ is known as a fundamental problem in nuclear many-body theories\,\cite{AGD1960}.
We thus have to calculate numerically the chemical potential $\mu_J$ for a nucleon $J$ from the energy density $\varepsilon$ of neutron star matter via
$
  \mu_J=\partial\varepsilon(\rho,\delta)/\partial\rho_J.
$

\begin{figure}[h!]
\centering
 \includegraphics[width=4.cm]{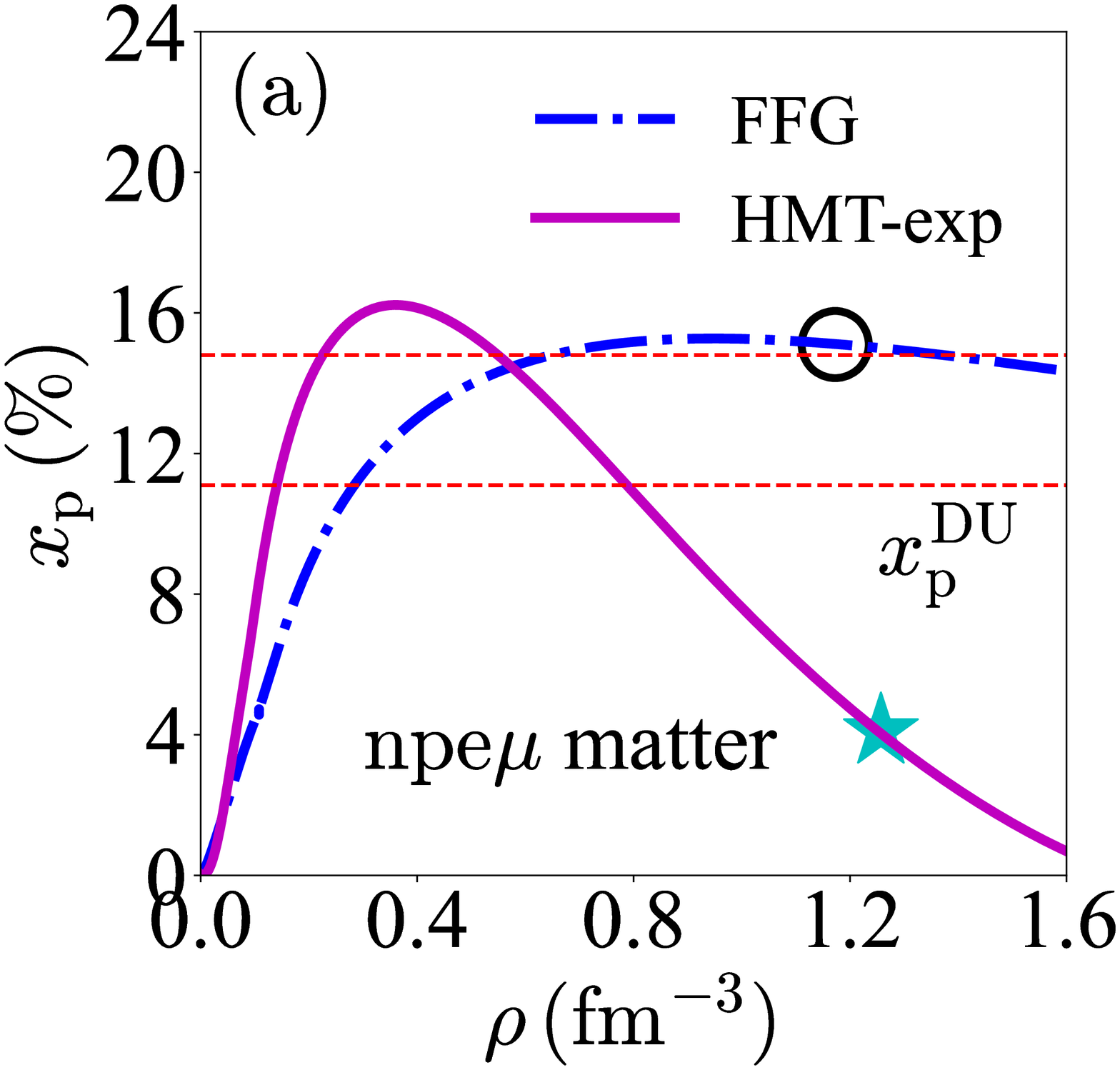}\quad
 \includegraphics[width=4.05cm]{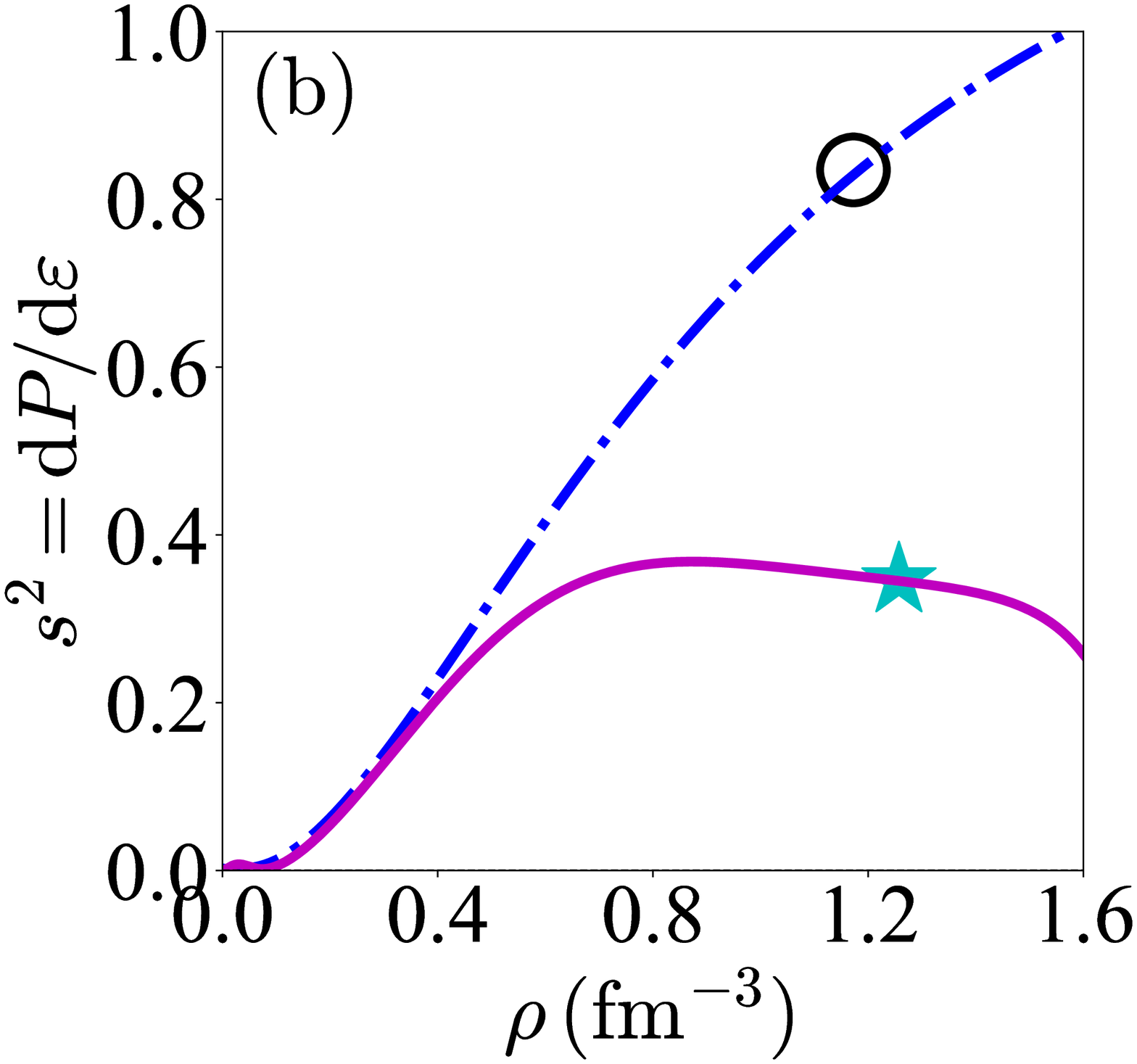}
  \caption{The proton fraction $x_{\rm{p}}$ (left) and the square of the speed of sound $s^2$ (right) in neutron star matter within the FFG and HMT-exp model, respectively.
  The black circle and the cyan star correspond to the central densities of the neutron star in the two models.}
  \label{fig_abMDI-xp}
\end{figure}

Shown in the left panel of FIG.\,\ref{fig_abMDI-xp} is the proton fraction $x_{\rm{p}}$ in neutron star matter as a function of density within the FFG and HMT-exp model, respectively.
Since the HMT-SCGF result is similar to the one using the FFG model, we thus in the following focus on the comparison between the predictions using the FFG and the HMT-exp models.
Obviously, the most important observation is that due to the reduction of the high-density symmetry energy in the HMT-exp model the proton fraction $x_{\rm{p}}$ in this model starts to decrease quickly above some critical density. 
It thus prompts the role played by the symmetry energy in the core of neutron stars as $\delta$ in the HMT-exp model is much closer to 1 than that in the FFG model.
Therefore, it may have important ramifications on properties of neutron stars. For example, it is well known that the  $x_{\rm{p}}$ determines the cooling mechanisms of protoneutron stars and the associated neutrino emissions\,\citep{LPPH}. 
More quantitatively, the critical proton fraction $x^{\rm{DU}}_{\rm{p}}$ enabling the direct URCA process (DU) for fast cooling is given by
$x^{\rm{DU}}_{\rm{p}}=1/[1+(1+x_{\rm{e}}^{1/3})^3]$
with the $x_{\rm{e}}\equiv \rho_{\rm{e}}/\rho_{\rm{p}}$ between 1 and 0.5 leading to $x^{\rm{DU}}_{\rm{p}}$ between 11.1\% to 14.8\%\,\citep{Kla06}, see the red dashed lines in the left panel of FIG.\,\ref{fig_abMDI-xp}.
The black circle and the cyan star in the left panel of FIG.\,\ref{fig_abMDI-xp} correspond to the central densities of neutron stars in the two models studied here. It is seen that around these central densities the direct URCA is allowed in the FFG model but forbidden in the HMT-exp model. 

The speed of sound squared $s^2=\d P/\d\varepsilon$ is a measure of the stiffness of nuclear EOS. Shown in the right panel of FIG.\,\ref{fig_abMDI-xp} are the predicted $s^2$ by the two models considered. It is seen that the EOS in the HMT-exp model is much softer than the FFG model prediction at densities above about $2.5\rho_0$. Since the ANM EOS $E(\rho,\delta)\approx E_0(\rho)+E_{\rm{sym}}(\rho)\delta^2+\cdots$, the reduced symmetry energy but increased $\delta^2$ at high densities together may effectively soften the EOS. However, its net effects depend on how strongly the SNM EOS $E_0(\rho)$ is stiffened by the SRC/HMT. The results shown here indicate clearly that the SRC/HMT effect on the symmetry energy is winning against that on the SNM EOS. 
This result is expected to have some dynamical effects on properties of neutron stars. Since the symmetry energy and SNM EOS in different density regions have different effects on properties of neutron stars, the SRC/HMT effects through these two terms of the ANM EOS have to be studied quantitatively as we shall do next. 

The mass-radius correlation of neutron stars is obtained from integrating the Tolman-Oppenheimer-Volkoff (TOV) equations\,\cite{Misner1973}
\begin{align}
\frac{\d P(r)}{\d r}=&-\frac{[\varepsilon(r)+P(r)][M(r)+4\pi r^3P(r)]}{r[r-2M(r)]},\\
\frac{\d M(r)}{\d r}=&4\pi r^2\varepsilon(r),
\end{align}
where $r$ is the radial distance from the center of the star, and
$M(r)$ is the mass enclosed within $r$ (adopting natural unit in which $c=G=1$).
The pressure of $\beta$-stable and charge neutral npe$\mu$ matter is given by $
P(\rho,\delta)=P_{\rm{N}}(\rho,\delta)
+P_{\rm{e}}(\rho,\delta)+P_{\mu}(\rho,\delta)$,
where $P_{\rm{N}}(\rho,\delta) $ is the nucleon pressure. 
The lepton pressure is further given as $
P_\ell(\rho,\delta)=\rho_\ell\mu_\ell-\varepsilon_\ell(\rho,\delta)
$, with $
\mu_\ell=[{k_\ell^2+m_\ell^2}]^{1/2}$ and $k_\ell=(3\pi^2\rho_\ell)^{1/3}$ for $\ell=\rm{e},\mu$, here $\varepsilon_{\ell}(\rho,\delta)=\eta_{\ell}\Phi_{\ell}(t_{\ell})$ (see (\ref{dd_e1}) for the definition of $\Phi_{\ell}$).
For the whole system, we have the self-consistency relation between the total pressure $P$ and the total energy density $\varepsilon$, i.e., 
$
P=\rho^2{\d(\varepsilon/\rho)}/{\d\rho}$.
The inner crust of neutron stars with densities ranging between $\rho_{\text{out}}=2.46\times
10^{-4}\,\rm{fm}^{-3}$ corresponding to the neutron dripline and the
core-crust transition density $\rho _{\text{t}}$ is the region where
some complex and exotic structures\,---\,the ``nuclear pasta'' may exist.  Because of our very poor knowledge
about this region we adopt the polytropic EOSs expressed in
terms of the pressure $P$ as a function of the total energy density
$\varepsilon$ as
$P=c+d\varepsilon^{4/3}$\,\cite{XuJ09,Hor03}. The constants $c$ and
$d$ are determined by the quantities at $\rho
_{\text{t}}$ and $\rho _{\text{out}}$\,\cite{XuJ09}, i.e.,
\begin{equation}
c=\frac{P_{\rm{out}}\varepsilon_{\rm{t}}^{4/3}-P_{\rm{t}}\varepsilon_{\rm{out}}^{4/3}}{\varepsilon_{\rm{t}}^{4/3}-\varepsilon_{\rm{out}}^{4/3}},
~~
d=\frac{P_{\rm{t}}-P_{\rm{out}}}{\varepsilon_{\rm{t}}^{4/3}-\varepsilon_{\rm{out}}^{4/3}}
.\end{equation}
For the outer
crust\,\cite{BPS71,Iida1997}, we use the conventional Baym-Pethick-Sutherland (BPS) EOS for the region with $6.93\times 10^{-13}\,\rm{fm}^{-3}\lesssim\rho \lesssim\rho _{\text{out}}$ and the
Feynman-Metropolis-Teller (FMT) EOS for $4.73\times 10^{-15}\,\rm{fm}^{-3}\lesssim\rho \lesssim6.93\times
10^{-13}\,\rm{fm}^{-3}$, respectively.
The total neutron star matter EOSs in the two models are shown in FIG.\,\ref{fig_abMDI-beta}.

\begin{figure}[h!]
\centering
 \includegraphics[width=6.5cm]{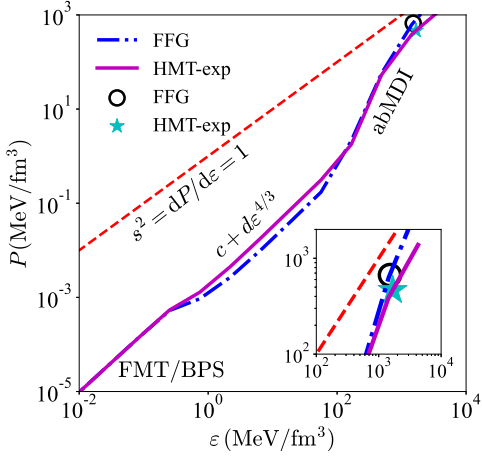}
  \caption{The EOS of neutron star matter in the FFG model and the HMT-exp model. See the text for details.}
  \label{fig_abMDI-beta}
\end{figure}

The resulting mass-radius correlation of neutron stars is shown in FIG.\,\ref{fig_abMDI-MR} for the FFG and the HMT-exp model, respectively. Firstly, we notice that both models give a radius of about 12\,km for canonical neutron stars of mass 1.4$M_{\odot}$ consistent with many analyses of the GW170817 observation, see, e.g., Refs.\,\cite{EPJA-review,Baiotti,Capano20,David,Kat20,AngLi} for reviews. 
This is mostly by design as our EDF parameters are fixed such that all the EOS characteristics at $\rho_0$ are consistent with existing constrains from both astrophysical observations and terrestrial experiments. 
Secondly, it is interesting to see that the two models indeed predict significantly different maximum masses and the corresponding radii. More quantitatively, the maximum mass of neutron stars in the FFG model is about $M_{\rm{NS}}^{\max}\approx1.98M_{\odot}$, with the corresponding radius being about 9.9\,km. While the $M_{\rm{NS}}^{\max}$ in the HMT-exp model is about $1.72M_{\odot}$ with a radius of about 10.3\,km. 
The reduction of the maximum mass $M_{\rm{NS}}^{\max}$ due to the SRC/HMT is thus about 13\%. 
The point corresponding to the maximum mass on the $P(\varepsilon)$ plane is also shown in FIG.\,\ref{fig_abMDI-beta}, by the black circle (cyan star) for the FFG (HMT-exp) model.
Thirdly, since the HMT largely affects the transition density $\rho_{\rm{t}}$ (see FIG.\,\ref{fig_ab_rhot}), the radii of low-mass neutron stars in the FFG model and the HMT-exp model show obvious differences.

The central density $\rho_{\rm{cen}}$ corresponding to $M_{\rm{NS}}^{\max}$ is about $\rho_{\rm{cen}}^{\rm{FFG}}\approx7.3\rho_0$ ($\rho_{\rm{cen}}^{\rm{HMT}\mbox{-}\rm{exp}}\approx7.9\rho_0$) for the FFG (HMT-exp) model (see the points shown in FIG.\,\ref{fig_abMDI-xp}).
Although the EDF constructed is non-relativistic in nature, it could be effectively applied to neutron stars and fulfill the principle of causality (set by the speed of sound $s^2=\d P/\d\varepsilon=1$).
In particular, we find that the square of the speed of sound in the FFG model at $\rho_{\rm{cen}}^{\rm{FFG}}$ is about $s^2_{\rm{FFG}}\approx0.84$, while that for the HMT-exp model is about $s_{\rm{HMT}\mbox{-}\rm{exp}}^2\approx0.35$, see the right panel of FIG.\,\ref{fig_abMDI-xp} for the density dependence of the $s^2$ in the two model. Thus, due to the softening of the symmetry energy above a critical density and the non-relativistic nature of the Gogny-like EDF, the high-density $s^2$ in the HMT-exp model is 
much smaller than that in the FFG model. In fact, it may not exceed 1 at even larger densities when the FFG becomes acausal (the density violating the principle of causality in the FFG model is about $9.8\rho_0$).

\begin{figure}[h!]
\centering
 \includegraphics[width=6.5cm]{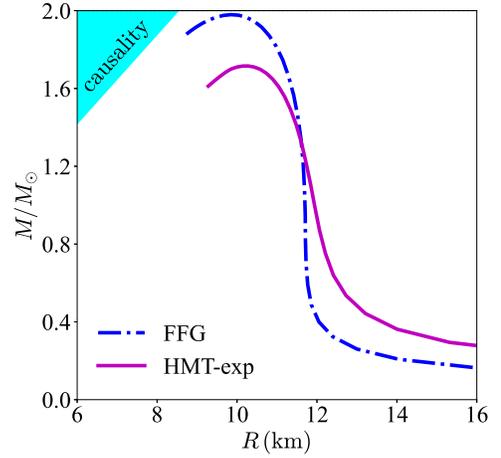}
  \caption{The mass-radius relation for neutron stars in the FFG and the HMT-exp models.}
  \label{fig_abMDI-MR}
\end{figure}

To this end, it is interesting to note that some earlier studies on the mass-radius relation of neutron stars adopting the nonlinear RMF models found that the SRC/HMT generally increases the maximum mass $M_{\rm{NS}}^{\max}$\,\cite{Cai16b,Lou22a,Lu2022,Hong22,Lou22,Lu21,Souza20}. In these models the SRC/HMT induced increase in SNM pressure dominates over the reduction of symmetry energy. 
One main reason is that in the nonlinear RMF models the symmetry energy itself plays a minor role in determining the maximum mass $M_{\rm{NS}}^{\max}$ as pointed out already a long time ago\,\cite{Ser79}. Moreover, the softening of the $E_{\rm{sym}}(\rho)$ in the RMF models is weaker than that in the current (non-relativistic) Gogny-like EDF. Of course, this then raises an interesting question: how can this kind of EDFs encapsulating the SRC/HMT effects support the currently observed most massive neutron star PSR J0740+6620 which has a mass of $2.08^{+0.07}_{-0.07}M_\odot$\,\cite{Fon21}? In fact, even without considering the SRC/HMT effects, it has been a challenging problem to find mechanisms to support heavy neutron stars (with masses around and above $2M_{\odot}$) within non-relativistic energy density functionals, see, e.g., Refs.\,\cite{Rios-G,Zha16,ZhouY2019,Gon18,Gon19,Lou20}. Our results presented above indicate that this problem may become more challenging and certainly deserves further investigations.

\setcounter{equation}{0}
\section{Summary and Outlook}\label{sec6}

In summary, we investigated SRC/HMT effects on the EOS and single-nucleon potential in cold neutron-rich matter within the Gogny-like EDF. The SRC effects are incorporated into the EDF by using a single-nucleon
momentum distribution function $n_{\v{k}}^J(\rho,\delta)$ that has an isospin-dependent high-momentum (low-momentum) tail (depletion) with its parameters determined by the available SRC experiments. We introduced a parametrization as a surrogate for the momentum-dependent kernel of the Gogny-like EDF to facilitate derivations of analytical expressions for all relevant physical quantities describing the EOS and single-nucleon potential in ANM. The surrogate is shown to be effective for problems with momentum (density) scale smaller than about 1 GeV/c (several times the saturation density). The resulting expressions for all terms in the ANM EOS and single-nucleon potential will facilitate further explorations of nuclear interactions in dense neutron-rich matter as well as their impacts on properties of neutron stars and heavy-ion reactions induced by high-energy radioactive beams. 
In particular, 
\begin{enumerate}
\item[(a)]We have derived the analytical expressions for the EOS of SNM $E_0(\rho)$,  the corresponding pressure $P_0(\rho)$,  the incompressibility coefficient $K_0(\rho)$, the nucleon effective k-mass $M_0^{\ast}(\rho)$, the isoscalar potential $U_0(\rho,|\v{k}|)$, the symmetry energy $E_{\rm{sym}}(\rho)$ together with its slope parameter $L(\rho)$, and the isovector (symmetry) potential $U_{\rm{sym}}(\rho,|\v{k}|)$.
Numerical constructions based on these analytical expressions are given through three models for the nucleon momentum distribution, namely the FFG model without the SRC-induced HMT in the $n_{\v{k}}^J(\rho,\delta)$, the
HMT-SCGF/HMT-exp model with different properties of the HMT, to fulfill certain empirical constraints on the EOS of ANM.  All three models contain proper potential parts.
\item[(b)] While the same set of available empirical constraints on the ANM EOS at $\rho_0$ are all maintained in the three models, the EOS $E_0(\rho)$ of SNM at high densities is made slightly harder once the SRC
effects are considered.  In addition, the symmetry energy in the presence of HMT becomes softer at large densities even to start decreasing above a critical density. Although qualitatively consistent with the nonlinear RMF model predictions, the reduction of the high-density symmetry energy $E_{\rm{sym}}(\rho)$ is much stronger while the enhancement of the SNM pressure $P_0$ is much weaker in the Gogny-like EDF with SRC/HMT. Consequently, as a result of the competition between these two effects, the maximum mass of a cold neutron star is found to be reduced compared with the prediction of the Gogny-like EDF without considering the SRC/HMT.
\item[(c)] The single-nucleon potential $U_0(\rho,|\v{k}|)$ in SNM at $\rho_0$ in all three models is consistent with the isoscalar nucleon optical potential from earlier analyses of various experimental data. The nucleon isovector (symmetry) potential $U_{\rm{sym}}(\rho,|\v{k}|)$ in ANM is largely enhanced due to the corresponding enhancement of the potential symmetry energy $E_{\rm{sym}}^{\rm{pot}}(\rho)$.
These are the direct consequences of the momentum-dependence of the nucleon potential in the Gogny-like EDF due to the finite ranges of nuclear interactions.  Such momentum-dependence is absent in the nonlinear RMF approach.
\end{enumerate}

The formulas, table of coupling constants, parameterizations and quantitative demonstrations of the relevant physical quantities given in this work provide a useful and convenient starting point to further investigate the nature of strong interactions at short distances and novel effects of SRC/HMT on properties of dense neutron-rich matter. In particular, the generally density- and momentum-dependent single-nucleon potential in neutron-rich matter is expected to play a significantly role in understanding the dynamics and observables of heavy-ion reactions induced by high-energy radioactive beams, the cooling mechanisms of protoneutron stars as well as properties of cold neutron stars at $\beta$-equilibrium and their mergers. For instance, with the coupling parameters given in Table   \ref{tab_para}, one can use the single-nucleon potential of Eq. (\ref{Gen-U}) in simulating heavy-ion reaction with the $f_J(\v{r},\v{k})$ self-consistently generated by the BUU transport model. After comparing the predicted observables with experimental data, one can then make better informed predictions for the EOS of hot, dense and neutron-rich matter. The latter is a necessary input for simulating mergers of two neutron stars based on principles of general relativity \cite{Ra1,Ra2}. In particular, it is important for understanding the high-frequency spectrum of gravitational waves from post-mergers that carry critical information about the nature of possible phase transitions and the separation boundary/gap between neutron stars and blackholes. Interestingly, it has already been found that the density dependence of nuclear symmetry energy play a significant role in answering many intriguing questions in simulating  neutron star mergers \cite{Most21}. 

As discussed earlier, our approach has some caveats and some challenging issues remain to be resolved. Fortunately, multimessenger nuclear astrophysics with high-precision X-ray and gravitational wave detectors as well as advanced rare isotope beam facilities are expected to provide us with various kinds of new and more accurate observational data. Combined analyses of these data using the same Gogny-like EDF with SRC/HMT may help us address some of the remaining issues. For example, the strong enhancement of the symmetry potential $U_{\rm{sym}}(\rho,|\v{k}|)$ due to the SRC/HMT especially at supra-saturation densities, e.g., $\rho\approx2\rho_0\mbox{$\sim$}3\rho_0$,  is expected to affect significantly the dynamics and some observables of nuclear reactions with high-energy radioactive beams. While the corresponding reduction of the total symmetry energy reduces significantly the maximum mass of neutron stars that the resulting EOS can support. A joint analysis of these effects using a unified Gogny-like EDF with SRC/HMT is expected to be fruitful. Besides the traditionally used forward predictions with the different model parameter sets, the expected new data and existing ones together will enable us to get new insights from inferences of relevant model parameters using machine leaning techniques, such as the Bayesian model selection \cite{MAL}. These studies are on the top of our working agenda. 

\section*{Acknowledgement} We would like to thank Lie-Wen Chen, Xiao-Tao He, Che Ming Ko and Xiao-Hua Li for useful discussions over the years on some of the issues studied in this work.
This work is supported in part by the U.S. Department of Energy, Office of Science,
under Award No. DE-SC0013702, the CUSTIPEN (China-
U.S. Theory Institute for Physics with Exotic Nuclei) under
US Department of Energy Grant No. DE-SC0009971.

\appendix

\renewcommand\theequation{\alph{section}\arabic{equation}}
\renewcommand\thesubsection{A-\arabic{subsection}}

\begin{widetext}

\section{Detailed Derivations on the Formulas}\label{app1}

In this appendix, we derive the analytical expressions for all the quantities appeared in the main text. In subsection \ref{sb_a_1}, the relevant expressions for the FFG model are given while in subsection \ref{sb_a_2} we develop the corresponding formulas for the HMT model.
The main formulas in subsection \ref{sb_a_1} are Eq.\,(\ref{FFG-E0}) for
$E_0(\rho)$, Eq.\,(\ref{FFG-p0}) for $P_0(\rho)/\rho$, Eq.\,(\ref{FFG-K0}) for
$K_0(\rho)$, Eq.\,(\ref{FFG-U0}) for $U_0(\rho,|\v{k}|)$ and
Eq.\,(\ref{FFG-M0}) for $M_0^{\ast}(\rho)/M$, Eq.\,(\ref{FFG-Esym}) for $E_{\rm{sym}}(\rho)$, Eq.\,(\ref{FFG-L}) for $L(\rho)$, and Eq.\,(\ref{FFG-Usym}) for $U_{\rm{sym}}(\rho,|\v{k}|)$.
The corresponding formulas for the HMT model are Eq.\,(\ref{HMT-E0}) for
$E_0(\rho)$, Eq.\,(\ref{HMT-p0}) for $P_0(\rho)/\rho$, Eq.\,(\ref{HMT-K0}) for
$K_0(\rho)$, Eq.\,(\ref{HMT-U0}) for $U_0(\rho,|\v{k}|)$ and
Eq.\,(\ref{HMT-M0}) for $M_0^{\ast}(\rho)/M$, Eq.\,(\ref{HMT-Esym}) for $E_{\rm{sym}}(\rho)$, Eq.\,(\ref{HMT-L}) for $L(\rho)$, and Eq.\,(\ref{HMT-Usym}) for $U_{\rm{sym}}(\rho,|\v{k}|)$.
In addition, one can check straightforwardly that if 
\begin{equation}
\phi_0=1,~~\phi_1=0,
\end{equation} are adopted, the HMT formulas are naturally reduced to the FFG formulas.
Finally, we give in subsection \ref{sb_a3} the expressions for the EOS $E(\rho,\delta)$ of ANM by collecting the relevant formulas.

\subsection{Derivations for Quantities in the FFG Model for $n_{\v{k}}^J(\rho,\delta)$}\label{sb_a_1}
We first derive the relevant expressions in the FFG model. In order
to obtain the single-nucleon potential, we should evaluate the
integral $ \int\d\v{k}'f_{J'}(\v{r},\v{k}')\Omega(\v{k},\v{k}')$,
with\begin{equation}
f_{J'}(\v{r},\v{k}')=\frac{1}{4\pi^3}n_{\v{k}}^{J'}(\rho,\delta)=\frac{1}{4\pi^3}\Theta(k_{\rm{F}}^{J'}-|\v{k}'|),~~\Omega(\v{k},\v{k}')
=1+{a}\left[\left(\frac{\v{k}\cdot\v{k}'}{\Lambda^2}\right)^2\right]^{1/4}
+{b}\left[\left(\frac{\v{k}\cdot\v{k}'}{\Lambda^2}\right)^2\right]^{1/6}
.\end{equation} 
By denoting the angle between $\v{k}$ and $\v{k}'$ as
$\vartheta$, i.e., \begin{equation}
\vartheta=\widehat{(\v{k},\v{k}')},\end{equation}
we then have,
\begin{align}
\int\d\v{k}'f_{J'}(\v{r},\v{k}')\Omega(\v{k},\v{k}')
=&\frac{1}{4\pi^3}\int_0^{2\pi}\d\phi\int_0^{Q'}k'^2\d
k'\int_0^{\pi}\sin\vartheta\d\vartheta\left[1+{a}\left[\left(\frac{kk'\cos\vartheta}{\Lambda^2}\right)^2\right]^{1/4}
+{b}\left[\left(\frac{kk'\cos\vartheta}{\Lambda^2}\right)^2\right]^{1/6}\right]\notag\\
=&\frac{1}{2\pi^2}\int_0^{Q'}k'^2\d
k'\int_0^{\pi}\sin\vartheta\d\vartheta\left[1+{a}\left[\left(\frac{kk'\cos\vartheta}{\Lambda^2}\right)^2\right]^{1/4}
+{b}\left[\left(\frac{kk'\cos\vartheta}{\Lambda^2}\right)^2\right]^{1/6}\right]\notag\\
=&\frac{1}{2\pi^2} \int_0^{Q'}k'^2\d
k'\left[2+\frac{4}{3}a\left(\frac{k^2k'^2}{\Lambda^2}\right)^{1/4}+\frac{3}{2}{b}\left(\frac{k^2k'^2}{\Lambda^2}\right)^{1/6}\right]\notag\\
=&\frac{1}{2\pi^2}\left[\frac{2}{3}Q'^3+\frac{8}{21}aQ'^3\left(\frac{k^2Q'^2}{\Lambda^4}\right)^{1/4}+
\frac{9}{20}bQ'^3\left(\frac{k^2Q'^2}{\Lambda^4}\right)^{1/6}\right]\notag\\
=&\frac{Q'^3}{3\pi^2}\left[1+\frac{4}{7}a\left(\frac{kQ'}{\Lambda^2}\right)^{1/2}
+\frac{27}{40}b\left(\frac{kQ'}{\Lambda^2}\right)^{1/3}\right],\label{FFG-1-inte}
\end{align}
here $ Q'\equiv k_{\rm{F}}^{J'}$. Then the single potential is given as,
\begin{align}
U_J^{\rm{mom}}(\rho,\delta,|\v{k}|)=&\frac{2C_{\ell}}
{\rho_0}\frac{k_{\rm{F}}^{J,3}}{3\pi^2}\left[1+\frac{4a}{7}\left(\frac{|\v{k}|k_{\rm{F}}^J}{\Lambda^2}\right)^{1/2}
+\frac{27b}{40}\left(\frac{|\v{k}|k_{\rm{F}}^J}{\Lambda^2}\right)^{1/3}\right]\notag\\
&+\frac{2C_{\rm{u}}}{\rho_0}\frac{k_{\rm{F}}^{\overline{J},3}}{3\pi^2}\left[1+\frac{4a}{7}\left(\frac{|\v{k}|k_{\rm{F}}^{\overline{J}}}{\Lambda^2}\right)^{1/2}
+\frac{27b}{40}\left(\frac{|\v{k}|k_{\rm{F}}^{\overline{J}}}{\Lambda^2}\right)^{1/3}\right]
,\label{FFG_UJ-mom}
\end{align}
here $\overline{\rm{n}}=\rm{p},\overline{\rm{p}}=\rm{n}$. According
to (\ref{FFG_UJ-mom}), we have
\begin{align}
U_J^{\rm{mom}}(\rho,\delta,|\v{k}|)\approx&
C_{\rm{tot}}{u}\left[1+\frac{4a}{7}\left(\frac{|\v{k}|k_{\rm{F}}}{\Lambda^2}\right)^2
+\frac{27b}{40}\left(\frac{|\v{k}|k_{\rm{F}}}{\Lambda^2}\right)^{1/3}\right]\notag\\
&+\frac{2C_{\ell}k_{\rm{F}}^3}{3\pi^2\rho_0}\left[1+\frac{2a}{3}\left(\frac{|\v{k}|k_{\rm{F}}}{\Lambda^2}\right)^2
+\frac{3b}{4}\left(\frac{|\v{k}|k_{\rm{F}}}{\Lambda^2}\right)^{1/3}\right]\tau_3^J\delta\notag\\
&-\frac{2C_{\rm{u}}k_{\rm{F}}^3}{3\pi^2\rho_0}\left[1+\frac{2a}{3}\left(\frac{|\v{k}|k_{\rm{F}}}{\Lambda^2}\right)^2
+\frac{3b}{4}\left(\frac{|\v{k}|k_{\rm{F}}}{\Lambda^2}\right)^{1/3}\right]\tau_3^J\delta,
\end{align}
here ${u}=\rho/\rho_0$ is the reduced density, and then,\begin{align}
U_0^{\rm{mom}}=&C_{\rm{tot}}{u}\left[1+\frac{4a}{7}\left(\frac{|\v{k}|k_{\rm{F}}}{\Lambda^2}\right)^2
+\frac{27b}{40}\left(\frac{|\v{k}|k_{\rm{F}}}{\Lambda^2}\right)^{1/3}\right],~~C_{\rm{tot}}=C_{\ell}+C_{\rm{u}},\\
U_{\rm{sym}}^{\rm{mom}}=&C_{\rm{d}}{u}\left[1+\frac{2a}{3}\left(\frac{|\v{k}|k_{\rm{F}}}{\Lambda^2}\right)^2
+\frac{3b}{4}\left(\frac{|\v{k}|k_{\rm{F}}}{\Lambda^2}\right)^{1/3}\right],~~C_{\rm{d}}=C_{\ell}-C_{\rm{u}}.
\end{align}
Taking into account the density-dependent part, we then have
\begin{align}
U_0(\rho,|\v{k}|)=&\frac{1}{2}A_{\rm{tot}}{u}+B{u}^{\sigma}+C_{\rm{tot}}{u}\left[1+\frac{4a}{7}
\left(\frac{|\v{k}|k_{\rm{F}}}{\Lambda^2}\right)^{1/2}
+\frac{27b}{40}\left(\frac{|\v{k}|k_{\rm{F}}}{\Lambda^2}\right)^{1/3}\right],\label{FFG-U0}
\\
U_{\rm{sym}}(\rho,|\v{k}|)=&\frac{1}{2}A_{\rm{d}}{u}-\frac{2Bx}{\sigma+1}{u}^{\sigma}
+C_{\rm{d}}{u}\left[1+\frac{2a}{3}\left(\frac{|\v{k}|k_{\rm{F}}}{\Lambda^2}\right)^2
+\frac{3b}{4}\left(\frac{|\v{k}|k_{\rm{F}}}{\Lambda^2}\right)^{1/3}\right].\label{FFG-Usym}
\end{align}
Similarly, here $A_{\rm{tot}}=A_{\ell}+A_{\rm{u}}$ and
$A_{\rm{d}}=A_{\ell}-A_{\rm{u}}$. Taking the momentum derivative of
(\ref{FFG-U0}) and then setting $|\v{k}|=k_{\rm{F}}$, we obtain the
effective mass
\begin{equation}\label{FFG-M0}{
M_0^{\ast}(\rho)/M=\left[1+\left.\frac{M}{k_{\rm{F}}}\frac{\partial
U_0}{\partial|\v{k}|}\right|_{|\v{k}|=k_{\rm{F}}}\right]^{-1}
=\left[1+C_{\rm{tot}}M{u}\left[\frac{2a}{7}\frac{1}{k_{\rm{F}}\Lambda}
+\frac{9b}{40}\left(\frac{1}{\Lambda^2k_{\rm{F}}^4}\right)^{1/3}\right]\right]^{-1}.}
\end{equation}

In order to obtain the equation of state, we should do the integral
(\ref{FFG-1-inte}) once more, i.e.,
\begin{align}
\int\d\v{k}f_J(\v{r},\v{k})\int\d\v{k}'f_{J'}(\v{r},\v{k}')\Omega(\v{k},\v{k}')
=&\int\d\v{k}f_J(\v{r},\v{k})\frac{Q'^3}{3\pi^2}\left[1+\frac{4}{7}a\left(\frac{kQ'}{\Lambda^2}\right)^{1/2}
+\frac{27}{40}b\left(\frac{kQ'}{\Lambda^2}\right)^{1/3}\right]\notag\\
=&\frac{1}{4\pi^3}\int_0^{2\pi}\d\varphi\int_0^{Q}k^2\d
k\int_0^{\pi}\sin\eta\d\eta\frac{Q'^3}{3\pi^2}\left[1+\frac{4}{7}a\left(\frac{kQ'}{\Lambda^2}\right)^{1/2}
+\frac{27}{40}b\left(\frac{kQ'}{\Lambda^2}\right)^{1/3}\right]\notag\\
=&\frac{Q'^3}{3\pi^4}\int_0^Qk^2\d
k\left[1+\frac{4}{7}a\left(\frac{kQ'}{\Lambda^2}\right)^{1/2}
+\frac{27}{40}b\left(\frac{kQ'}{\Lambda^2}\right)^{1/3}\right]\notag\\
=&\frac{Q'^3}{3\pi^4}\left[\frac{1}{3}Q^3+\frac{8a}{49}Q^3\left(\frac{QQ'}{\Lambda^2}\right)^{1/2}
+\frac{81b}{400}Q^3\left(\frac{QQ'}{\Lambda^2}\right)^{1/3}\right]
\notag\\
=&\frac{Q^3Q'^3}{9\pi^4}\left[1+\frac{24a}{49}\left(\frac{QQ'}{\Lambda^2}\right)^{1/2}
+\frac{243b}{400}\left(\frac{QQ'}{\Lambda^2}\right)^{1/3}\right],
\end{align}
here $\varphi$ and $\eta$ are the space angles of $\v{k}$ and
$Q=k_{\rm{F}}^J$. Thus the momentum part of the equation of state is
\begin{align}
E^{\rm{mom}}(\rho,\delta)=&\frac{C_\ell
k_{\rm{F}}^{\rm{p},6}}{9\pi^4\rho\rho_0}\left[1+\frac{24a}{49}\frac{k_{\rm{F}}^{\rm{p}}}{\Lambda}
+\frac{243b}{400}\left(\frac{k_{\rm{F}}^{\rm{p}}}{\Lambda}\right)^{2/3}\right]
+\frac{C_\ell
k_{\rm{F}}^{\rm{n},6}}{9\pi^4\rho\rho_0}\left[1+\frac{24a}{49}\frac{k_{\rm{F}}^{\rm{n}}}{\Lambda}
+\frac{243b}{400}\left(\frac{k_{\rm{F}}^{\rm{n}}}{\Lambda}\right)^{2/3}\right]\notag\\
&+\frac{2C_{\rm{u}}
k_{\rm{F}}^{\rm{p},3}k_{\rm{F}}^{\rm{n},3}}{9\pi^4\rho\rho_0}\left[1+\frac{24a}{49}\left(\frac{k_{\rm{F}}^{\rm{p}}k_{\rm{F}}^{\rm{n}}}{\Lambda^2}
\right)^{1/2}+\frac{243b}{400}\left(\frac{k_{\rm{F}}^{\rm{p}}k_{\rm{F}}^{\rm{n}}}{\Lambda^2}\right)^{1/3}\right]\notag\\
=&\frac{1}{4}C_\ell{u}(1-\delta)^2\left[1+\frac{24a}{49}\frac{k_{\rm{F}}^{\rm{p}}}{\Lambda}
+\frac{243b}{400}\left(\frac{k_{\rm{F}}^{\rm{p}}}{\Lambda}\right)^{2/3}\right]
+\frac{1}{4}C_\ell{u}(1+\delta)^2\left[1+\frac{24a}{49}\frac{k_{\rm{F}}^{\rm{n}}}{\Lambda}
+\frac{243b}{400}\left(\frac{k_{\rm{F}}^{\rm{n}}}{\Lambda}\right)^{2/3}\right]\notag\\
&+\frac{1}{2}C_{\rm{u}}{u}(1-\delta^2)\left[1+\frac{24a}{49}\left(\frac{k_{\rm{F}}^{\rm{p}}k_{\rm{F}}^{\rm{n}}}{\Lambda^2}
\right)^{1/2}+\frac{243b}{400}\left(\frac{k_{\rm{F}}^{\rm{p}}k_{\rm{F}}^{\rm{n}}}{\Lambda^2}\right)^{1/3}\right].
\end{align}
After doing the Taylor's expansion with $\delta$, it is
straightforward to obtain
\begin{equation}
E_0^{\rm{mom}}(\rho)=\frac{1}{2}C_{\rm{tot}}{u}\left[1+\frac{24a}{49}\frac{k_{\rm{F}}}{\Lambda}
+\frac{243b}{400}\left(\frac{k_{\rm{F}}}{\Lambda}\right)^{2/3}\right],
\end{equation}
and
\begin{equation}
E_{\rm{sym}}^{\rm{mom}}(\rho)=\frac{1}{2}C_\ell{u}\left[1+\frac{16a}{21}\frac{k_{\rm{F}}}{\Lambda}
+\frac{33b}{40}\left(\frac{k_{\rm{F}}}{\Lambda}\right)^{2/3}\right]
-\frac{1}{2}C_{\rm{u}}{u}\left[1+\frac{4a}{7}\frac{k_{\rm{F}}}{\Lambda}
+\frac{27b}{40}\left(\frac{k_{\rm{F}}}{\Lambda}\right)^{2/3}\right].
\end{equation}
Consequently,
\begin{equation}\label{FFG-E0}
{E_0(\rho)=\frac{3k_{\rm{F}}^2}{10M}+\frac{1}{4}A_{\rm{tot}}{u}+\frac{B}{\sigma+1}{u}^{\sigma}
+\frac{1}{2}C_{\rm{tot}}{u}\left[1+\frac{24a}{49}\frac{k_{\rm{F}}}{\Lambda}
+\frac{243b}{400}\left(\frac{k_{\rm{F}}}{\Lambda}\right)^{2/3}\right],}
\end{equation}
and,
\begin{equation}\label{FFG-Esym}
{E_{\rm{sym}}(\rho)=\frac{k_{\rm{F}}^2}{6M}+\frac{1}{4}A_{\rm{d}}{u}-\frac{Bx}{\sigma+1}{u}^{\sigma}+
\frac{1}{2}C_\ell{u}\left[1+\frac{16a}{21}\frac{k_{\rm{F}}}{\Lambda}
+\frac{33b}{40}\left(\frac{k_{\rm{F}}}{\Lambda}\right)^{2/3}\right]
-\frac{1}{2}C_{\rm{u}}{u}\left[1+\frac{4a}{7}\frac{k_{\rm{F}}}{\Lambda}
+\frac{27b}{40}\left(\frac{k_{\rm{F}}}{\Lambda}\right)^{2/3}\right].}
\end{equation}

In order to derive the expressions of $P_0$, $K_0$ and $L$, we
notice that the above density structure of the equation of state,
and then have $ \rho{\partial
\rho^{4/3}}/{\partial\rho}\sim({4}/{3})\rho^{4/3},\rho{\partial
\rho^{11/9}}/{\partial\rho}\sim({11}/{9})\rho^{11/9}$. Obviously,
for example
\begin{equation}\label{FFG-p0}
{P_0(\rho)/\rho=\frac{k_{\rm{F}}^2}{5M}+\frac{1}{4}A_{\rm{tot}}{u}
+\frac{B\sigma}{\sigma+1}{u}^{\sigma}+\frac{1}{2}C_{\rm{tot}}{u}\left[1+\frac{32a}{49}\frac{k_{\rm{F}}}{\Lambda}
+\frac{297b}{400}\left(\frac{k_{\rm{F}}}{\Lambda}\right)^{2/3}\right],}
\end{equation}
where $
{32}/{49}={4}/{3}\times{24}/{49},{297}/{400}={11}/{9}\times{243}/{400}$.
Similarly,
\begin{equation}\label{FFG-K0}
{K_0(\rho)=-\frac{3k_{\rm{F}}^2}{5M}+\frac{9B\sigma(\sigma-1)}{\sigma+1}{u}^{\sigma}
+\frac{1}{2}C_{\rm{tot}}{u}\left[\frac{96a}{49}\frac{k_{\rm{F}}}{\Lambda}
+\frac{297b}{200}\left(\frac{k_{\rm{F}}}{\Lambda}\right)^{2/3}\right],}
\end{equation}
and
\begin{equation}\label{FFG-L}
{L(\rho)=\frac{k_{\rm{F}}^2}{3M}+\frac{3}{4}A_{\rm{d}}{u}
-\frac{3Bx\sigma}{\sigma+1}{u}^{\sigma}+\frac{3}{2}C_{\ell}{u}
\left[1+\frac{64a}{63}\frac{k_{\rm{F}}}{\Lambda}
+\frac{121b}{120}\left(\frac{k_{\rm{F}}}{\Lambda}\right)^{2/3}\right]
-\frac{3}{2}C_{\rm{u}}{u}\left[1+\frac{16a}{21}\frac{k_{\rm{F}}}{\Lambda}
+\frac{33b}{20}\left(\frac{k_{\rm{F}}}{\Lambda}\right)^{2/3}\right].}
\end{equation}
Expressions (\ref{FFG-E0}) for $E_0(\rho)$, (\ref{FFG-p0}) for
$P_0(\rho)$, (\ref{FFG-K0}) for $K_0(\rho)$, (\ref{FFG-U0}) for
$U_0(\rho,|\v{k}|)$, (\ref{FFG-M0}) for $M_0^{\ast}(\rho)$,
(\ref{FFG-Esym}) for $E_{\rm{sym}}(\rho)$, (\ref{FFG-L}) for
$L(\rho)$ and (\ref{FFG-Usym}) for $U_{\rm{sym}}(\rho,|\v{k}|)$ are
the main results of this part.

\subsection{Derivations for Quantities in the HMT Model for $n_{\v{k}}^J(\rho,\delta)$}\label{sb_a_2}

In the case of HMT, the derivations become more involved, here we
list the main steps.
First, we have (here for simplicity, we denote $C=C_J$, etc.),
\begin{align}
&\int\d\v{k}f_J(\v{r},\v{k})\int\d\v{k}'f_{J'}(\v{r},\v{k}')\Omega(\v{k},\v{k}')\notag\\
=& \frac{\Delta'Q'^3}{3\pi^4}\int_0^{Q}\Delta k^2\d
k\left[1+\frac{4a}{7}\left(\frac{kQ'}{\Lambda^2}\right)^{1/2}
+\frac{27b}{40}\left(\frac{kQ'}{\Lambda^2}\right)^{1/3}\right]\tag{\v{I}}\\
&+\frac{\Delta'Q'^3}{3\pi^4}\int_Q^{\phi Q}\frac{CQ^4}{k^4}k^2\d
k\left[1+\frac{4a}{7}\left(\frac{kQ'}{\Lambda^2}\right)^{1/2}
+\frac{27b}{40}\left(\frac{kQ'}{\Lambda^2}\right)^{1/3}\right]\tag{\v{II}}\\
&+\frac{C'Q'^3}{\pi^4}\int_0^{Q}\Delta k^2\d k
\left[1+\frac{4a}{3}\left(\frac{kQ'}{\Lambda^2}\right)^{1/2}
+\frac{9b}{8}\left(\frac{kQ'}{\Lambda^2}\right)^{1/3}
-\frac{1}{\phi'}\left[1+\frac{4a}{3}\left(\frac{k\phi'Q'}{\Lambda^2}\right)^{1/2}
+\frac{9b}{8}\left(\frac{k\phi'Q'}{\Lambda^2}\right)^{1/3}\right]\right]\tag{\v{III}}\\
&+\frac{C'Q'^3}{\pi^4}\int_Q^{\phi Q}\frac{CQ^4}{k^4}k^2\d
k\left[1+\frac{4a}{3}\left(\frac{kQ'}{\Lambda^2}\right)^{1/2}
+\frac{9b}{8}\left(\frac{kQ'}{\Lambda^2}\right)^{1/3}
-\frac{1}{\phi'}\left[1+\frac{4a}{3}\left(\frac{k\phi'Q'}{\Lambda^2}\right)^{1/2}
+\frac{9b}{8}\left(\frac{k\phi'Q'}{\Lambda^2}\right)^{1/3}\right]\right]\tag{\v{IV}},
\end{align}
where we use the following relation
\begin{align}
\int\d\v{k}'f_{J'}(\v{r},\v{k}')\Omega(\v{k},\v{k}') =&
\frac{\Delta'Q'^3}{3\pi^2}
\left[1+\frac{4a}{7}\left(\frac{kQ'}{\Lambda^2}\right)^{1/2}
+\frac{27b}{40}\left(\frac{kQ'}{\Lambda^2}\right)^{1/3}\right]\notag\\
&\hspace*{-1cm}+\frac{C'Q'^3}{\pi^2}
\left[1+\frac{4a}{3}\left(\frac{kQ'}{\Lambda^2}\right)^{1/2}
+\frac{9b}{8}\left(\frac{kQ'}{\Lambda^2}\right)^{1/3}
-\frac{1}{\phi'}\left[1+\frac{4a}{3}\left(\frac{k\phi'Q'}{\Lambda^2}\right)^{1/2}
+\frac{9b}{8}\left(\frac{k\phi'Q'}{\Lambda^2}\right)^{1/3}\right]\right].
\end{align}
Furthermore,
\begin{align}
\v{I}=& \frac{\Delta'Q'^3}{3\pi^4}\int_0^{Q}\Delta k^2\d
k\left[1+\frac{4a}{7}\left(\frac{kQ'}{\Lambda^2}\right)^{1/2}
+\frac{27b}{40}\left(\frac{kQ'}{\Lambda^2}\right)^{1/3}\right]\notag\\
=&\frac{\Delta\Delta'Q^3Q'^3}{9\pi^4}\left[1+\frac{24a}{49}\left(\frac{QQ'}{\Lambda^2}\right)^{1/2}
+\frac{243b}{400}\left(\frac{QQ'}{\Lambda^2}\right)^{1/3}\right],\\
\v{II}=&\frac{\Delta'Q'^3}{3\pi^4}\int_Q^{\phi Q}\frac{CQ^4}{k^4}k^2\d
k\left[1+\frac{4a}{7}\left(\frac{kQ'}{\Lambda^2}\right)^{1/2}
+\frac{27b}{40}\left(\frac{kQ'}{\Lambda^2}\right)^{1/3}\right]\notag\\
=&\frac{C\Delta'Q^3Q'^3}{3\pi^4}\left[1
+\frac{8a}{7}\left(\frac{QQ'}{\Lambda^2}\right)^{1/2}
+\frac{81b}{80}\left(\frac{QQ'}{\Lambda^2}\right)^{1/3}
-\frac{1}{\phi}\left[1 +\frac{8a}{7}\left(\frac{\phi
QQ'}{\Lambda^2}\right)^{1/2} +\frac{81b}{80}\left(\frac{\phi
QQ'}{\Lambda^2}\right)^{1/3}\right]\right],\\
\v{III}=&\frac{C'Q'^3}{\pi^4}\int_0^{Q}\Delta k^2\d k
\left[1+\frac{4a}{3}\left(\frac{kQ'}{\Lambda^2}\right)^{1/2}
+\frac{9b}{8}\left(\frac{kQ'}{\Lambda^2}\right)^{1/3}
-\frac{1}{\phi'}\left[1+\frac{4a}{3}\left(\frac{k\phi'Q'}{\Lambda^2}\right)^{1/2}
+\frac{9b}{8}\left(\frac{k\phi'Q'}{\Lambda^2}\right)^{1/3}\right]\right]\notag\\
=&\frac{\Delta C'Q^3Q'^3}{3\pi^4}\left[1
+\frac{8a}{7}\left(\frac{QQ'}{\Lambda^2}\right)^{1/2}
+\frac{81b}{80}\left(\frac{QQ'}{\Lambda^2}\right)^{1/3}
-\frac{1}{\phi'}\left[1 +\frac{8a}{7}\left(\frac{\phi'
QQ'}{\Lambda^2}\right)^{1/2} +\frac{81b}{80}\left(\frac{\phi'
QQ'}{\Lambda^2}\right)^{1/3}\right]\right],
\end{align}
and
\begin{align}
\v{IV}=&\frac{C'Q'^3}{\pi^4}\int_Q^{\phi
Q}\frac{CQ^4}{k^4}k^2\d
k\left[1+\frac{4a}{3}\left(\frac{kQ'}{\Lambda^2}\right)^{1/2}
+\frac{9b}{8}\left(\frac{kQ'}{\Lambda^2}\right)^{1/3}
-\frac{1}{\phi'}\left[1+\frac{4a}{3}\left(\frac{k\phi'Q'}{\Lambda^2}\right)^{1/2}
+\frac{9b}{8}\left(\frac{k\phi'Q'}{\Lambda^2}\right)^{1/3}\right]\right]\notag\\
=&\frac{CC'Q^3Q'^3}{\pi^4}\left[
1+\frac{8a}{3}\left(\frac{QQ'}{\Lambda^2}\right)^{1/2}
+\frac{27b}{16}\left(\frac{QQ'}{\Lambda^2}\right)^{1/3}\right.\notag\\
&\hspace*{1cm}-\frac{1}{\phi}\left[1+\frac{8a}{3}\left(\frac{\phi
QQ'}{\Lambda^2}\right)^{1/2} +\frac{27b}{16}\left(\frac{\phi
QQ'}{\Lambda^2}\right)^{1/3}\right]-\frac{1}{\phi'}\left[1+\frac{8a}{3}\left(\frac{\phi'
QQ'}{\Lambda^2}\right)^{1/2} +\frac{27b}{16}\left(\frac{\phi'
QQ'}{\Lambda^2}\right)^{1/3}\right]\notag \\
&\hspace*{1cm}\left.+
\frac{1}{\phi\phi'}\left[1+\frac{8a}{3}\left(\frac{\phi\phi'
QQ'}{\Lambda^2}\right)^{1/2} +\frac{27b}{16}\left(\frac{\phi\phi'
QQ'}{\Lambda^2}\right)^{1/3}\right]\right].
\end{align}

In SNM, $ C=C'=C_0,\Delta'=\Delta=\Delta_0,\phi=\phi'=\phi_0$. And
at the same time, part $\v{II}$ equals to part $\v{III}$, thus
\begin{align}
\iint\cdots=&\frac{\Delta_0^2k_{\rm{F}}^6}{9\pi^4}\left[1+\frac{24a}{49}\frac{k_{\rm{F}}}{\Lambda}
+\frac{243b}{400}\left(\frac{k_{\rm{F}}}{\Lambda}\right)^{2/3}\right]\notag\\
&
+\frac{2C_0\Delta_0k_{\rm{F}}^6}{3\pi^4}\left[1+\frac{8a}{7}\frac{k_{\rm{F}}}{\Lambda}
+\frac{81b}{80}\left(\frac{k_{\rm{F}}}{\Lambda}\right)^{2/3}
-\frac{1}{\phi_0}\left[1+\frac{8a\phi_0^{1/2}}{7}\frac{k_{\rm{F}}}{\Lambda}
+\frac{81b\phi_0^{1/3}}{80}\left(\frac{k_{\rm{F}}}{\Lambda}\right)^{2/3}\right]\right]\notag\\
&+\frac{C_0^2k_{\rm{F}}^6}{\pi^4}\left[
1+\frac{8a}{3}\frac{k_{\rm{F}}}{\Lambda}
+\frac{27b}{16}\left(\frac{k_{\rm{F}}}{\Lambda}\right)^{2/3}
-\frac{2}{\phi_0}\left[1+\frac{8a\phi_0^{1/2}}{3}\frac{k_{\rm{F}}}{\Lambda}
+\frac{27b\phi_0^{1/3}}{16}\left(\frac{k_{\rm{F}}}{\Lambda}\right)^{2/3}\right]\right.\notag\\
&\hspace*{3cm}\left.+\frac{1}{\phi_0^2}\left[1+\frac{8a}{3}\frac{\phi_0k_{\rm{F}}}{\Lambda}
+\frac{27b}{16}\left(\frac{\phi_0k_{\rm{F}}}{\Lambda}\right)^{2/3}\right]
\right].
\end{align}
Consequently,
\begin{align}
E_0^{\rm{mom}}(\rho)=
&\frac{2C_{\rm{tot}}}{\rho\rho_0}\iint\cdots\notag\\
=&\frac{1}{2}C_{\rm{tot}}\Delta_0^2{u}\left[1+\frac{24a}{49}\frac{k_{\rm{F}}}{\Lambda}
+\frac{243b}{400}\left(\frac{k_{\rm{F}}}{\Lambda}\right)^{2/3}\right]\notag\\
&+3C_{\rm{tot}}C_0\Delta_0{u}\left[1+\frac{8a}{7}\frac{k_{\rm{F}}}{\Lambda}
+\frac{81b}{80}\left(\frac{k_{\rm{F}}}{\Lambda}\right)^{2/3}
-\frac{1}{\phi_0}\left[1+\frac{8a\phi_0^{1/2}}{7}\frac{k_{\rm{F}}}{\Lambda}
+\frac{81b\phi_0^{1/3}}{80}\left(\frac{k_{\rm{F}}}{\Lambda}\right)^{2/3}\right]\right]\notag\\
&+\frac{9}{2}C_{\rm{tot}}C_0^2{u}\left[
1+\frac{8a}{3}\frac{k_{\rm{F}}}{\Lambda}
+\frac{27b}{16}\left(\frac{k_{\rm{F}}}{\Lambda}\right)^{2/3}
-\frac{2}{\phi_0}\left[1+\frac{8a\phi_0^{1/2}}{3}\frac{k_{\rm{F}}}{\Lambda}
+\frac{27b\phi_0^{1/3}}{16}\left(\frac{k_{\rm{F}}}{\Lambda}\right)^{2/3}\right]\right.\notag\\
&\hspace*{3cm}\left.+\frac{1}{\phi_0^2}\left[1+\frac{8a}{3}\frac{\phi_0k_{\rm{F}}}{\Lambda}
+\frac{27b}{16}\left(\frac{\phi_0k_{\rm{F}}}{\Lambda}\right)^{2/3}\right]\right].
\end{align}
It can be rewritten as
\begin{equation}
E_0^{\rm{mom}}(\rho)=\frac{1}{2}C_{\rm{tot}}{u}
\left[\Pi_0+\Pi_1^2\frac{24a}{49}\frac{k_{\rm{F}}}{\Lambda}+\Pi_2^2\frac{243b}{400}\left(\frac{k_{\rm{F}}}{\Lambda}\right)^{2/3}\right]
,\end{equation} after some straightforward calculations, we obtain
\begin{equation}
\Pi_0=1,~~\Pi_1=\Delta_0+7C_0\left(1-\frac{1}{\sqrt{\phi_0}}\right),~~
\Pi_2=\Delta_0+5C_0\left(1-\frac{1}{\phi_0^{2/3}}\right).
\end{equation}
Thus,
\begin{equation}
E_0^{\rm{mom}}(\rho)=\frac{1}{2}C_{\rm{tot}}{u}
\left[1+\frac{24a}{49}\left[\Delta_0+7C_0\left(1-\frac{1}{\sqrt{\phi_0}}\right)\right]^2\frac{k_{\rm{F}}}{\Lambda}+
\frac{243b}{400}\left[\Delta_0+5C_0\left(1-\frac{1}{\phi_0^{2/3}}\right)\right]^2\left(\frac{k_{\rm{F}}}{\Lambda}\right)^{2/3}\right].
\end{equation}
Adding the kinetic and the density-dependent terms, we then have
\begin{align}\label{HMT-E0}
E_0(\rho)=&\frac{3k_{\rm{F}}^2}{10M}\left(5\phi_0+\frac{3}{\phi_0}-8\right)
+\frac{1}{4}A_{\rm{tot}}{u}+\frac{B}{\sigma+1}{u}^{\sigma}\notag\\
&+\frac{1}{2}C_{\rm{tot}}{u}
\left[1+\frac{24a}{49}\left[\Delta_0+7C_0\left(1-\frac{1}{\sqrt{\phi_0}}\right)\right]^2\frac{k_{\rm{F}}}{\Lambda}+
\frac{243b}{400}\left[\Delta_0+5C_0\left(1-\frac{1}{\phi_0^{2/3}}\right)\right]^2\left(\frac{k_{\rm{F}}}{\Lambda}\right)^{2/3}\right].
\end{align}
The factor $5\phi_0+{3}/{\phi_0}-8$ characterizes the HMT.

Similarly,
\begin{align}\label{HMT-p0}
P_0(\rho)/\rho=&\frac{k_{\rm{F}}^2}{5M}\left(5\phi_0+\frac{3}{\phi_0}-8\right)
+\frac{1}{4}A_{\rm{tot}}{u}
+\frac{B\sigma}{\sigma+1}{u}^{\sigma}\notag\\
&+\frac{1}{2}C_{\rm{tot}}{u}
\left[1+\frac{32a}{49}\left[\Delta_0+7C_0\left(1-\frac{1}{\sqrt{\phi_0}}\right)\right]^2\frac{k_{\rm{F}}}{\Lambda}+
\frac{297b}{400}\left[\Delta_0+5C_0\left(1-\frac{1}{\phi_0^{2/3}}\right)\right]^2\left(\frac{k_{\rm{F}}}{\Lambda}\right)^{2/3}\right],
\end{align}
and
\begin{align}\label{HMT-K0}
K_0(\rho)=&-\frac{3k_{\rm{F}}^2}{5M}\left(5\phi_0+\frac{3}{\phi_0}-8\right)
+\frac{9B\sigma(\sigma-1)}{\sigma+1}{u}^{\sigma}\notag\\
&+\frac{1}{2}C_{\rm{tot}}{u}
\left[\frac{96a}{49}\left[\Delta_0+7C_0\left(1-\frac{1}{\sqrt{\phi_0}}\right)\right]^2\frac{k_{\rm{F}}}{\Lambda}+
\frac{297b}{200}\left[\Delta_0+5C_0\left(1-\frac{1}{\phi_0^{2/3}}\right)\right]^2\left(\frac{k_{\rm{F}}}{\Lambda}\right)^{2/3}\right].
\end{align}
The nucleon potential in the HMT model is,
\begin{align}
U_0^{\rm{mom}}(\rho,|\v{k}|)=
&\frac{2C_{\rm{tot}}}{\rho_0}\frac{\Delta_0k_{\rm{F}}^3}{3\pi^2}\left[1+\frac{4a}{7}\left(\frac{|\v{k}|k_{\rm{F}}}{\Lambda^2}\right)^{1/2}+
\frac{27b}{40}\left(\frac{|\v{k}|k_{\rm{F}}}{\Lambda^2}\right)^{1/3}\right]\notag\\
&+\frac{2C_{\rm{tot}}}{\rho_0}\frac{C_0k_{\rm{F}}^3}{\pi^2} \left[
1+\frac{4a}{3}\left(\frac{|\v{k}|k_{\rm{F}}}{\Lambda^2}\right)^{1/2}+
\frac{9b}{8}\left(\frac{|\v{k}|k_{\rm{F}}}{\Lambda^2}\right)^{1/3}
-\frac{1}{\phi_0} \left[
1+\frac{4a}{3}\left(\frac{|\v{k}|\phi_0k_{\rm{F}}}{\Lambda^2}\right)^{1/2}+
\frac{9b}{8}\left(\frac{|\v{k}|\phi_0k_{\rm{F}}}{\Lambda^2}\right)^{1/3}\right]\right]\notag\\
=&\Delta_0 C_{\rm{tot}}{u}
\left[1+\frac{4a}{7}\left(\frac{|\v{k}|k_{\rm{F}}}{\Lambda^2}\right)^{1/2}+
\frac{27b}{40}\left(\frac{|\v{k}|k_{\rm{F}}}{\Lambda^2}\right)^{1/3}\right]\notag\\
&+3C_0C_{\rm{tot}}{u}\left[
1+\frac{4a}{3}\left(\frac{|\v{k}|k_{\rm{F}}}{\Lambda^2}\right)^{1/2}+
\frac{9b}{8}\left(\frac{|\v{k}|k_{\rm{F}}}{\Lambda^2}\right)^{1/3}
-\frac{1}{\phi_0} \left[
1+\frac{4a}{3}\left(\frac{|\v{k}|\phi_0k_{\rm{F}}}{\Lambda^2}\right)^{1/2}+
\frac{9b}{8}\left(\frac{|\v{k}|\phi_0k_{\rm{F}}}{\Lambda^2}\right)^{1/3}\right]\right]\notag\\
=&C_{\rm{tot}}{u}\left[1+\frac{4a}{7}
\left[\Delta_0+7C_0\left(1-\frac{1}{\sqrt{\phi_0}}\right)\right]\left(\frac{|\v{k}|k_{\rm{F}}}{\Lambda^2}\right)^{1/2}
+\frac{27b}{40}\left[\Delta_0+5C_0\left(1-\frac{1}{\phi_0^{2/3}}\right)\right]\left(\frac{|\v{k}|k_{\rm{F}}}{\Lambda^2}
\right)^{1/3} \right],
\end{align}
i.e.,
\begin{align}\label{HMT-U0}
U_0(\rho,|\v{k}|)=&\frac{1}{2}A_{\rm{tot}}{u}+B{u}^{\sigma}\notag\\
&+C_{\rm{tot}}{u}\left[1+\frac{4a}{7}
\left[\Delta_0+7C_0\left(1-\frac{1}{\sqrt{\phi_0}}\right)\right]\left(\frac{|\v{k}|k_{\rm{F}}}{\Lambda^2}\right)^{1/2}
+\frac{27b}{40}\left[\Delta_0+5C_0\left(1-\frac{1}{\phi_0^{2/3}}\right)\right]\left(\frac{|\v{k}|k_{\rm{F}}}{\Lambda^2}
\right)^{1/3} \right].
\end{align}
The effective mass becomes,
\begin{align}\label{HMT-M0}
M_0^{\ast}(\rho)/M
=\left[1+C_{\rm{tot}}M{u}\left[\frac{2a}{7}\left[\Delta_0+7C_0\left(1-\frac{1}{\sqrt{\phi_0}}\right)\right]\frac{1}{k_{\rm{F}}\Lambda}
+\frac{9b}{40}\left[\Delta_0+5C_0\left(1-\frac{1}{\phi_0^{2/3}}\right)\right]\left(\frac{1}{\Lambda^2k_{\rm{F}}^4}\right)^{1/3}\right]\right]^{-1}.
\end{align}

In order to obtain the symmetry energy, for the first part,
\begin{align}
E_{\rm{I}}^{\rm{mom}}(\rho,\delta) =&
\frac{C_\ell\Delta_{\rm{p}}^2k_{\rm{F}}^{\rm{p},6}}{9\pi^4\rho\rho_0}\left[1+\frac{24a}{49}\frac{k_{\rm{F}}^{\rm{p}}}{\Lambda}+
\frac{243b}{400}\left(\frac{k_{\rm{F}}^{\rm{p}}}{\Lambda}\right)^{2/3}\right]
+\frac{C_\ell\Delta_{\rm{n}}^2k_{\rm{F}}^{\rm{n},6}}{9\pi^4\rho\rho_0}\left[1+\frac{24a}{49}\frac{k_{\rm{F}}^{\rm{n}}}{\Lambda}+
\frac{243b}{400}\left(\frac{k_{\rm{F}}^{\rm{n}}}{\Lambda}\right)^{2/3}\right]\notag\\
&+\frac{2C_{\rm{u}}\Delta_{\rm{p}}\Delta_{\rm{n}}k_{\rm{F}}^{\rm{p},3}k_{\rm{F}}^{\rm{n},3}}{9\pi^4\rho\rho_0}\left[1+\frac{24a}{49}\left(\frac{k_{\rm{F}}^{\rm{p}}k_{\rm{F}}^{\rm{n}}}{\Lambda^2}
\right)^{1/2}+
\frac{243b}{400}\left(\frac{k_{\rm{F}}^{\rm{p}}k_{\rm{F}}^{\rm{n}}}{\Lambda^2}\right)^{1/3}\right]\notag\\
=&\frac{1}{4}C_\ell\Delta_{\rm{p}}^2(1-\delta)^2{u}\left[1+\frac{24a}{49}\frac{k_{\rm{F}}^{\rm{p}}}{\Lambda}+
\frac{243b}{400}\left(\frac{k_{\rm{F}}^{\rm{p}}}{\Lambda}\right)^{2/3}\right]
+\frac{1}{4}C_\ell\Delta_{\rm{n}}^2(1+\delta)^2{u}\left[1+\frac{24a}{49}\frac{k_{\rm{F}}^{\rm{n}}}{\Lambda}+
\frac{243b}{400}\left(\frac{k_{\rm{F}}^{\rm{n}}}{\Lambda}\right)^{2/3}\right]\notag\\
&+\frac{1}{2}C_{\rm{u}}\Delta_{\rm{n}}\Delta_{\rm{p}}(1-\delta^2){u}\left[1+\frac{24a}{49}\left(\frac{k_{\rm{F}}^{\rm{p}}k_{\rm{F}}^{\rm{n}}}{\Lambda^2}
\right)^{1/2}+
\frac{243b}{400}\left(\frac{k_{\rm{F}}^{\rm{p}}k_{\rm{F}}^{\rm{n}}}{\Lambda^2}\right)^{1/3}\right],
\end{align}
which could be worked out to give,
\begin{equation}\label{HMT-Esym-I}
E_{\rm{sym,I}}^{\rm{mom}}(\rho)=\frac{1}{4}C_\ell\Delta_0^2{u}\left[
Y_{10}+Y_{11}a\frac{k_{\rm{F}}}{\Lambda}+Y_{12}b\left(\frac{k_{\rm{F}}}{\Lambda}\right)^{2/3}
\right] +\frac{1}{4}C_{\rm{u}}\Delta_0^2{u} \left[
Z_{10}+Z_{11}a\frac{k_{\rm{F}}}{\Lambda}+Z_{12}b\left(\frac{k_{\rm{F}}}{\Lambda}\right)^{2/3}
\right],
\end{equation}
where
\begin{equation}\label{Def_Delta012}
\Delta_0=1-3C_0\left(1-\frac{1}{\phi_0}\right),~~\Delta_0\Delta_1=-\frac{3C_0(C_1(\phi_0-1)+\phi_1)}{\phi_0},
~~\Delta_0\Delta_2=-\frac{3C_0\phi_1(C_1-\phi_1)}{\phi_0},
\end{equation}
and,
\begin{align}
Y_{10} =& 2 \Delta_1^{2}+8 \Delta_1 +4 \Delta_2 +2,~~Y_{11}=\frac{32}{21}+\frac{32}{7} \Delta_1 +\frac{48}{49} \Delta_1^{2}+\frac{96}{49} \Delta_2,~~Y_{1}=\frac{33}{20}+\frac{27}{5} \Delta_1 +\frac{243}{200} \Delta_1^{2}+\frac{243}{100} \Delta_2,\label{def_Y1}\\
Z_{10}=&-2 \Delta_1^{2}+4 \Delta_2 -2,~~Z_{11}=-\frac{8}{7}-\frac{48 \Delta_1^{2}}{49}+\frac{96 \Delta_2}{49},~~Z_{12}=-\frac{27}{20}-\frac{243 \Delta_1^{2}}{200}+\frac{243 \Delta_2}{100}\label{def_Z1}.
\end{align}
Notice that, in the FFG limit, $\phi_0=1$ and $\phi_1=0$, then $
\Delta_0=1,\Delta_1=\Delta_2=0$, consequently,
\begin{equation}
Y_{10}\to2,~~Y_{11}\to\frac{32}{21},~~Y_{12}\to\frac{33}{20},~~Z_{10}\to-2,~~Z_{11}\to-\frac{8}{7},~~Z_{12}\to-\frac{27}{20},
\end{equation}
just the FFG results, see (\ref{FFG-Esym}).

Similarly,
\begin{align}
E_{\rm{II}}^{\rm{mom}}(\rho,\delta)= &\frac{C_\ell
C_{\rm{p}}\Delta_{\rm{p}}k_{\rm{F}}^{\rm{p},6}}{3\pi^4\rho\rho_0}
\left[1+\frac{8a}{7}\frac{k_{\rm{F}}^{\rm{p}}}{\Lambda}+
\frac{81b}{80}\left(\frac{k_{\rm{F}}^{\rm{p}}}{\Lambda}\right)^{2/3}-\frac{1}{\phi_{\rm{p}}}
\left[1+\frac{8a\phi_{\rm{p}}^{1/2}}{7}\frac{k_{\rm{F}}^{\rm{p}}}{\Lambda}+
\frac{81b\phi_{\rm{p}}^{1/3}}{80}\left(\frac{k_{\rm{F}}^{\rm{p}}}{\Lambda}\right)^{2/3}\right]\right]\notag\\
&+\frac{C_\ell
C_{\rm{n}}\Delta_{\rm{n}}k_{\rm{F}}^{\rm{n},6}}{3\pi^4\rho\rho_0}
\left[1+\frac{8a}{7}\frac{k_{\rm{F}}^{\rm{n}}}{\Lambda}+
\frac{81b}{80}\left(\frac{k_{\rm{F}}^{\rm{n}}}{\Lambda}\right)^{2/3}-\frac{1}{\phi_{\rm{n}}}
\left[1+\frac{8a\phi_{\rm{n}}^{1/2}}{7}\frac{k_{\rm{F}}^{\rm{n}}}{\Lambda}+
\frac{81b\phi_{\rm{n}}^{1/3}}{80}\left(\frac{k_{\rm{F}}^{\rm{n}}}{\Lambda}\right)^{2/3}\right]\right]\notag\\
&\hspace*{-2cm}+\frac{C_{\rm{u}}C_{\rm{p}}\Delta_{\rm{n}}k_{\rm{F}}^{\rm{p},3}k_{\rm{F}}^{\rm{n},3}}{3\pi^4\rho\rho_0}
\left[1+\frac{8a}{7}\left(\frac{k_{\rm{F}}^{\rm{p}}k_{\rm{F}}^{\rm{n}}}{\Lambda^2}\right)^{1/2}+
\frac{81b}{80}\left(\frac{k_{\rm{F}}^{\rm{p}}k_{\rm{F}}^{\rm{n}}}{\Lambda^2}\right)^{1/3}
-\frac{1}{\phi_{\rm{p}}}\left[1+\frac{8a}{7}\left(\frac{\phi_{\rm{p}}k_{\rm{F}}^{\rm{p}}k_{\rm{F}}^{\rm{n}}}{\Lambda^2}\right)^{1/2}+
\frac{81b}{80}\left(\frac{\phi_{\rm{p}}k_{\rm{F}}^{\rm{p}}k_{\rm{F}}^{\rm{n}}}{\Lambda^2}\right)^{1/3}\right]\right]\notag\\
&\hspace*{-2cm}+\frac{C_{\rm{u}}C_{\rm{n}}\Delta_{\rm{p}}k_{\rm{F}}^{\rm{p},3}k_{\rm{F}}^{\rm{n},3}}{3\pi^4\rho\rho_0}
\left[1+\frac{8a}{7}\left(\frac{k_{\rm{F}}^{\rm{p}}k_{\rm{F}}^{\rm{n}}}{\Lambda^2}\right)^{1/2}+
\frac{81b}{80}\left(\frac{k_{\rm{F}}^{\rm{p}}k_{\rm{F}}^{\rm{n}}}{\Lambda^2}\right)^{1/3}
-\frac{1}{\phi_{\rm{n}}}\left[1+\frac{8a}{7}\left(\frac{\phi_{\rm{n}}k_{\rm{F}}^{\rm{p}}k_{\rm{F}}^{\rm{n}}}{\Lambda^2}\right)^{1/2}+
\frac{81b}{80}\left(\frac{\phi_{\rm{n}}k_{\rm{F}}^{\rm{p}}k_{\rm{F}}^{\rm{n}}}{\Lambda^2}\right)^{1/3}\right]\right],
\end{align}
or,
\begin{align}
E_{\rm{II}}^{\rm{mom}}(\rho,\delta)=&\frac{3}{4}C_\ell\Delta_{\rm{p}}C_{\rm{p}}(1-\delta)^2{u}\left[1+\frac{8a}{7}\frac{k_{\rm{F}}^{\rm{p}}}{\Lambda}+
\frac{81b}{80}\left(\frac{k_{\rm{F}}^{\rm{p}}}{\Lambda}\right)^{2/3}-\frac{1}{\phi_{\rm{p}}}
\left[1+\frac{8a\phi_{\rm{p}}^{1/2}}{7}\frac{k_{\rm{F}}^{\rm{p}}}{\Lambda}+
\frac{81b\phi_{\rm{p}}^{1/3}}{80}\left(\frac{k_{\rm{F}}^{\rm{p}}}{\Lambda}\right)^{2/3}\right]\right]\notag\\
&+\frac{3}{4}C_\ell\Delta_{\rm{n}}C_{\rm{n}}(1+\delta)^2{u}\left[1+\frac{8a}{7}\frac{k_{\rm{F}}^{\rm{n}}}{\Lambda}+
\frac{81b}{80}\left(\frac{k_{\rm{F}}^{\rm{n}}}{\Lambda}\right)^{2/3}-\frac{1}{\phi_{\rm{n}}}
\left[1+\frac{8a\phi_{\rm{n}}^{1/2}}{7}\frac{k_{\rm{F}}^{\rm{n}}}{\Lambda}+
\frac{81b\phi_{\rm{n}}^{1/3}}{80}\left(\frac{k_{\rm{F}}^{\rm{n}}}{\Lambda}\right)^{2/3}\right]\right]\notag\\
&+\frac{3}{4}C_{\rm{u}}\Delta_{\rm{n}}C_{\rm{p}}(1-\delta^2){u}\left[1+\frac{8a}{7}\left(\frac{k_{\rm{F}}^{\rm{p}}k_{\rm{F}}^{\rm{n}}}{\Lambda^2}\right)^{1/2}+
\frac{81b}{80}\left(\frac{k_{\rm{F}}^{\rm{p}}k_{\rm{F}}^{\rm{n}}}{\Lambda^2}\right)^{1/3}\right.\notag\\
&\hspace*{4cm}\left.-\frac{1}{\phi_{\rm{p}}}\left[1+\frac{8a}{7}\left(\frac{\phi_{\rm{p}}k_{\rm{F}}^{\rm{p}}k_{\rm{F}}^{\rm{n}}}{\Lambda^2}\right)^{1/2}+
\frac{81b}{80}\left(\frac{\phi_{\rm{p}}k_{\rm{F}}^{\rm{p}}k_{\rm{F}}^{\rm{n}}}{\Lambda^2}\right)^{1/3}\right]\right]\notag\\
&+\frac{3}{4}C_{\rm{u}}\Delta_{\rm{p}}C_{\rm{n}}(1-\delta^2){u}
\left[1+\frac{8a}{7}\left(\frac{k_{\rm{F}}^{\rm{p}}k_{\rm{F}}^{\rm{n}}}{\Lambda^2}\right)^{1/2}+
\frac{81b}{80}\left(\frac{k_{\rm{F}}^{\rm{p}}k_{\rm{F}}^{\rm{n}}}{\Lambda^2}\right)^{1/3}\right.\notag\\
&\hspace*{4cm}
\left.-\frac{1}{\phi_{\rm{n}}}\left[1+\frac{8a}{7}\left(\frac{\phi_{\rm{n}}k_{\rm{F}}^{\rm{p}}k_{\rm{F}}^{\rm{n}}}{\Lambda^2}\right)^{1/2}+
\frac{81b}{80}\left(\frac{\phi_{\rm{n}}k_{\rm{F}}^{\rm{p}}k_{\rm{F}}^{\rm{n}}}{\Lambda^2}\right)^{1/3}\right]\right],
\end{align}
which can be wrote out to order $\delta^2$, i.e.,
\begin{equation}\label{HMT-Esym-II}
E_{\rm{sym,II}}^{\rm{mom}}(\rho)=\frac{3}{4}C_\ell\Delta_0C_0{u}\left[
Y_{20}+Y_{21}a\frac{k_{\rm{F}}}{\Lambda}+Y_{22}b\left(\frac{k_{\rm{F}}}{\Lambda}\right)^{2/3}
\right] +\frac{3}{4}C_{\rm{u}}\Delta_0C_0{u} \left[
Z_{20}+Z_{21}a\frac{k_{\rm{F}}}{\Lambda}+Z_{22}b\left(\frac{k_{\rm{F}}}{\Lambda}\right)^{2/3}
\right],
\end{equation}
with
\begin{align}
Y_{20}=&4\Delta_1 +2\Delta_1 {C_1} +2+4{C_1} +2\Delta_2 -\frac{4 \Delta_1}{\phi_0}-\frac{2 \Delta_1 {C_1}}{\phi_0}+\frac{2 \Delta_1 \phi_1}{\phi_0}\notag\\
&-\frac{4 {C_1}}{\phi_0}+\frac{2 {C_1} \phi_1}{\phi_0}-\frac{2 \Delta_2}{\phi_0}-\frac{2 \phi_1^{2}}{\phi_0}-\frac{2}{\phi_0}+\frac{4 \phi_1}{\phi_0},\label{def_Y20}\\
Y_{21}=&\frac{16 \Delta_1 {C_1}}{7}-\frac{16 \Delta_1 {C_1}}{7 \sqrt{\phi_0}}+\frac{16 {C_1}}{3}+\frac{8 {C_1} \phi_1}{7 \sqrt{\phi_0}}+\frac{16 \Delta_1}{3}+\frac{8 \Delta_1 \phi_1}{7 \sqrt{\phi_0}}\notag\\
&+\frac{16 \Delta_2}{7}-\frac{6 \phi_1^{2}}{7 \sqrt{\phi_0}}-\frac{16 {C_1}}{3 \sqrt{\phi_0}}-\frac{16 \Delta_1}{3 \sqrt{\phi_0}}-\frac{16 \Delta_2}{7 \sqrt{\phi_0}}+\frac{32}{9}+\frac{8 \phi_1}{3 \sqrt{\phi_0}}-\frac{32}{9 \sqrt{\phi_0}},\label{def_Y21}\\
Y_{22}=&\frac{81 \Delta_1 {C_1}}{40}+\frac{9 {C_1}}{2}+\frac{9 \Delta_1}{2}+\frac{81 \Delta_2}{40}+\frac{27 \Delta_1 \phi_1}{20 \phi_0^{{2}/{3}}}+\frac{11}{4}-\frac{9 \phi_1^{2}}{8 \phi_0^{{2}/{3}}}-\frac{9 {C_1}}{2 \phi_0^{{2}/{3}}}\notag\\
&-\frac{81 \Delta_1 {C_1}}{40 \phi_0^{{2}/{3}}}-\frac{9 \Delta_1}{2 \phi_0^{{2}/{3}}}-\frac{81 \Delta_2}{40 \phi_0^{{2}/{3}}}+\frac{27 {C_1} \phi_1}{20 \phi_0^{{2}/{3}}}+\frac{3 \phi_1}{\phi_0^{{2}/{3}}}-\frac{11}{4 \phi_0^{{2}/{3}}},\label{def_Y22}\\
Z_{20}=&\frac{2}{\phi_0}-2+2 \Delta_2 -2 \Delta_1 {C_1} +\frac{2 \Delta_1 {C_1}}{\phi_0}-\frac{2 \Delta_1 \phi_1}{\phi_0}-\frac{2 \Delta_2}{\phi_0}+\frac{2 {C_1} \phi_1}{\phi_0}-\frac{2 \phi_1^{2}}{\phi_0},\label{def_Z20}\\
Z_{21}=&-\frac{16 \Delta_1 {C_1}}{7}+\frac{16 \Delta_1 {C_1}}{7 \sqrt{\phi_0}}+\frac{8 {C_1} \phi_1}{7 \sqrt{\phi_0}}-\frac{8 \Delta_1 \phi_1}{7 \sqrt{\phi_0}}+\frac{16 \Delta_2}{7}-\frac{6 \phi_1^{2}}{7 \sqrt{\phi_0}}-\frac{16 \Delta_2}{7 \sqrt{\phi_0}}-\frac{8}{3}+\frac{8}{3 \sqrt{\phi_0}},\label{def_Z21}\\
Z_{22}=&-\frac{81 \Delta_1 {C_1}}{40}+\frac{81 \Delta_2}{40}+\frac{81 \Delta_1 {C_1}}{40 \phi_0^{{2}/{3}}}+\frac{27 {C_1} \phi_1}{20 \phi_0^{{2}/{3}}}-\frac{27 \Delta_1 \phi_1}{20 \phi_0^{{2}/{3}}}-\frac{9}{4}-\frac{9 \phi_1^{2}}{8 \phi_0^{{2}/{3}}}-\frac{81 \Delta_2}{40 \phi_0^{{2}/{3}}}+\frac{9}{4 \phi_0^{{2}/{3}}}.\label{def_Z22}
\end{align}
It is still obvious that in the FFG model,
\begin{equation}
Y_{2j}\to0,~~Z_{2j}\to0,~~j=0,1,2,\end{equation} in this limit,
$E_{\rm{sym,II}}^{\rm{mom}}(\rho)\to0$.

According to the isospin
symmetry, the symmetry energy from the third part is identical to
that of the second part, i.e.,
\begin{equation}\label{HMT-Esym-III}
E_{\rm{sym,III}}^{\rm{mom}}(\rho)=E_{\rm{sym,II}}^{\rm{mom}}(\rho).
\end{equation}
The fourth part can be evaluated in a similar manner as
\begin{align}
E_{\rm{IV}}^{\rm{mom}}(\rho,\delta)= &\frac{C_\ell
C_{\rm{p}}^2k_{\rm{F}}^{\rm{p},6}}{\pi^4\rho\rho_0}\left[
1+\frac{8a}{3}\frac{k_{\rm{F}}^{\rm{p}}}{\Lambda}+\frac{27b}{16}\left(\frac{k_{\rm{F}}^{\rm{p}}}{\Lambda}\right)^{2/3}
-\frac{2}{\phi_{\rm{p}}}\left[1+\frac{8a\phi_{\rm{p}}^{1/2}}{3}\frac{k_{\rm{F}}^{\rm{p}}}{\Lambda}+\frac{27b\phi_{\rm{p}}^{1/3}}{16}\left(\frac{k_{\rm{F}}^{\rm{p}}}{\Lambda}\right)^{2/3}
\right]\right.\notag\\
&\hspace*{4cm}\left.+\frac{1}{\phi_{\rm{p}}^2}
\left[1+\frac{8a}{3}\frac{\phi_{\rm{p}}k_{\rm{F}}^{\rm{p}}}{\Lambda}+\frac{27b}{16}\left(\frac{\phi_{\rm{p}}k_{\rm{F}}^{\rm{p}}}{\Lambda}\right)^{2/3}\right]
 \right]\notag\\
&+\frac{C_\ell
C_{\rm{n}}^2k_{\rm{F}}^{\rm{n},6}}{\pi^4\rho\rho_0}\left[
1+\frac{8a}{3}\frac{k_{\rm{F}}^{\rm{n}}}{\Lambda}+\frac{27b}{16}\left(\frac{k_{\rm{F}}^{\rm{n}}}{\Lambda}\right)^{2/3}
-\frac{2}{\phi_{\rm{n}}}\left[1+\frac{8a\phi_{\rm{n}}^{1/2}}{3}\frac{k_{\rm{F}}^{\rm{n}}}{\Lambda}+\frac{27b\phi_{\rm{n}}^{1/3}}{16}\left(\frac{k_{\rm{F}}^{\rm{n}}}{\Lambda}\right)^{2/3}
\right]\right.\notag\\
&\hspace*{4cm}\left.+\frac{1}{\phi_{\rm{n}}^2}
\left[1+\frac{8a}{3}\frac{\phi_{\rm{n}}k_{\rm{F}}^{\rm{n}}}{\Lambda}+\frac{27b}{16}\left(\frac{\phi_{\rm{n}}k_{\rm{F}}^{\rm{n}}}{\Lambda}\right)^{2/3}\right]\right]\notag\\
&\hspace*{-2cm}+\frac{2C_{\rm{u}}C_{\rm{p}}C_{\rm{n}}k_{\rm{F}}^{\rm{n},3}k_{\rm{F}}^{\rm{p},3}}{\pi^4\rho\rho_0}
\left[
1+\frac{8a}{3}\left(\frac{k_{\rm{F}}^{\rm{p}}k_{\rm{F}}^{\rm{n}}}{\Lambda^2}\right)^{1/2}+\frac{27b}{16}\left(\frac{k_{\rm{F}}^{\rm{p}}k_{\rm{F}}^{\rm{n}}}{\Lambda^2}\right)^{1/3}
-\frac{1}{\phi_{\rm{p}}}\left[1+\frac{8a}{3}\left(\frac{\phi_{\rm{p}}k_{\rm{F}}^{\rm{p}}k_{\rm{F}}^{\rm{n}}}{\Lambda^2}\right)^{1/2}+
\frac{27b}{16}\left(\frac{\phi_{\rm{p}}k_{\rm{F}}^{\rm{p}}k_{\rm{F}}^{\rm{n}}}{\Lambda^2}\right)^{1/3}\right]\right.\notag\\
&\hspace*{2cm}
-\frac{1}{\phi_{\rm{n}}}\left[1+\frac{8a}{3}\left(\phi_{\rm{n}}\frac{k_{\rm{F}}^{\rm{p}}k_{\rm{F}}^{\rm{n}}}{\Lambda^2}\right)^{1/2}+
\frac{27b}{16}\left(\frac{\phi_{\rm{n}}k_{\rm{F}}^{\rm{p}}k_{\rm{F}}^{\rm{n}}}{\Lambda^2}\right)^{1/3}\right]\notag\\
&\hspace*{4cm}\left.
+\frac{1}{\phi_{\rm{p}}\phi_{\rm{n}}}\left[1+\frac{8a}{3}\left(\frac{\phi_{\rm{p}}\phi_{\rm{n}}k_{\rm{F}}^{\rm{p}}k_{\rm{F}}^{\rm{n}}}{\Lambda^2}\right)^{1/2}+
\frac{27b}{16}\left(\frac{\phi_{\rm{p}}\phi_{\rm{n}}k_{\rm{F}}^{\rm{p}}k_{\rm{F}}^{\rm{n}}}{\Lambda^2}\right)^{1/3}\right]
 \right],
\end{align}
then
\begin{equation}\label{HMT-Esym-IV}
E_{\rm{sym,IV}}^{\rm{mom}}(\rho)=\frac{9}{4}C_\ell C_0^2{u}\left[
Y_{30}+Y_{31}a\frac{k_{\rm{F}}}{\Lambda}+Y_{32}b\left(\frac{k_{\rm{F}}}{\Lambda}\right)^{2/3}
\right]+\frac{9}{4} C_{\rm{u}}C_0^2{u}\left[
Z_{30}+Z_{31}a\frac{k_{\rm{F}}}{\Lambda}+Z_{32}b\left(\frac{k_{\rm{F}}}{\Lambda}\right)^{2/3}
\right],
\end{equation}
with,
\begin{align}
Y_{30}=&2{C_1^2}-\frac{4 {C_1^2}}{\phi_0}+\frac{2 {C_1^2}}{\phi_0^{2}}+8 {C_1} +\frac{8 {C_1} \phi_1}{\phi_0}-\frac{16 {C_1}}{\phi_0}-\frac{8 {C_1} \phi_1}{\phi_0^{2}}+\frac{8 {C_1}}{\phi_0^{2}}\notag\\
&+2-\frac{4 \phi_1^{2}}{\phi_0}+\frac{8 \phi_1}{\phi_0}-\frac{4}{\phi_0}+\frac{2}{\phi_0^{2}}+\frac{6 \phi_1^{2}}{\phi_0^{2}}-\frac{8 \phi_1}{\phi_0^{2}},\label{def_Y30}\\
Y_{31}=&\frac{224}{9}{} {C_1} +\frac{224}{27}+\frac{16}{3}{} {C_1^2}+\phi_0^{-{1}/{2}}\left(-\frac{32}{3}{} {C_1^2}+\frac{32}{3}{} {C_1} {} \phi_1 -4\phi_1^{2}-\frac{448}{9}{} {C_1} +\frac{112}{9}{} \phi_1 -\frac{448}{27}\right)\notag\\
&+\frac{1}{\phi_0}\left(\frac{16}{3}{} {C_1^2}-\frac{32}{3}{} {C_1} {} \phi_1 +\frac{16}{3}{} \phi_1^{2}+\frac{224}{9}{} {C_1} -\frac{112}{9}{} \phi_1 +\frac{224}{27}\right),\label{def_Y31}\\
Y_{32}=&\frac{27 {C_1^2}}{8}+15 {C_1} -\frac{27 {C_1^2}}{4 \phi_0^{{2}/{3}}}+\frac{9 {C_1} \phi_1}{\phi_0^{{2}/{3}}}+\frac{55}{12}-\frac{15 \phi_1^{2}}{4 \phi_0^{{2}/{3}}}-\frac{30 {C_1}}{\phi_0^{{2}/{3}}}+\frac{10 \phi_1}{\phi_0^{{2}/{3}}}\notag\\
&+\frac{27 {C_1^2}}{8 \phi_0^{{4}/{3}}}-\frac{9 {C_1} \phi_1}{\phi_0^{{4}/{3}}}-\frac{55}{6 \phi_0^{{2}/{3}}}+\frac{21 \phi_1^{2}}{4 \phi_0^{{4}/{3}}}+\frac{15 {C_1}}{\phi_0^{{4}/{3}}}-\frac{10 \phi_1}{\phi_0^{{4}/{3}}}+\frac{55}{12 \phi_0^{{4}/{3}}},\label{def_Y32}\\
Z_{30}=&-\frac{2}{\phi_0^{2}}-2 {C_1^2}-2+\frac{4 {C_1^2}}{\phi_0}-\frac{4 \phi_1^{2}}{\phi_0}+\frac{4}{\phi_0}-\frac{2 {C_1^2}}{\phi_0^{2}}+\frac{2 \phi_1^{2}}{\phi_0^{2}}\label{def_Z30},\\
Z_{31}=&-\frac{16 {C_1^2}}{3}-\frac{16 {C_1^2}}{3 \phi_0}+\frac{32 {C_1^2}}{3 \sqrt{\phi_0}}+\frac{8 \phi_1^{2}}{3 \phi_0}-\frac{4 \phi_1^{2}}{\sqrt{\phi_0}}-\frac{56}{9}-\frac{56}{9 \phi_0}+\frac{112}{9 \sqrt{\phi_0}}\label{def_Z31},\\
Z_{32}=&-\frac{27 {C_1^2}}{8}+\frac{27 {C_1^2}}{4 \phi_0^{{2}/{3}}}-\frac{15}{4}-\frac{15 \phi_1^{2}}{4 \phi_0^{{2}/{3}}}-\frac{27 {C_1^2}}{8 \phi_0^{{4}/{3}}}+\frac{15}{2 \phi_0^{{2}/{3}}}+\frac{9 \phi_1^{2}}{4 \phi_0^{{4}/{3}}}-\frac{15}{4 \phi_0^{{4}/{3}}} .\label{def_Z32}
\end{align}

Combining (\ref{HMT-Esym-I}),  (\ref{HMT-Esym-II}),
(\ref{HMT-Esym-III}), and  (\ref{HMT-Esym-IV}), we obtain
\begin{align}\label{HMT-Esym}
E_{\rm{sym}}(\rho)=&\frac{k_{\rm{F}}^2}{6M}\left[1+C_0(1+3C_1)\left(5\phi_0+\frac{3}{\phi_0}-8\right)
+3C_0\phi_1\left(1+\frac{5}{3}C_1\right)\left(5\phi_0-\frac{3}{\phi_0}\right)+\frac{27C_0\phi_1^2}{5\phi_0}\right]
\notag\\
&+\frac{1}{4}C_\ell\Delta_0^2{u}\left[
Y_{10}+Y_{11}a\frac{k_{\rm{F}}}{\Lambda}+Y_{12}b\left(\frac{k_{\rm{F}}}{\Lambda}\right)^{2/3}
\right] +\frac{1}{4}C_{\rm{u}}\Delta_0^2{u} \left[
Z_{10}+Z_{11}a\frac{k_{\rm{F}}}{\Lambda}+Z_{12}b\left(\frac{k_{\rm{F}}}{\Lambda}\right)^{2/3}
\right]\notag\\
&+\frac{3}{2}C_\ell\Delta_0C_0{u}\left[
Y_{20}+Y_{21}a\frac{k_{\rm{F}}}{\Lambda}+Y_{22}b\left(\frac{k_{\rm{F}}}{\Lambda}\right)^{2/3}
\right] +\frac{3}{2}C_{\rm{u}}\Delta_0C_0{u} \left[
Z_{20}+Z_{21}a\frac{k_{\rm{F}}}{\Lambda}+Z_{22}b\left(\frac{k_{\rm{F}}}{\Lambda}\right)^{2/3}
\right]\notag\\
&+\frac{9}{4}C_\ell C_0^2{u}\left[
Y_{30}+Y_{31}a\frac{k_{\rm{F}}}{\Lambda}+Y_{32}b\left(\frac{k_{\rm{F}}}{\Lambda}\right)^{2/3}
\right]+\frac{9}{4} C_{\rm{u}}C_0^2{u}\left[
Z_{30}+Z_{31}a\frac{k_{\rm{F}}}{\Lambda}+Z_{32}b\left(\frac{k_{\rm{F}}}{\Lambda}\right)^{2/3}
\right]\notag\\
&+\frac{1}{4}A_{\rm{d}}{u}-\frac{Bx}{\sigma+1}{u}^{\sigma},\end{align}
and its slope parameter given by
\begin{align}\label{HMT-L}
L(\rho)=&\frac{k_{\rm{F}}^2}{3M}\left[1+C_0(1+3C_1)\left(5\phi_0+\frac{3}{\phi_0}-8\right)
+3C_0\phi_1\left(1+\frac{5}{3}C_1\right)\left(5\phi_0-\frac{3}{\phi_0}\right)+\frac{27C_0\phi_1^2}{5\phi_0}\right]\notag\\
&+\frac{3}{4}C_\ell\Delta_0^2{u}\left[
Y_{10}+\frac{4}{3}Y_{11}a\frac{k_{\rm{F}}}{\Lambda}+\frac{11}{9}Y_{12}b\left(\frac{k_{\rm{F}}}{\Lambda}\right)^{2/3}
\right]+\frac{3}{4}C_{\rm{u}}\Delta_0^2{u} \left[
Z_{10}+\frac{4}{3}Z_{11}a\frac{k_{\rm{F}}}{\Lambda}+\frac{11}{9}Z_{12}b\left(\frac{k_{\rm{F}}}{\Lambda}\right)^{2/3}
\right]\notag\\
&+\frac{9}{2}C_\ell\Delta_0C_0{u}\left[
Y_{20}+\frac{4}{3}Y_{21}a\frac{k_{\rm{F}}}{\Lambda}+\frac{11}{9}Y_{22}b\left(\frac{k_{\rm{F}}}{\Lambda}\right)^{2/3}
\right] +\frac{9}{2}C_{\rm{u}}\Delta_0C_0{u} \left[
Z_{20}+\frac{4}{3}Z_{21}a\frac{k_{\rm{F}}}{\Lambda}+\frac{11}{9}Z_{22}b\left(\frac{k_{\rm{F}}}{\Lambda}\right)^{2/3}
\right]\notag\\
&+\frac{27}{4}C_\ell C_0^2{u}\left[
Y_{30}+\frac{4}{3}Y_{31}a\frac{k_{\rm{F}}}{\Lambda}+\frac{11}{9}Y_{32}b\left(\frac{k_{\rm{F}}}{\Lambda}\right)^{2/3}
\right]+\frac{27}{4} C_{\rm{u}} C_0^2{u}\left[
Z_{30}+\frac{4}{3}Z_{31}a\frac{k_{\rm{F}}}{\Lambda}+\frac{11}{9}Z_{32}b\left(\frac{k_{\rm{F}}}{\Lambda}\right)^{2/3}
\right]\notag\\
&+\frac{3}{4}A_{\rm{d}}{u}-\frac{3Bx\sigma}{\sigma+1}{u}^{\sigma}.
\end{align}

Finally, the symmetry potential is given by
\begin{align}
U_{\rm{sym}}(\rho,|\v{k}|)=&\frac{1}{2}A_{\rm{d}}{u}-\frac{2Bx}{\sigma+1}{u}^{\sigma}+
C_{\rm{d}}{u}\left[\gamma_0+\gamma_1a\left(\frac{|\v{k}|k_{\rm{F}}}{\Lambda^2}\right)^{1/2}
+\gamma_2b\left(\frac{|\v{k}|k_{\rm{F}}}{\Lambda^2}\right)^{1/3}\right],\label{HMT-Usym}
\end{align}
with the coefficients $\gamma_0,\gamma_1$ and $\gamma_2$ given by,
\begin{align}
\gamma_0=&3C_0C_1+\Delta_0\Delta_1+3C_0+\Delta_0-\frac{3C_0}{\phi_0}\left(C_1-\phi_1+1\right),\label{def_gamma0}\\
\gamma_1=&-\frac{4 {C_0} {C_1}}{\sqrt{\phi_0}}+4 {C_0} {C_1} +\frac{2 {C_0} \phi_1}{\sqrt{\phi_0}}+\frac{4 \Delta_0 \Delta_1}{7}-\frac{14 {C_0}}{3 \sqrt{\phi_0}}+\frac{14 {C_0}}{3}+\frac{2 \Delta_0}{3},\label{def_gamma1}\\
\gamma_2=&-\frac{27 {C_0} {C_1}}{8 \phi_0^{{2}/{3}}}+\frac{27 {C_0} {C_1}}{8}+\frac{9 {C_0} \phi_1}{4 \phi_0^{{2}/{3}}}+\frac{27 \Delta_0 \Delta_1}{40}-\frac{15 {C_0}}{4 \phi_0^{{2}/{3}}}+\frac{15 {C_0}}{4}+\frac{3 \Delta_0}{4}.\label{def_gamma2}
\end{align}
For the FFG model, we then have $\gamma_0=1,\gamma_1=2/3$ and $\gamma_2=4/3$ (see (\ref{FFG-Usym})).
Expressions (\ref{HMT-E0}) for $E_0(\rho)$, (\ref{HMT-p0}) for
$P_0(\rho)$, (\ref{HMT-K0}) for $K_0(\rho)$, (\ref{HMT-U0}) for
$U_0(\rho,|\v{k}|)$, (\ref{HMT-M0}) for $M_0^{\ast}(\rho)$,
(\ref{HMT-Esym}) for $E_{\rm{sym}}(\rho)$, (\ref{HMT-L}) for
$L(\rho)$ and (\ref{HMT-Usym}) for $U_{\rm{sym}}(\rho,|\v{k}|)$ are
the main results for the HMT model.

\subsection{EOS $E(\rho,\delta)$ of ANM}\label{sb_a3}

We finally write out the expressions for the EOS $E(\rho,\delta)$ of ANM by collecting all the relevant formulas, for future potential uses.
For the HMT model, we have
\begin{align}
E(\rho,\delta)=&E^{\rm{kin}}(\rho,\delta)+E^{\rm{m}}(\rho,\delta)+E^{\rm{d}}(\rho,\delta),\label{HMT-Erd}
\end{align}
with the kinetic part given by,
\begin{align}
E^{\rm{kin}}(\rho,\delta)
=&\frac{3}{5}\frac{k_{\rm{F}}^2}{2M}
\frac{1}{2}\left[\left(1+C_{\rm{p}}\left(5\phi_{\rm{p}}+\frac{3}{\phi_{\rm{p}}}-8\right)\right)(1-\delta)^{5/3}+
\left(1+C_{\rm{n}}\left(5\phi_{\rm{n}}+\frac{3}{\phi_{\rm{n}}}-8\right)\right)(1+\delta)^{5/3}
\right],\label{HMT_Erd1}
\end{align}
and the terms originated from the momentum-dependent and the density-dependent parts of the potential are,
\begin{align}
E^{\rm{m}}(\rho,\delta)
=&\frac{1}{4}C_\ell\Delta_{\rm{p}}^2(1-\delta)^2\left(\frac{\rho}{\rho_0}\right)\left[1+\frac{24a}{49}\frac{k_{\rm{F}}^{\rm{p}}}{\Lambda}+
\frac{243b}{400}\left(\frac{k_{\rm{F}}^{\rm{p}}}{\Lambda}\right)^{2/3}\right]
+\frac{1}{4}C_\ell\Delta_{\rm{n}}^2(1+\delta)^2\left(\frac{\rho}{\rho_0}\right)\left[1+\frac{24a}{49}\frac{k_{\rm{F}}^{\rm{n}}}{\Lambda}+
\frac{243b}{400}\left(\frac{k_{\rm{F}}^{\rm{n}}}{\Lambda}\right)^{2/3}\right]\notag\\
&+\frac{1}{2}C_{\rm{u}}\Delta_{\rm{n}}\Delta_{\rm{p}}(1-\delta^2)\left(\frac{\rho}{\rho_0}\right)\left[1+\frac{24a}{49}\left(\frac{k_{\rm{F}}^{\rm{p}}k_{\rm{F}}^{\rm{n}}}{\Lambda^2}
\right)^{1/2}+
\frac{243b}{400}\left(\frac{k_{\rm{F}}^{\rm{p}}k_{\rm{F}}^{\rm{n}}}{\Lambda^2}\right)^{1/3}\right]\notag\\
&+\frac{3}{2}C_\ell\Delta_{\rm{p}}C_{\rm{p}}(1-\delta)^2\left(\frac{\rho}{\rho_0}\right)\left[1+\frac{8a}{7}\frac{k_{\rm{F}}^{\rm{p}}}{\Lambda}+
\frac{81b}{80}\left(\frac{k_{\rm{F}}^{\rm{p}}}{\Lambda}\right)^{2/3}-\frac{1}{\phi_{\rm{p}}}
\left[1+\frac{8a\phi_{\rm{p}}^{1/2}}{7}\frac{k_{\rm{F}}^{\rm{p}}}{\Lambda}+
\frac{81b\phi_{\rm{p}}^{1/3}}{80}\left(\frac{k_{\rm{F}}^{\rm{p}}}{\Lambda}\right)^{2/3}\right]\right]\notag\\
&+\frac{3}{2}C_\ell\Delta_{\rm{n}}C_{\rm{n}}(1+\delta)^2\left(\frac{\rho}{\rho_0}\right)\left[1+\frac{8a}{7}\frac{k_{\rm{F}}^{\rm{n}}}{\Lambda}+
\frac{81b}{80}\left(\frac{k_{\rm{F}}^{\rm{n}}}{\Lambda}\right)^{2/3}-\frac{1}{\phi_{\rm{n}}}
\left[1+\frac{8a\phi_{\rm{n}}^{1/2}}{7}\frac{k_{\rm{F}}^{\rm{n}}}{\Lambda}+
\frac{81b\phi_{\rm{n}}^{1/3}}{80}\left(\frac{k_{\rm{F}}^{\rm{n}}}{\Lambda}\right)^{2/3}\right]\right]\notag\\
&+\frac{3}{2}C_{\rm{u}}\Delta_{\rm{n}}C_{\rm{p}}(1-\delta^2)\left(\frac{\rho}{\rho_0}\right)\left[1+\frac{8a}{7}\left(\frac{k_{\rm{F}}^{\rm{p}}k_{\rm{F}}^{\rm{n}}}{\Lambda^2}\right)^{1/2}+
\frac{81b}{80}\left(\frac{k_{\rm{F}}^{\rm{p}}k_{\rm{F}}^{\rm{n}}}{\Lambda^2}\right)^{1/3}-\frac{1}{\phi_{\rm{p}}}\left[1+\frac{8a}{7}\left(\frac{\phi_{\rm{p}}k_{\rm{F}}^{\rm{p}}k_{\rm{F}}^{\rm{n}}}{\Lambda^2}\right)^{1/2}+
\frac{81b}{80}\left(\frac{\phi_{\rm{p}}k_{\rm{F}}^{\rm{p}}k_{\rm{F}}^{\rm{n}}}{\Lambda^2}\right)^{1/3}\right]\right]\notag\\
&+\frac{3}{2}C_{\rm{u}}\Delta_{\rm{p}}C_{\rm{n}}(1-\delta^2)\left(\frac{\rho}{\rho_0}\right)
\left[1+\frac{8a}{7}\left(\frac{k_{\rm{F}}^{\rm{p}}k_{\rm{F}}^{\rm{n}}}{\Lambda^2}\right)^{1/2}+
\frac{81b}{80}\left(\frac{k_{\rm{F}}^{\rm{p}}k_{\rm{F}}^{\rm{n}}}{\Lambda^2}\right)^{1/3}-\frac{1}{\phi_{\rm{n}}}\left[1+\frac{8a}{7}\left(\frac{\phi_{\rm{n}}k_{\rm{F}}^{\rm{p}}k_{\rm{F}}^{\rm{n}}}{\Lambda^2}\right)^{1/2}+
\frac{81b}{80}\left(\frac{\phi_{\rm{n}}k_{\rm{F}}^{\rm{p}}k_{\rm{F}}^{\rm{n}}}{\Lambda^2}\right)^{1/3}\right]\right]\notag\\
&+\frac{9}{4}C_{\ell}C_{\rm{p}}^2(1-\delta)^2\left(\frac{\rho}{\rho_0}\right)\left[
1+\frac{8a}{3}\frac{k_{\rm{F}}^{\rm{p}}}{\Lambda}+\frac{27b}{16}\left(\frac{k_{\rm{F}}^{\rm{p}}}{\Lambda}\right)^{2/3}
-\frac{2}{\phi_{\rm{p}}}\left[1+\frac{8a\phi_{\rm{p}}^{1/2}}{3}\frac{k_{\rm{F}}^{\rm{p}}}{\Lambda}+\frac{27b\phi_{\rm{p}}^{1/3}}{16}\left(\frac{k_{\rm{F}}^{\rm{p}}}{\Lambda}\right)^{2/3}
\right]\right.\notag\\
&\hspace*{6.cm}\left.+\frac{1}{\phi_{\rm{p}}^2}
\left[1+\frac{8a}{3}\frac{\phi_{\rm{p}}k_{\rm{F}}^{\rm{p}}}{\Lambda}+\frac{27b}{16}\left(\frac{\phi_{\rm{p}}k_{\rm{F}}^{\rm{p}}}{\Lambda}\right)^{2/3}\right]
 \right]\notag\\
&+\frac{9}{4}C_{\ell}C_{\rm{n}}^2(1+\delta)^2\left(\frac{\rho}{\rho_0}\right)
\left[
1+\frac{8a}{3}\frac{k_{\rm{F}}^{\rm{n}}}{\Lambda}+\frac{27b}{16}\left(\frac{k_{\rm{F}}^{\rm{n}}}{\Lambda}\right)^{2/3}
-\frac{2}{\phi_{\rm{n}}}\left[1+\frac{8a\phi_{\rm{n}}^{1/2}}{3}\frac{k_{\rm{F}}^{\rm{n}}}{\Lambda}+\frac{27b\phi_{\rm{n}}^{1/3}}{16}\left(\frac{k_{\rm{F}}^{\rm{n}}}{\Lambda}\right)^{2/3}
\right]\right.\notag\\
&\hspace*{6.cm}\left.+\frac{1}{\phi_{\rm{n}}^2}
\left[1+\frac{8a}{3}\frac{\phi_{\rm{n}}k_{\rm{F}}^{\rm{n}}}{\Lambda}+\frac{27b}{16}\left(\frac{\phi_{\rm{n}}k_{\rm{F}}^{\rm{n}}}{\Lambda}\right)^{2/3}\right]\right]\notag\\
&+\frac{9}{2}C_{\rm{u}}C_{\rm{p}}C_{\rm{n}}(1-\delta^2)\left(\frac{\rho}{\rho_0}\right)
\left[
1+\frac{8a}{3}\left(\frac{k_{\rm{F}}^{\rm{p}}k_{\rm{F}}^{\rm{n}}}{\Lambda^2}\right)^{1/2}+\frac{27b}{16}\left(\frac{k_{\rm{F}}^{\rm{p}}k_{\rm{F}}^{\rm{n}}}{\Lambda^2}\right)^{1/3}\right.\notag\\
&\hspace*{6.cm}-\frac{1}{\phi_{\rm{p}}}\left[1+\frac{8a}{3}\left(\frac{\phi_{\rm{p}}k_{\rm{F}}^{\rm{p}}k_{\rm{F}}^{\rm{n}}}{\Lambda^2}\right)^{1/2}+
\frac{27b}{16}\left(\frac{\phi_{\rm{p}}k_{\rm{F}}^{\rm{p}}k_{\rm{F}}^{\rm{n}}}{\Lambda^2}\right)^{1/3}\right]\notag\\
&\hspace*{6.cm}
-\frac{1}{\phi_{\rm{n}}}\left[1+\frac{8a}{3}\left(\phi_{\rm{n}}\frac{k_{\rm{F}}^{\rm{p}}k_{\rm{F}}^{\rm{n}}}{\Lambda^2}\right)^{1/2}+
\frac{27b}{16}\left(\frac{\phi_{\rm{n}}k_{\rm{F}}^{\rm{p}}k_{\rm{F}}^{\rm{n}}}{\Lambda^2}\right)^{1/3}\right]\notag\\
&\hspace*{6.cm}\left.
+\frac{1}{\phi_{\rm{p}}\phi_{\rm{n}}}\left[1+\frac{8a}{3}\left(\frac{\phi_{\rm{p}}\phi_{\rm{n}}k_{\rm{F}}^{\rm{p}}k_{\rm{F}}^{\rm{n}}}{\Lambda^2}\right)^{1/2}+
\frac{27b}{16}\left(\frac{\phi_{\rm{p}}\phi_{\rm{n}}k_{\rm{F}}^{\rm{p}}k_{\rm{F}}^{\rm{n}}}{\Lambda^2}\right)^{1/3}\right]
 \right],\label{HMT_Erd2}\\
E^{\rm{d}}(\rho,\delta)=&\frac{A_\ell(\rho_{\rm{p}}^2+\rho_{\rm{n}}^2)}{2\rho\rho_0}
+\frac{A_{\rm{u}}\rho_{\rm{p}}\rho_{\rm{n}}}{\rho\rho_0}+\frac{B}{\sigma+1}\left(\frac{\rho}{\rho_0}\right)^{\sigma}\left(1-x\delta^2\right).\label{HMT_Erd3}
\end{align}

For the FFG model, the expression is largely simplified, i.e.,
\begin{align}
E^{\rm{kin}}(\rho,\delta)=&\frac{3}{5}\frac{k_{\rm{F}}^2}{2M}
\frac{1}{2}\left[(1-\delta)^{5/3}+
(1+\delta)^{5/3}
\right],\\
E^{\rm{m}}(\rho,\delta)=&\frac{1}{4}C_\ell\left(\frac{\rho}{\rho_0}\right)\left[(1-\delta)^2\left[1+\frac{24a}{49}\frac{k_{\rm{F}}^{\rm{p}}}{\Lambda}+
\frac{243b}{400}\left(\frac{k_{\rm{F}}^{\rm{p}}}{\Lambda}\right)^{2/3}\right]
+(1+\delta)^2\left[1+\frac{24a}{49}\frac{k_{\rm{F}}^{\rm{n}}}{\Lambda}+
\frac{243b}{400}\left(\frac{k_{\rm{F}}^{\rm{n}}}{\Lambda}\right)^{2/3}\right]\right]\notag\\
&+\frac{1}{2}C_{\rm{u}}(1-\delta^2)\left(\frac{\rho}{\rho_0}\right)\left[1+\frac{24a}{49}\left(\frac{k_{\rm{F}}^{\rm{p}}k_{\rm{F}}^{\rm{n}}}{\Lambda^2}
\right)^{1/2}+
\frac{243b}{400}\left(\frac{k_{\rm{F}}^{\rm{p}}k_{\rm{F}}^{\rm{n}}}{\Lambda^2}\right)^{1/3}\right],
\end{align}
where the density-dependent part takes the same form as that in the HMT model.
Consequently, we have
\begin{align}
E(\rho,\delta)=&\frac{3}{5}\frac{k_{\rm{F}}^2}{2M}
\frac{1}{2}\left[(1-\delta)^{5/3}+
(1+\delta)^{5/3}
\right]+\frac{A_\ell(\rho_{\rm{p}}^2+\rho_{\rm{n}}^2)}{2\rho\rho_0}
+\frac{A_{\rm{u}}\rho_{\rm{p}}\rho_{\rm{n}}}{\rho\rho_0}+\frac{B}{\sigma+1}\left(\frac{\rho}{\rho_0}\right)^{\sigma}\left(1-x\delta^2\right)\notag\\
&+\frac{1}{4}C_\ell\left(\frac{\rho}{\rho_0}\right)\left[(1-\delta)^2\left[1+\frac{24a}{49}\frac{k_{\rm{F}}^{\rm{p}}}{\Lambda}+
\frac{243b}{400}\left(\frac{k_{\rm{F}}^{\rm{p}}}{\Lambda}\right)^{2/3}\right]
+(1+\delta)^2\left[1+\frac{24a}{49}\frac{k_{\rm{F}}^{\rm{n}}}{\Lambda}+
\frac{243b}{400}\left(\frac{k_{\rm{F}}^{\rm{n}}}{\Lambda}\right)^{2/3}\right]\right]\notag\\
&+\frac{1}{2}C_{\rm{u}}(1-\delta^2)\left(\frac{\rho}{\rho_0}\right)\left[1+\frac{24a}{49}\left(\frac{k_{\rm{F}}^{\rm{p}}k_{\rm{F}}^{\rm{n}}}{\Lambda^2}
\right)^{1/2}+
\frac{243b}{400}\left(\frac{k_{\rm{F}}^{\rm{p}}k_{\rm{F}}^{\rm{n}}}{\Lambda^2}\right)^{1/3}\right].\label{FFG-Erd}
\end{align}

\section{$\Lambda$-evolution of the Model Parameters}\label{app2}

In our fitting scheme, $E_0(\rho_0),P_0(\rho_0)/\rho_0=0,K_0(\rho_0),M_0^{\ast}(\rho_0)$
together with $U_0(\rho_0,0)$ are fixed. Consequently,
the five parameters
$A_{\rm{tot}},B,C_{\rm{tot}},\sigma$ and $a$ can be (implicitly) determined
in terms of $E_0(\rho_0),\rho_0,K_0(\rho_0),M_0^{\ast}(\rho_0)$ and
$U_0(\rho_0,0)$. 
The problem of the dependence of physical
quantities on the cutoff $\Lambda$ can be put in the following manner:
What is the value of the single-nucleon potential at an arbitrary
scale $W$? Considering the expressions of the above
quantities, e.g., in the FFG model (the one in the HMT model is very similar), see Eqs.\,(\ref{FFG-E0-mm}) for
$E_0(\rho)$, (\ref{FFG-p0}) for $P_0(\rho)/\rho$, (\ref{FFG-K0}) for
$K_0(\rho)$, (\ref{FFG-U0}) for $U_0(\rho,|\v{k}|)$ and
(\ref{FFG-M0}) for $M_0^{\ast}(\rho)/M$, combining the above five
constraints at the saturation density and then taking derivatives on
both sides of these equations with respect to the cutoff $\Lambda$,
we obtain,
\begin{align}
&\frac{1}{4}\frac{\d A_{\rm{tot}}}{\d\Lambda}+ \frac{1}{1+\sigma}
\frac{\d B}{\d \Lambda}
-\frac{B}{(1+\sigma)^2}\frac{\d\sigma}{\d\Lambda}\notag\\
&\hspace*{2cm}+\frac{1}{2}\frac{\d
C_{\rm{tot}}}{\d\Lambda}\left(1+\frac{24a}{49}\theta+\frac{243b}{400}\theta^{2/3}\right)
+\frac{1}{2}\frac{C_{\rm{tot}}}{\Lambda}\left[\left(\Lambda\frac{\d
a}{\d\Lambda}-a\right)\frac{24}{49}\theta-\frac{81b}{200}\theta^{2/3}\right]=0,\\
&\frac{1}{4}\frac{\d A_{\rm{tot}}}{\d\Lambda}+ \frac{1}{1+\sigma}
\left(\sigma\frac{\d B}{\d
\Lambda}+B\frac{\d\sigma}{\d\Lambda}\right)
-\frac{B\sigma}{(1+\sigma)^2}\frac{\d\sigma}{\d\Lambda}\notag\\
&\hspace*{2cm}+\frac{1}{2}\frac{\d
C_{\rm{tot}}}{\d\Lambda}\left(1+\frac{32a}{49}\theta+\frac{297b}{400}\theta^{2/3}\right)
+\frac{1}{2}\frac{C_{\rm{tot}}}{\Lambda}\left[\left(\Lambda\frac{\d
a}{\d\Lambda}-a\right)\frac{32}{49}\theta-\frac{99b}{200}\theta^{2/3}\right]=0,\\
&\frac{9}{1+\sigma}\left[\sigma(\sigma-1)\frac{\d
B}{\d\Lambda}+B(2\sigma-1)\frac{\d\sigma}{\d\Lambda}\right]-\frac{9B\sigma(\sigma-1)}{(1+\sigma)^2}
\frac{\d\sigma}{\d\Lambda}\notag\\
&\hspace*{2cm}+\frac{1}{2}\frac{\d
C_{\rm{tot}}}{\d\Lambda}\left(\frac{96a}{49}\theta+\frac{297b}{200}\theta^{2/3}\right)
+\frac{1}{2}\frac{C_{\rm{tot}}}{\Lambda}\left[\left(\Lambda\frac{\d
a}{\d\Lambda}-a\right)\frac{96}{49}\theta-\frac{99b}{100}\theta^{2/3}\right]=0,\\
&\frac{1}{2}\frac{\d A_{\rm{tot}}}{\d\Lambda}+\frac{\d B}{\d\Lambda}
+\frac{\d C_{\rm{tot}}}{\d\Lambda}=0,\\
&\frac{\d
C_{\rm{tot}}}{\d\Lambda}\left(\frac{2a}{7}\frac{1}{k_{\rm{F}}\Lambda}+\frac{9b}{40}\frac{1}{k_{\rm{F}}^{4/3}\Lambda^{2/3}}\right)
+\frac{C_{\rm{tot}}}{\Lambda}\left[\left(\Lambda\frac{\d
a}{\d\Lambda}-a\right)\frac{2}{7}\frac{1}{k_{\rm{F}}\Lambda}-\frac{3b}{20}\frac{1}{k_{\rm{F}}^{4/3}\Lambda^{2/3}}\right]=0,
\end{align}
where $\theta$ is the $k_{\rm{F}}/\Lambda$ at $\rho_0$. 
The solutions of these equations are
\begin{equation}
\frac{\d B}{\d\Lambda}=0,~~\frac{\d\sigma}{\d\Lambda}=0,
~~
\frac{\d C_{\rm{tot}}}{\d\Lambda}-\frac{2C_{\rm{tot}}}{3\Lambda}=0, ~~
a\frac{\d
C_{\rm{tot}}}{\d\Lambda}+\frac{C_{\rm{tot}}}{\Lambda}\left(\Lambda\frac{\d
a}{\d\Lambda}-a\right)=0,
\end{equation}
from which one obtains,
\begin{equation}\label{SolRG}
\Lambda\frac{\d C_{\rm{tot}}}{\d\Lambda}=\frac{2}{3}C_{\rm{tot}},~~
\Lambda\frac{\d a}{\d\Lambda}=\frac{1}{3}a,
\end{equation}
together with $\d A_{\rm{tot}}/\d\Lambda=-2\d C_{\rm{tot}}/\d\Lambda$. 
The $\Lambda$-dependence of the single-nucleon potential at any momentum
scale $W$ can be obtained as,
\begin{align}
\frac{\d U_0(\rho_0,W)}{\d\Lambda} =&\frac{1}{2}\frac{\d
A_{\rm{tot}}}{\d\Lambda}+\frac{\d C_{\rm{tot}}}{\d\Lambda}
+\frac{\d
C_{\rm{tot}}}{\d\Lambda}\left[\frac{4a}{7}\frac{(Wk_{\rm{F}})^{1/2}}{\Lambda}+\frac{27b}{40}\frac{(Wk_{\rm{F}})^{1/3}}{\Lambda^{2/3}}\right]\notag\\
&+\frac{C_{\rm{tot}}}{\Lambda}\left[\left(\Lambda\frac{\d
a}{\d\Lambda}-a\right)\frac{4}{7}\frac{(Wk_{\rm{F}})^{1/2}}{\Lambda}
-\frac{9b}{20}\frac{(Wk_{\rm{F}})^{1/3}}{\Lambda^{2/3}}\right],
\end{align}
which could be finally shown to be \begin{equation}
\frac{\d U_0(\rho_0,W)}{\d\Lambda}=0.
\end{equation}

Solutions of Eq.\,(\ref{SolRG}) are further given as \begin{equation}\label{CC-aa}
C_{\rm{tot}}(\Lambda)=C_{\rm{tot}}^0\left(\frac{\Lambda}{\Lambda_0}\right)^{2/3},~~
a(\Lambda)=a^0\left(\frac{\Lambda}{\Lambda_0}\right)^{1/3}\end{equation} with $C_{\rm{tot}}^0,a^0$
and $\Lambda_0$ being three constants (independent of $\Lambda$).
These result means that the value of the single
nucleon potential at momentum scale $W$ is independent of the high
momentum cutoff $\Lambda$. This situation is similar as the
scale-invariance of the physical quantities studied in some quantum
field theories, i.e., the renormalization group equation control the
energy-dependence of the coupling constants while the physical
quantities such as the cross section is cutoff
independent\,\cite{Wil74}. Very similarly, the EOS of SNM can be proved to be
independent of $\Lambda$, e.g., the $E_0(\rho)$ at any reference
density $\rho_{\rm{f}}$ is independent of the cutoff $\Lambda$.
From (\ref{CC-aa}), we find although $C_{\rm{tot}}$ and $a$ both depend on the cutoff $\Lambda$, the ratio $C_{\rm{tot}}/a^2=\rm{const}.$ does not depend on $\Lambda$.
In fact, the ratio $C_{\rm{tot}}/a^2$ depends on the physical quantities like $E_0(\rho_0),\rho_0$, etc., and on the structure of the EDF (i.e., on which the model for $n_{\v{k}}^J(\rho,\delta)$ is used).

For the symmetry energy, the treatment is almost parallel, for example,
we start by the expression for the symmetry energy (\ref{HMT-Esym}), the $\Lambda$-derivative of $E_{\rm{sym}}(\rho)$ at $\rho=\rho_0$ is given by,
\begin{align}
\frac{\d E_{\rm{sym}}(\rho_0)}{\d\Lambda}
=&\frac{1}{4}\Delta_0^2u\left(\frac{\d C_{\ell}}{\d \Lambda}\kappa_1+C_{\ell}\frac{\d\kappa_1}{\d\Lambda}\right)
+\frac{1}{4}\Delta_0^2u\left(\frac{\d C_{\rm{u}}}{\d \Lambda}\kappa_2+C_{\rm{u}}\frac{\d\kappa_2}{\d\Lambda}\right)\notag\\
&+\frac{3}{2}\Delta_0C_0u\left(\frac{\d C_{\ell}}{\d \Lambda}\kappa_3+C_{\ell}\frac{\d\kappa_3}{\d\Lambda}\right)
+\frac{3}{2}\Delta_0C_0u\left(\frac{\d C_{\rm{u}}}{\d \Lambda}\kappa_4+C_{\rm{u}}\frac{\d\kappa_4}{\d\Lambda}\right)\notag\\
&+\frac{9}{4}C_0^2u\left(\frac{\d C_{\ell}}{\d \Lambda}\kappa_5+C_{\ell}\frac{\d\kappa_5}{\d\Lambda}\right)
+\frac{9}{4}C_0^2u\left(\frac{\d C_{\rm{u}}}{\d \Lambda}\kappa_6+C_{\rm{u}}\frac{\d\kappa_6}{\d\Lambda}\right)\notag\\
&
+\frac{1}{4}u\frac{\d A_{\rm{d}}}{\d\Lambda}-\frac{B}{\sigma+1}u^{\sigma}\frac{\d x}{\d\Lambda},
\end{align}
which should be zero (as it is fixed in the fitting scheme),
where $\kappa_1\sim \kappa_6$ are defined in (\ref{HMT-Esym}).
When writing out the last term here, we consider the fact that both $B$ and $\sigma$ are $\Lambda$-independent (as shown in the SNM case).
Each term in the bracket could be used to deduce the $\Lambda$-dependence of the coefficient $C_{\ell}$ or $C_{\rm{u}}$.
For example, we have
\begin{equation}
\frac{\d C_{\ell}}{\d\Lambda}\kappa_1+C_{\ell}\frac{\d\kappa_1}{\d\Lambda}
=\frac{\d C_{\ell}}{\d\Lambda}Y_{10}+Y_{11}a\frac{k_{\rm{F}}}{\Lambda}
\left(\frac{\d C_{\ell}}{\d\Lambda}-\frac{C_{\ell}}{\Lambda}+\frac{C_{\ell}}{\Lambda}\frac{\Lambda}{a}\frac{\d a}{\d\Lambda}\right)
+Y_{12}b\left(\frac{k_{\rm{F}}}{\Lambda}\right)^{2/3}\left(\frac{\d C_{\ell}}{\d\Lambda}-\frac{2}{3}\frac{C_{\ell}}{\Lambda}\right).
\end{equation}
From the last two terms, we find that
\begin{equation}
\Lambda\frac{\d C_{\ell}}{\d\Lambda}=\frac{2}{3}C_{\ell},~~a\frac{\d a}{\d\Lambda}=\frac{1}{3}a,
\end{equation}
the similar relation holds for $C_{\rm{u}}$,
which are same as those from the analysis on the EOS of SNM.

By putting the above expressions into $\d E_{\rm{sym}}(\rho_0)/\d\Lambda$, we then have
\begin{align}
\frac{\d E_{\rm{sym}}(\rho_0)}{\d\Lambda}
=&u\frac{\d C_{\ell}}{\d\Lambda}\left(\frac{1}{4}\Delta_0^2Y_{10}+\frac{3}{2}\Delta_0C_0Y_{20}+\frac{9}{4}C_0^2Y_{30}\right)
+u\frac{\d C_{\rm{u}}}{\d\Lambda}\left(\frac{1}{4}\Delta_0^2Z_{10}+\frac{3}{2}\Delta_0C_0Z_{20}+\frac{9}{4}C_0^2Z_{30}\right)\notag\\
&+\frac{1}{4}u\frac{\d A_{\rm{d}}}{\d\Lambda}-\frac{B}{\sigma+1}u^{\sigma}\frac{\d x}{\d\Lambda}.
\end{align}
Here the density scaling of the $x$-term (i.e., $u^{\sigma}$) is different from others (i.e., $u$), thus
\begin{equation}
\frac{\d x}{\d\Lambda}=0.
\end{equation}
Moreover, we obtain the $\Lambda$-dependence of the $A_{\rm{d}}$ coefficient by setting $\d E_{\rm{sym}}(\rho_0)/\d\Lambda=0$,  i.e.,
\begin{align}
\Lambda\frac{\d A_{\rm{d}}}{\d\Lambda}
=&\Lambda\frac{\d C_{\ell}}{\d\Lambda}\left(\Delta_0^2Y_{10}+6\Delta_0C_0Y_{20}+9C_0^2Y_{30}\right)
+\Lambda\frac{\d C_{\rm{u}}}{\d\Lambda}\left(\Delta_0^2Z_{10}+6\Delta_0C_0Z_{20}+9C_0^2Z_{30}\right)\notag\\
=&\frac{2 }{3}\left[C_{\ell}\left(\Delta_0^2Y_{10}+6\Delta_0C_0Y_{20}+9C_0^2Y_{30}\right)
+C_{\rm{u}}\left(\Delta_0^2Z_{10}+6\Delta_0C_0Z_{20}+9C_0^2Z_{30}\right)\right].
\end{align}
Similarly, one can demonstrate that the symmetry energy at any reference density $\rho_{\rm{f}}$ does not changed when $\Lambda$ varies, i.e.,
\begin{equation}
\frac{\d E_{\rm{sym}}(\rho_{\rm{f}})}{\d\Lambda}=0.
\end{equation}
The slope parameter $L$ and the symmetry potential $U_{\rm{sym}}$ share the same property.

\end{widetext}


\begin{references}

\bibitem{Gale1987}C. Gale, G. Bertsch, and S. Das Gupta, Phys. Rev. C \textbf{35}, 1666 (1987).
\bibitem{Bertsch1988}G. Bertsch and S. Das Gupta, Phys. Rep. \textbf{160}, 189 (1988).
\bibitem{Pra88}M. Prakash, T.S. Kuo and S. Das Gupta, Phys.  Rev.  C \textbf{37}, 2253 (1988).
\bibitem{Welke}G. Welke \textit{et al.}, Phys. Rev. C \textbf{38}, 2101 (1988).
\bibitem{Gale90}C. Gale \textit{et al.}, Phys. Rev. C \textbf{41}, 1545 (1990).
\bibitem{Pan}Q. Pan and P. Danielewicz, Phys. Rev. Lett. \textbf{70}, 2062 (1993); \textbf{70}, 3523 (1993).
\bibitem{Zhang94}J. Zhang, S. Das Gupta and C. Gale, Phys. Rev. C \textbf{50}, 1617 (1994).
\bibitem{Aichelin1990}J. Aichelin, Phys. Rep. \textbf{202}, 233 (1991).
\bibitem{Hart} C. Hartnack \textit{et al.}, Eur. Phys. J. A \textbf{1}, 151 (1998).
\bibitem{Pawel} P. Danielewicz, Nucl. Phys.  \textbf{A673}, 375 (2000).


\bibitem{Das2003}C. Das \textit{et al.}, Phys. Rev. C \textbf{67}, 034611 (2003).

\bibitem{LiBA04} B.A. Li \textit{et al.}, Phys. Rev. C {\bf 69}, 011603 (2004); Nucl. Phys. {\bf A735}, 563 (2004).


\bibitem{LiBA98} B.A. Li, C.M. Ko, and  W. Bauer, Int. J. Mod. Phys. E \textbf{7}, 147 (1998).
\bibitem{Dan02} P. Danielewicz, R. Lacey, and W.G. Lynch, Science \textbf{298}, 1592 (2002).
\bibitem{Bar05} V. Baran \textit{et al.}, Phys. Rep. \textbf{410}, 335 (2005).
\bibitem{Ste05} A. Steiner \textit{et al.}, Phys. Rep. \textbf{411}, 325 (2005).
\bibitem{Che07a} L.W. Chen \textit{et al.}, Front. Phys. China \textbf{2}, 327 (2007).
\bibitem{LCK08} B.A. Li, L.W. Chen, and C.M. Ko, Phys. Rep. \textbf{464}, 113 (2008).
\bibitem{Tsa12} B.M. Tsang \textit{et al.}, Phys. Rev. C \textbf{86}, 105803 (2012).
\bibitem{Che14} L.W. Chen \textit{et al.}, Eur. Phys. J. A \textbf{50}, 29 (2014).
\bibitem{Gle00} N. Glendenning, \textit{Compact Stars}, 2nd edition, Spinger-Verlag New York, Inc., 2000.
\bibitem{Lat04} J. Lattimer and M. Prakash, Science \textbf{304}, 536 (2004); Phys. Rep. \textbf{442}, 109 (2007).
\bibitem{Lat12} J. Lattimer, Annu. Rev. Nucl. Part. Sci. \textbf{62}, 485 (2012).
\bibitem{Lat14}J. Lattimer and A. Steiner, Eur. Phys. J. A \textbf{50}, 40 (2014).
\bibitem{Oze16} F. $\ddot{\rm{O}}$zel and P. Freire, Ann. Rev.
Astron.  Astrophys. \textbf{54}, 401 (2016).

\bibitem{Oer17} M. Oertel \textit{et al.}, Rev. Mod. Phys.  \textbf{89}, 015007 (2017).

\bibitem{Hor14} C. Horowitz \textit{et al.}, J. Phys. G  \textbf{41}, 093001 (2014).

\bibitem{EPJA}  ``Topical issue on nuclear symmetry energy'', Eds., B.A. Li, A. Ramos, G. Verde, and I. Vida\~na, Eur. Phys. J. A {\bf 50}, 9, (2014).

\bibitem{Cen09} M. Centelles \textit{et al.}, Phys. Rev. Lett. \textbf{102}, 122502 (2009).

\bibitem{Zha13} Z. Zhang and L.W. Chen, Phys. Lett. \textbf{B726}, 234 (2013).

\bibitem{Maz13} X. Maza \textit{et al.}, Phys. Rev. C \textbf{88}, 024316 (2013).

\bibitem{Zha14} Z. Zhang and L.W. Chen, Phys. Rev. C \textbf{90},
064317 (2014).
\bibitem{Zha15} Z. Zhang and L.W. Chen, Phys. Rev. C \textbf{92}, 031301(R) (2015).

\bibitem{Dan14} P. Danielewicz and J. Lee, Nucl. Phys.  \textbf{A922}, 1 (2014).


\bibitem{Ste10} A. Steiner, J. Lattimer, and E.F. Brown, Astrophys. J. \textbf{722}, 33 (2010).

\bibitem{LiBA21}B.A. Li \textit{et al.}, Universe \textbf{7}, 182 (2021).

\bibitem{Huth22}S. Huth \textit{et al.}, Nature \textbf{606}, 276 (2022).

\bibitem{ZL21}N.B. Zhang and B.A.  Li, Astrophys. J. \textbf{921}, 111 (2021).

\bibitem{Fr22}C. Mondal and F. Gulminelli, Phys. Rev. D \textbf{105},  083016 (2022).

\bibitem{Riley21}T. Riley \textit{et al.},  Astrophys. J  {\bf 918}, L27 (2021).

\bibitem{Miller21} M. Miller \textit{et al.}  Astrophys. J  {\bf 918}, L28 (2021).

\bibitem{Xia09} Z.G. Xiao \textit{et al.}, Phys. Rev. Lett. \textbf{102}, 062502 (2009).

\bibitem{EPJA-review} B.A. Li, P. G. Krastev, D.H. Wen, and N.B. Zhang, Eur. Phys. J. A {\bf 55}, 117 (2019).

\bibitem{Xu16} J. Xu \textit{et al.}, Phys. Rev. C \textbf{93}, 044609 (2016).

\bibitem{Zhang18}Y.X. Zhang \textit{et al.}, Phys. Rev. C \textbf{97}, 034625 (2018).

\bibitem{Ono19}A. Ono \textit{et al.}, Phys. Rev. C \textbf{100}, 044617 (2019).

\bibitem{Colonna21}M. Colonna \textit{et al.}, Phys. Rev. C \textbf{104}, 024603 (2021).

\bibitem{TMEP}H. Wolter \textit{et al.}, Prog. Part. Nucl. Phys. {\bf 125}, 103962 (2022).

\bibitem{LiChen05} B.A. Li and L.W. Chen, Phys. Rev. C {\bf 72}, 064611 (2005).

\bibitem{Cozma17}M. Cozma, Eur. Phys. J. A \textbf{54}, 40 (2018).

\bibitem{WangLi}Y.J. Wang and Q.F. Li, Front. Phys. {\bf 15}, 44302 (2020). 

\bibitem{Cozma21}M. Cozma and M. Tsang, Eur. Phys. J. A \textbf{57}, 309 (2021).

\bibitem{Cai15a}B.J. Cai \textit{et al.}, Phys. Rev. C \textbf{92}, 015802 (2015).
\bibitem{Li15}B.A. Li, W.J. Guo, and Z.Z. Shi, Phys. Rev. C \textbf{91}, 044601 (2015).
\bibitem{Li15a}B.A. Li, Phys. Rev. C \textbf{92}, 034603 (2015).
\bibitem{Hen15b} O. Hen \textit{et al.}, Phys. Rev. C \textbf{91}, 025803 (2015).
\bibitem{Yon17} G.C. Yong, Phys. Lett. \textbf{B765}, 104 (2017).

\bibitem{LiBA13} B.A. Li and X. Han, Phys. Lett. \textbf{B727}, 276 (2013).

\bibitem{CXu10}C. Xu and B.A. Li, Phys. Rev. C81, 064612 (2010).
\bibitem{Lee11} H.K.  Lee, B.Y.  Park and M. Rho, Phys. Rev. C \textbf{83}, 025206 (2011).
\bibitem{Lee14}H.K.  Lee and M. Rho, Eur. Phys. J. A \textbf{50}, 14 (2014).
\bibitem{Wang12} Y.N. Wang \textit{et al.}, Prog. Theor. Phys. \textbf{127}, 739 (2012).

\bibitem{Bethe} H. Bethe, Ann. Rev. Nucl. Part. Sci.  \textbf{21}, 93 (1971).
\bibitem{Ant88} A. Antonov, P. Hodgson, and I.Zh. Petkov, \textit{Nucleon Momentum and Density Distribution in Nuclei}, Clarendon Press, Oxford, 1988.
\bibitem{Arr12}J. Arrington \textit{et al.}, Prog. Part. Nucl. Phys. \textbf{67}, 898 (2012).
\bibitem{Cio15}C. Ciofi degli Atti, Phys. Rep. \textbf{590}, 1 (2015).

\bibitem{Wei15} R. Weiss, B. Bazak, and N. Barnea, Phys. Rev. Lett. \textbf{114}, 012501 (2015).
\bibitem{Wei15a} R. Weiss, B. Bazak, and N. Barnea, Phys. Rev. C \textbf{92}, 054311 (2015).

\bibitem{Cruz2018}R.Cruz-Torres \textit{et al.}, Phys. Lett. \textbf{B785}, 304 (2018).
\bibitem{Weiss2019-1}R. Weiss \textit{et al.}, Phys. Lett. \textbf{790}, 484 (2019).
\bibitem{Weiss2019-2}R. Weiss \textit{et al.}, Phys. Lett. \textbf{791}, 242 (2019).
\bibitem{Weiss2021}R. Weiss \textit{et al.}, Phys. Rev. C \textbf{103}, 031301(L) (2021).




\bibitem{Hen14} O. Hen \textit{et al.}, Science \textbf{346}, 614 (2015).
\bibitem{Hen15} O. Hen \textit{et al.}, Phys. Rev. C \textbf{92}, 045205 (2015).
\bibitem{Col15} C. Colle \textit{et al.}, Phys. Rev. C \textbf{92}, 024604 (2015).


\bibitem{Egi06}K. Egiyan \textit{et al.}, Phys. Rev. Lett. \textbf{96}, 082501 (2006).
\bibitem{kk1} E. Piasetzky \textit{et al.}, Phys. Rev. Lett. \textbf{97}, 162504 (2006).
 \bibitem{kk2} R. Shneor \textit{et al.}, Phys. Rev. Lett. \textbf{99}, 072501 (2007).
 
 \bibitem{kk3} R. Subedi \textit{et al.}, Science \textbf{320}, 1467 (2008).
 
 \bibitem{kk4} L. Weinstein \textit{et al.}, Phys. Rev. Lett. \textbf{106}, 052301 (2011).
 \bibitem{kk5}  I. Korover \textit{et al.}, Phys. Rev. Lett. {\bf 113}, 022501 (2014).

\bibitem{Hen2017RMP}O. Hen \textit{et al.}, Rev. Mod. Phys. \textbf{89}, 045002 (2017).


\bibitem{Duer2018}M. Duer \textit{et al.}, Nature \textbf{560}, 617 (2018).
\bibitem{Schmookler2019}B. Schmookler \textit{et al.}, Nature \textbf{566}, 354 (2019).
\bibitem{Schmidt2020}A. Schmidt \textit{et al.}, Nature \textbf{578}, 540 (2020).
\bibitem{Li22} S. Li \textit{et al.}, Nature \textbf{609},  41 (2022).

\bibitem{Rio09} A. Rios, A. Polls, and W. Dickhoff, Phys. Rev. C \textbf{79}, 064308 (2009).
\bibitem{Rio14} A. Rios, A. Polls, and W. Dickhoff, Phys. Rev. C \textbf{89}, 044303 (2014).

\bibitem{ZHLi} Z.H. Li and H. Schulze, Phys. Rev. C{\bf 94}, 024322 (2016),

\bibitem{VMC} R. Wiringa \textit{et al.}, Phys. Rev. C {\bf 89}, 024305 (2014).

\bibitem{Wir16} See, e.g., R. Wiringa, \url{http://www.phy.anl.gov/theory/research/momenta/}

\bibitem{Cai16c} B.J. Cai, B.A. Li, and L.W. Chen,  Phys. Rev. C \textbf{94}, 061302 (2016).


\bibitem{CXu11} C. Xu and B.A. Li, arXiv:1104.2075 (2011).
\bibitem{CXu13} C. Xu, A. Li, B.A. Li, J. of Phys: Conference Series \textbf{420}, 012190 (2013).
\bibitem{Vid11} I. Vida$\tilde{\textrm{n}}$a, A. Polls, and C. Provid$\hat{\textrm{e}}$ncia, Phys. Rev. C \textbf{84}, 062801(R) (2011).
\bibitem{Lov11} A. Lovato \textit{et al.}, Phys. Rev. C \textbf{83}, 054003 (2011).
\bibitem{Car12} A. Carbone, A. Polls, A. Rios, Eur. Phys. Lett. \textbf{97}, 22001 (2012).

\bibitem{Car14} A. Carbone \textit{et al.}, Eur. Phys. A, (2014) {50}: 13.


\bibitem{Cai15} B.J. Cai and B.A. Li, Phys. Rev. C \textbf{92}, 011601(R) (2015).


\bibitem{Cai16b} B.J. Cai and B.A. Li, Phys. Rev. C \textbf{93},
014619 (2016).

\bibitem{Lou22a}O. Louren\c{c}o \textit{et al.}, Phys. Rev. D \textbf{106},  043010 (2022).

\bibitem{Lu2022}H. Lu, Z.Z. Ren, and D. Bai, Nucl. Phys.  \textbf{A1021}, 122408 (2022).


\bibitem{Bur22} S. Burrello and S. Typel, Eur. Phys. J. A \textbf{58}, 120 (2022).

\bibitem{Hong22} B. Hong, Z.Z. Ren, and X.L. Mu, Chin. Phys. C \textbf{6}, 065104 (2022).


\bibitem{Lou22}O. Lourenco, T. Frederico, and M. Dutra, Phys. Rev.  D \textbf{105}, 023008 (2022).


\bibitem{Lu21} H. Lu, Z.Z. Ren and D. Bai, Nucl. Phys.  \textbf{A1011}, 122200 (2021).

\bibitem{Souza20} A. Souza \textit{et al.}, Phys. Rev. C \textbf{101}, 065202 (2020).

\bibitem{Coleman15}P. Coleman, \textit{Introduction to Many-body Physics}, Cambridge University Press, 2015, Section 6.8 (FIG.\,6.11).


\bibitem{CLC18}B.J. Cai, B.A. Li, and L.W. Chen, AIP Conference Proceedings \textbf{2038}, 020041 (2018).


\bibitem{Cai14} B.J. Cai and L.W. Chen,  Nucl. Sci. Tech. \textbf{28}, 185 (2017).
\bibitem{Sel14}R. Sellahewa and A. Rios, Phys. Rev. C \textbf{90}, 054327 (2014).

\bibitem{ZL18}N.B. Zhang and B.A. Li, Astrophys. J. \textbf{859}, 90 (2018).
\bibitem{ZL19}N.B. Zhang and B.A. Li, Astrophys.  J. \textbf{879}, 99 (2019).
\bibitem{ZL22}N.B. Zhang and B.A. Li, arXiv:2208.00321 (2022).

\bibitem{Hag21}K. Hagel and J. Natowitz, arXiv:2111.09399 (2021).



\bibitem{Tan08}S.N. Tan, Ann. Phys. \textbf{323}, 2952 (2008); \textbf{323}, 2971 (2008); \textbf{323}, 2987 (2008).

\bibitem{Sch05} A. Schwenk and C. Pethick, Phys. Rev. Lett. \textbf{95}, 160401 (2005).
\bibitem{Epe09a} E. Epelbaum, H. Krebs, D. Lee, and Ulf-G. Meissner, Eur. Phys. A \textbf{40}, 199 (2009).
\bibitem{Tew13} I. Tews \textit{et al.}, Phys. Rev. Lett. \textbf{110}, 032504 (2013).
\bibitem{mm}
 T. Kr$\ddot{\textrm{u}}$ger \textit{et al.}, Phys. Rev. C \textbf{88}, 025802 (2013).
\bibitem{Gez13} A. Gezerlis \textit{et al.}, Phys. Rev. Lett. \textbf{111}, 032501 (2013).
\bibitem{Gez10} A. Gezerlis and J. Calson, Phys. Rev. C \textbf{81}, 025803 (2010).



\bibitem{Stew10} J. Stewart \textit{et al.}, Phys. Rev. Lett. \textbf{104}, 235301 (2010).
\bibitem{Kuh10} E. Kuhnle \textit{et al.}, Phys. Rev. Lett. \textbf{105}, 070402 (2010).

\bibitem{Cai16a}B.J. Cai and B.A. Li, Phys. Lett. \textbf{B759}, 79
(2016).





\bibitem{LiBA15} B.A. Li and L.W. Chen, Mod. Phys. Lett. A \textbf{30}, 1530010 (2015).

\bibitem{LiBA16} B.A. Li \textit{et al.},
Nucl. Sci. Tech. \textbf{27}, 141 (2016).


\bibitem{Hen16}O. Hen \textit{et al.}, arXiv:1608.00487 (2016).


\bibitem{CaiLi2022} B.J. Cai and B.A. Li, Phys. Rev. C \textbf{105}, 064607 (2022).

\bibitem{CaiLi2022a}B.J. Cai and B.A. Li, Ann. Phys. \textbf{444}, 169062 (2022).


\bibitem{LiBA2018PPNP}B.A. Li \textit{et al.}, Prog. Part. Nucl. Phys. \textbf{99}, 29 (2018).
\bibitem{Souza2020} A. Souza \textit{et al.}, arXiv:2004.10390 (2020).



\bibitem{YongGC2017PRC}G.C. Yong and B.A. Li, Phys. Rev. C \textbf{96}, 064614 (2017).


\bibitem{YongGC2018PLB} G.C. Yong, Phys. Lett. \textbf{B776}, 447 (2018).


\bibitem{Wang2017PRC}Z. Wang \textit{et al.}, Phys. Rev. C \textbf{96}, 054603 (2017).


\bibitem{Guo2021PRC}W.M. Guo, B.A. Li,  and G.C. Yong, Phys. Rev. C \textbf{104}, 034603 (2021).

\bibitem{Yong2022}G.C. Yong, Phys. Rev. C \textbf{105}, 011601(L) (2022).
\bibitem{Yang2019}Z.X. Yang \textit{et al.}, Phys. Rev. C \textbf{100}, 054325 (2019).




\bibitem{Bulgac2022}
A. Bulgac, arXiv:2203.12079 (2022).

\bibitem{Bulgac2022a}
A. Bulgac, arXiv:2203.04843 (2022).

\bibitem{Miller2019}G. Miller \textit{et al.}, Phys. Lett. \textbf{B793}, 360 (2019).





\bibitem{Che05} L.W. Chen, C.M. Ko, and B.A. Li, Phys. Rev. Lett.
\textbf{94}, 032701 (2005).

\bibitem{XuJ10} J. Xu and C.M. Ko, Phys. Rev. C \textbf{82}, 044311
(2010).

\bibitem{XuJ15} J. Xu, L.W. Chen, and B.A. Li, Phys. Rev. C
\textbf{91}, 014611 (2015).

\bibitem{Sto86}H. St\"ocker and W. Greiner, Phys. Rep. 137, 277 (1986).

\bibitem{GOG}J.  Decharge and D. Gogny, Phys. Rev. C \textbf{21}, 1568 (1980).

\bibitem{SURR}A. Forrester, A. Sobester, and A. Keane, \textit{Engineering Design via Surrogate Modelling: A Practical Guide}, John Wiley $\&$ Sons, 2008.


\bibitem{Ham90} S. Hama \textit{et al.}, Phys. Rev. C \textbf{41}, 2737 (1990).

\bibitem{Lan62} A.M. Lane, Nucl. Phys. \textbf{35}, 676 (1962).

\bibitem{mu04} Kh.S.A. Hassaneen and H. Muether, Phys. Rev. C {\bf 70},  054308 (2004).

\bibitem{Li04} B.A. Li, Phys. Rev. C {\bf 69}, 064602 (2004).

\bibitem{Ron06} Z.Y. Ma, J. Rong, B.Q. Chen, Z.Y. Zhu, and H.Q. Song, Phys. Lett. \textbf{B604}, 170 (2004).

\bibitem{Dal1} E.N.E. van Dalen, C. Fuchs and A. Faessler, Phys. Rev. Lett. \textbf{95}, 022302 (2005).

\bibitem{zuo05} W. Zuo, L.G. Cao, B.A. Li, U. Lombardo, and C.W. Shen, Phys. Rev. C {\bf 72}, 014005 (2005).


\bibitem{Beh11} B. Behera, T.R. Routray, and S.K. Tripathy, J. Phys. G \textbf{38}, 115104 (2011).


\bibitem{LiX13}X.H. Li \textit{et al.}, Phys. Lett. \textbf{B721}, 101 (2013).

\bibitem{LiX15}X.H. Li \textit{et al.}, Phys. Lett. \textbf{B743}, 408 (2015).


\bibitem{XuC10} C. Xu, B.A. Li, and L.W. Chen, Phys. Rev. C \textbf{82},
054607 (2010).

\bibitem{XuC11} C. Xu \textit{et al.}, Nucl. Phys. \textbf{%
A865}, 1 (2011).


\bibitem{CheR12}R. Chen \textit{et al.},
 Phys. Rev. C \textbf{85}, 024305 (2012).

\bibitem{Hug58} N.M. Hugenholtz, L. Van Hove, Physica \textbf{24}, 363 (1958).

\bibitem{You99} D. Youngblood, H. Clark, and Y.-W. Lui, Phys. Rev. Lett. \textbf{82}, 691 (1999).
\bibitem{Shl06} S. Shlomo, V. Kolomietz, and G. Col\`{o}, Eur. Phys. J. A \textbf{30}, 23 (2006).
\bibitem{Pie10} J. Piekarewicz, J. Phys. G \textbf{37}, 064038 (2010).
\bibitem{Che12} L.W. Chen and J.Z. Gu, J. Phys. G \textbf{39}, 035104 (2012).
\bibitem{Col14} G. Col\`{o}, U. Garg and H. Sagawa, Eur. Phys. J. A50, 26 (2014).

\bibitem{Stone2014PRC}J. Stone, N. Stone and S. Moszkowski,
Phys. Rev. C \textbf{89}, 044316 (2014).
\bibitem{Garg2018PPNP}U. Garg and G. Col\`{o}, Prog. Part. Nucl. Phys. \textbf{101}, 55 (2018).


\bibitem{LiBA2021PRC}
B.A. Li and W.J. Xie, Phys. Rev. C \textbf{104}, 034610 (2021).
\bibitem{XuJ2021PRC}
J. Xu, Z. Zhang, and B.A. Li, Phys. Rev. C \textbf{104}, 054324 (2021).
\bibitem{ZhangZ2021CPC}
Z. Zhang, X.B. Feng, and L.W. Chen, Chin. Phys. C \textbf{45}, 064104 (2021).

\bibitem{Chen17}L.W. Chen, Nucl. Phys. Rev. \textbf{34}, 21 (2017).

\bibitem{Cai12} B.J. Cai and L.W. Chen, Phys. Rev. C \textbf{85}, 024302 (2012).
\bibitem{Sei14} W. Seif and D. Basu, Phys. Rev. C \textbf{89}, 028801 (2014).
\bibitem{Gon17} C. Gonzalez-Boquera \textit{et al.}, Phys. Rev. C \textbf{96}, 065806 (2017).
\bibitem{PuJ17} J. Pu, Z. Zhang, and L.W. Chen, Phys. Rev. C \textbf{96}, 054311 (2017).
\bibitem{CWZC2022}B.J. Cai \textit{et al.}, arXiv:2208.10438 (2022).

\bibitem{Dri20}C. Drischler \textit{et al.}, Phys. Rev. Lett. \textbf{125}, 202702 (2020).
\bibitem{CL2021}B.J. Cai and B.A. Li, Phys. Rev. C \textbf{103}, 054611 (2021).


\bibitem{LWChen2009}L.W. Chen \textit{et al.}, Phys. Rev. C \textbf{80}, 014322 (2009).

\bibitem{Hol16} J. Holt, N. Kaiser, and G. Miller, Phys. Rev. C
\textbf{93}, 064603 (2016).


\bibitem{Kut93} M. Kutschera and W. W\'{o}jcik, Phys. Rev. C \textbf{47}, 1077 (1993).

\bibitem{Kut94} M. Kutschera, Phys. Lett.  \textbf{B340}, 1 (1994).

\bibitem{LiBA02} B.A. Li, Phys. Rev. Lett. \textbf{88}, 192701 (2002).

\bibitem{Szm06} A. Szmagli\'{n}ski, W. W\'{o}jcik, and M. Kutschera, Acta Phys. Pol. B \textbf{37}, 277 (2006).

\bibitem{wen} D.H. Wen, B.A. Li, and L.W. Chen, Phys. Rev. Lett. \textbf{103}, 211102 (2009).

\bibitem{Kho96} D.T. Khoa, W. von Oertzen, and A. Ogloblin, Nucl. Phys. \textbf{A602}, 98 (1996).

\bibitem{Bas07} D. Basu, P. Chowdhury, and C. Samanta, Nucl. Phys.  \textbf{A811}, 140 (2008).

\bibitem{Ban00} S. Banik and D. Bandyopadhyay, J. Phys. G \textbf{26}, 1495 (2000).

\bibitem{Kubis1} S. Kubis and M. Kutschera, Nucl. Phys. \textbf{A720}, 189 (2003).


\bibitem{Fri05} S. Fritsch, N. Kaiser, and W. Weise, Nucl. Phys.
\textbf{A750}, 259 (2005).


\bibitem{Hor01} C. Horowitz  and J. Piekarewicz, Phys. Rev. Lett. \textbf{86}, 5647 (2001).
\bibitem{Pro06} C. Provid\^{e}ncia \textit{et al.}, Phys. Rev. C \textbf{73}, 025805 (2006).

\bibitem{Duc08a} C. Ducoin, J. Margueron, and P. Chomaz, Nucl. Phys. \textbf{A809}, 30 (2008).

\bibitem{Duc08b} C. Ducoin \textit{et al.}, Phys. Rev. C \textbf{78}, 055801 (2008).

\bibitem{XuJ09} J. Xu \textit{et al.}, Phys. Rev. C \textbf{%
79}, 035802 (2009); Astrophys. J. \textbf{697}, 1549 (2009).

\bibitem{NBZ19}N.B. Zhang and B.A. Li, J. Phys. G: Nucl. Part. Phys. {\bf 46}, 014002 (2019).

\bibitem{Kub07} S. Kubis, Phys. Rev. C \textbf{76}, 035801 (2007); Phys.
Rev. C \textbf{70}, 065804 (2004).

\bibitem{Gear11} M. Gearheart \textit{et al.}, Mon.  Not.  Roy.  Astron.  Soc. \textbf{418}, 2343 (2011).

\bibitem{Wen12}D.H. Wen, W.G. Newton, and B.A. Li, Phys. Rev. C \textbf{85}, 025801 (2012).

\bibitem{Newton12} W.G. Newton, M. Gearheart,  and B.A. Li,  Astrophys. J. Supp. Ser. \textbf{204}, 9 (2013).

\bibitem{Newton14} W.G. Newton \textit{et al.},  Eur. Phys. J. A \textbf{50}, 41 (2014).

\bibitem{Hooker15} J. Hooker, W.G. Newton and B.A. Li,  Mon.  Not.  Roy.  Astron.  Soc. \textbf{449},  3559 (2015).

\bibitem{Zhou21} X. Zhou, A. Li and B.A. Li, Astrophys. J. \textbf{910}, 62 (2021).

\bibitem{Cac08}E. Cackett \textit{et al.},  Astrophys. J. \textbf{687}, L87 (2008).

\bibitem{AGD1960} A. Abrikosov, L. Gorkov, and I. Dzyaloshinski, 
\textit{Methods of Quantum Field Theory in Statistical Physics}, Dover Press, 1963, 
Chap.2, Chap.5.

\bibitem{LPPH} J. Lattimer \textit{et al.}, Phys. Rev. Lett. {\bf 66}, 2701 (1991).

\bibitem{Kla06} T. Kl\"{a}hn \textit{et al.}, Phys. Rev. C {\bf  74}, 035802 (2006).

\bibitem{Misner1973}C. Misner, K. Thorne, and J. Wheeler, \textit{Gravitation},  Princeton University Press, 2017, Section 23.5.

\bibitem{Hor03} J. Carriere, C.J. Horowitz, and J. Piekarewicz, Astrophys.
J. \textbf{593}, 463 (2003).

\bibitem{BPS71} G. Baym, C. Pethick, and P. Sutherland, Astrophys. J. \textbf{%
170}, 299 (1971).
\bibitem{Iida1997}
K. Iida and K. Sato, Astrophys. J. \textbf{477},
294 (1997).

\bibitem{Baiotti}L. Baiotti, Prog. Part. Nucl. Phys. {\bf 109}, 103714 (2019).

\bibitem{Capano20} C. Capano \textit{et al.}, Nat. Astron. {\bf 4}, 625 (2020).

\bibitem{David} D. Blaschke \textit{et al.},  Universe {\bf 6}, 81 (2020).

\bibitem{Kat20} K. Chatziioannou, Gen. Rel. Grav. {\bf 52}, 109 (2020).

\bibitem{AngLi}  A. Li \textit{et al.},  J. High. Ener. Astrophys.  {\bf 28}, 19 (2020).

\bibitem{Ser79}B.D. Serot, Phys. Lett. \textbf{B86}, 146 (1979).


\bibitem{Fon21} E. Fonseca \textit{et al.}, Astrophys. J. Lett.  {\bf 915}, L12 (2021).


\bibitem{Rios-G} Roshan Sellahewa and Arnau Rios,  Phys. Rev. C {\bf 90}, 054327 (2014).

\bibitem{Zha16}Z. Zhang and L.W. Chen, Phys. Rev. C \textbf{94}, 064326 (2016).
\bibitem{ZhouY2019} Y. Zhou, L.W. Chen, and Z. Zhang, Phys. Rev. D \textbf{99}, 121301(R) (2019).
\bibitem{Gon18}C. Gonzalez-Boquera \textit{et al.}, Phys. Lett. \textbf{B779}, 195 (2018).



\bibitem{Gon19}C. Gonzalez-Boquera \textit{et al.}, Phys. Rev. C \textbf{100}, 015806 (2019).

\bibitem{Lou20}O. Lourenco \textit{et al.}, Phys. Lett. \textbf{B803}, 135306 (2020).

\bibitem{Wil74} K. Wilson and J. Kogut, Phys. Rep. \textbf{12}, 75 (1974).

\bibitem{Ra1}C. A. Raithel,  F. $\ddot{\rm{O}}$zel and D. Psaltis, Astro. Phys. J {\bf 875}, 1 (2019).

\bibitem{Ra2}C.A. Raithel, D. Psaltis and  F. $\ddot{\rm{O}}$zel, Phys. Rev. D {\bf 104}, 6 (2021).

\bibitem{Most21}
E.~R.~Most and C.~A.~Raithel,
Phys. Rev. D \textbf{104}, 124012 (2021).
\bibitem{MAL}
A.~Boehnlein \textit{et al.}, Rev. Mod. Phys. \textbf{94}, 031003 (2022).
\end{references}
\end{document}